# OBSERVATIONS AND MODELS OF ECLIPSING BINARY SYSTEMS

by

Jeffrey L. Coughlin

Adviser Dr. Richard M. Williamon

A thesis submitted to the Faculty of Emory College
of Emory University in partial fulfillment
of the requirements of the degree of
Bachelor of Sciences with Honors

Department of Physics

2007

## ABSTRACT

Eclipsing binary systems form the fundamental basis of Astronomy in the sense that they are the primary means to determine fundamental stellar astrophysical quantities such as mass, radius, and temperature. Furthermore, they allow us to study the internal dynamos and resulting magnetic cycles of stars that we would normally only be able to study for one star, our Sun. The systems themselves are extremely interesting objects, consisting of a multitude of configurations that are tied together by a complex evolutionary history. Finally, they allow us to test theories of stellar structure and even General Relativity. Thus the accurate observation and modeling of these systems is of great importance to the field.

The first three chapters of this thesis are devoted to acquainting a reader with a general science background, but no knowledge of Astronomy, to eclipsing binaries and the field in general, and should provide the reader with an adequate background to understand the rest of the thesis. The subsequent eight chapters are each devoted to the analysis of eight separate systems that I have studied while at Emory, with each chapter arranged as would be generally found in a journal article. The collected data, models, and derived parameters for each system are analyzed in context to previous findings and general trends seen throughout the thesis. An evolutionary scenario for the formation of A and W type W Uma systems, with two types of near-contact systems as precursors and intermediates, is proposed.

## ACKNOWLEDGEMENTS

My sincerest thanks to my advisor, Richard Williamon, for providing me with a truly exceptional undergraduate background and inspiring me to pursue a career in Astronomy. You have always left your door wide open to my endless questions. This is the greatest quality I could have ever asked for in an adviser.

My thanks to Horace Dale for teaching me all I know about photometry, data processing, instrumentation, and the many black arts of Astronomy that are normally missed in an undergraduate education. You are a truly a renaissance man, and have taught me to never neglect the smallest detail nor task.

My thanks to J. Scott Shaw for patiently mentoring me this past summer, and introducing me to the “joys” of IRAF.

My thanks to Jason Boss for setting up the Astromac for my modeling.

My thanks to all the Astronomy students I have taught for showing me the joy of revealing the wonders of the universe to others and inspiring me to become a professor.

My thanks to the Physics professors with whom I had morning classes, for being so understanding the many times I was either half-awake or skipped class to sleep having been up all night at the telescope.

My thanks to Lowell Observatory for time on their 42” telescope, and to Scholarly Inquiry and Research at Emory (SIRE) for funding my travel to Lowell.

Finally, my thanks to Jamie for keeping me company so many times at the telescope, drawing Jeff Jr., and always accompanying me to Java Monkey to work on my thesis.

## CONTENTS

## LIST OF FIGURES

## LIST OF TABLES

# I. INTRODUCTION TO ECLIPSING BINARIES AND ASTRONOMICAL MEASUREMENTS

## 1.1 Definition of Eclipsing Binary Stars

Although we happen to live in a solar system with a single star, nearly half of all stars occur in pairs, called binary stars. Governed by the same physics that control the movement of Earth around the Sun, two stars in a binary system orbit a common center of mass, called a barycenter, in the same orbital plane. Occasionally this orbital plane is oriented such that, as viewed from Earth, the two stars eclipse, or pass in front of each other. When this arrangement is satisfied the system is referred to as an eclipsing binary. Due to the wave nature of light, and since binary stars are close together and very distant, these systems appear as point sources. Thus, one cannot actually visually differentiate between a single star and a binary system. However, when the system is eclipsing, the amount of light received at Earth continually varies as the stars go in and out of eclipse. The plot of an eclipsing binary's brightness over a complete orbital cycle is referred to as its light curve. By studying this curve, a great deal of basic astrophysical quantities can be determined.

## 1.2 System Parameters

An eclipsing binary system has a number of basic physical quantities that will affect the shape of its light curve. The period of the system is easily measured and directly determined by the mass of each component and the orbital separation, the distance between the two component's centers of mass, via the gravitational force. The mass ratio of a system is mass of one component divided by the mass of the other component. The radius of each star, along with the period, will directly determine the width of eclipses in the light curve. Although the radius of a star is usually well-correlated with its mass, in close binary systems mutual tidal forces and\or mass transfer can cause inflation. Surface temperature is the main determinant of eclipse depth, with the ratio of primary and secondary eclipse depths roughly equal to the ratio of surface temperatures to the fourth power, via the Stefan-Boltzman law. The inclination of a system is defined to be 90 degrees if viewed exactly edge on, and zero if face on. A lower inclination will result in shallower eclipse depths as the eclipses are no longer complete. Although most close binaries tend to have perfectly circular orbits due to tidal synchronization, some systems may have a measure of eccentricity. Mass and radius measurements are usually represented in multiples of solar mass and radii, represented by $M_\odot$ and $R_\odot$ respectively. An illustration of both a partially and completely eclipsing system is shown in Figure 1-1.

Just as our Sun has sunspots, so do other stars have starspots. These are regions of increased magnetic activity which act to constrain and cool the encapsulated gas, resulting in a region of lower temperature, and thus brightness. These are responsible for asymmetries in the light curve, and are defined by their location in longitude and latitude, size, and temperature ratio with respect to the surrounding surface temperature.

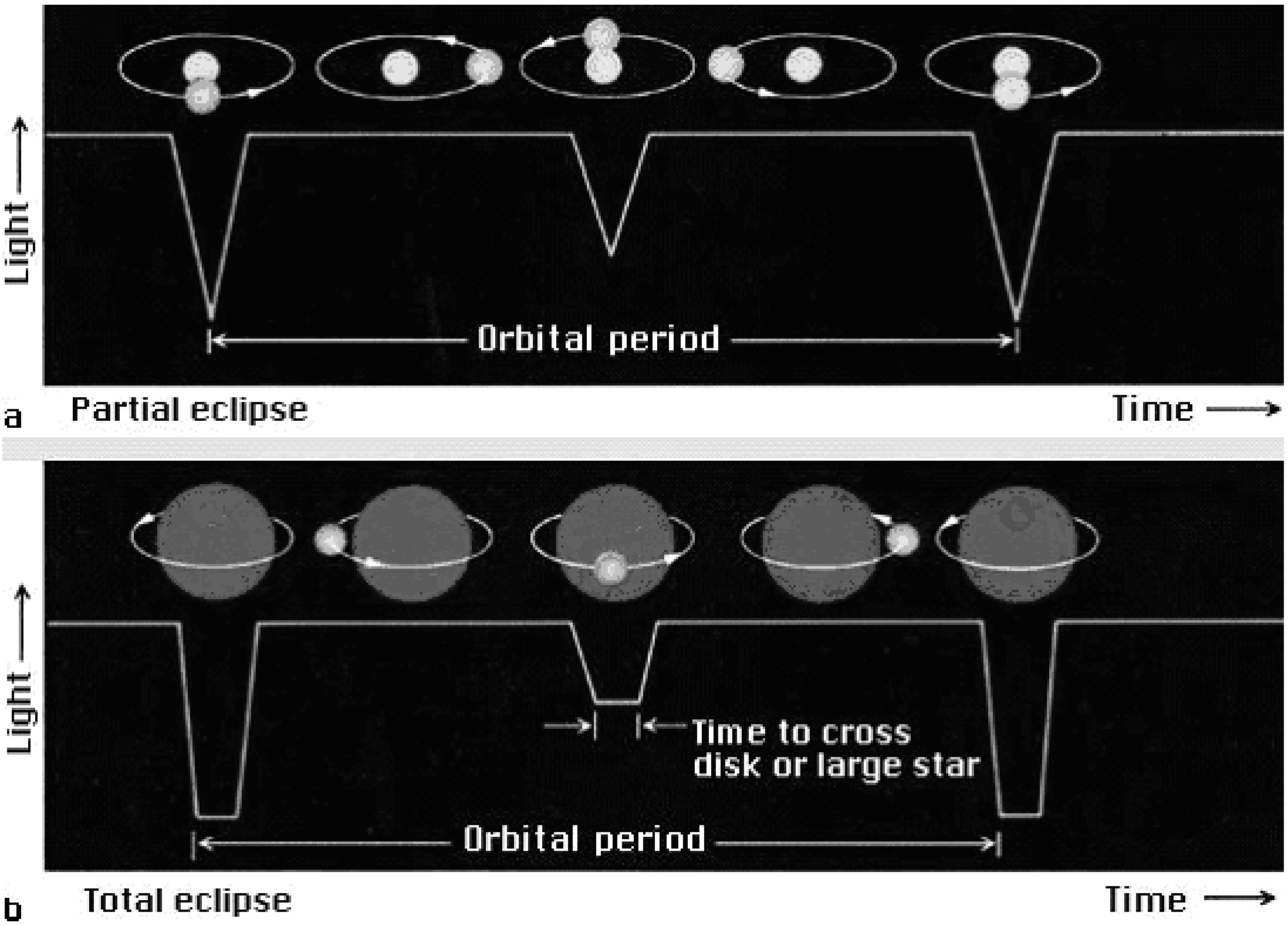

Fig 1-1: Illustration of both a partially and totally eclipsing binary system.

### 1.3 Naming Conventions and System Classification

When variable stars were first discovered, they were wrongly assumed to be extremely rare, and so the adopted naming convention was to designate the first variable star found in a constellation with the letter R, and subsequently discovered variables S, T, U, ect., followed by the shorthand name of the constellation. For example, the first variable star discovered in the constellation Andromedae was R And, the second S And, and so on until Z And was discovered. When this was reached they started with RR, RS, RT…SS, ST, SU, and so on until ZZ. They then started with AA, AB, AC…BB, BC, BD, never using the letter J to avoid confusion with I. When QZ was reached, the 334$^{th}$ combination, it was finally realized just how many variable stars exist, and so the numerical designations V335 And, V336 And, ect. were finally employed.

The broadest way to classify the physical configuration of a binary is to determine if it is a detached, semi-detached, or contact system. In a co-rotating frame of reference, one can map out the potential of the gravitational force for the system, as shown in Figure 1-2. Points for which the derivative of the potential, and thus the force, is zero are called the Lagrangian points, designated by $L_1$, $L_2$, ect. Around each star there is a teardrop shaped boundary, called the Roche Lobe, along which there is also no net force; they meet at $L_1$. If both stars are inside their Roche Lobes, then the system is classified as detached, and both components are fully physically separated and non-interacting. If just one of the stars fills its Roche Lobe, then matter will flow though the inner Lagrangian point, $L_1$, onto the other star, and the system is classified as semi-detached. If both stars fill or overfill their Roche Lobe, then they are referred to as a contact system, as their surfaces are in physical contact and they now share a common outer envelope.

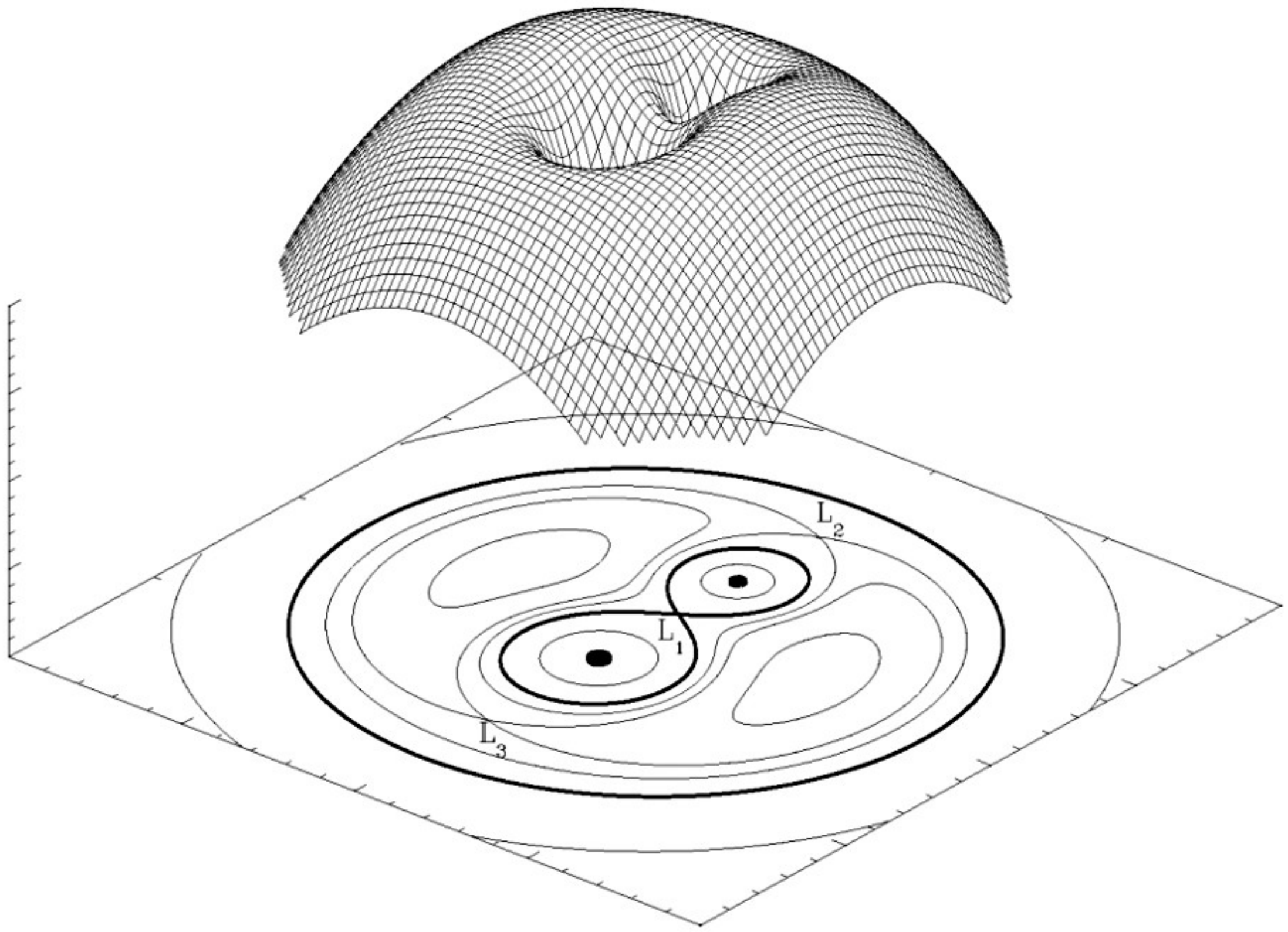

Figure 1-2: Distribution of the gravitational potential in a co-rotating reference frame of a detached binary star system with a mass ratio of 2. The inner bold lines are the Roche Lobes for each star.

If one wishes to further classify a system according to several distinguishing characteristics, it is customary to refer to it as a type of the original star found to possess those qualities. For example, the star W Uma was the first found to be a short-period contact system, and so stars that possess those qualities are referred to as W Uma type stars. Other widely used types are RS CVn type stars, which are detached systems with high spot and flaring activity, Algol type stars, which are widely detached systems, and Beta Lyrae type stars, which are massive semi-detached systems.

## 1.4 Measurements of Brightness and Time

Brightness can either be measured in units of magnitude or flux. Magnitude is an absolute unit that is both backwards and logarithmic, such that an increase of five magnitudes corresponds exactly to a hundred times drop in brightness. The bright star Vega is used as a zero point, so that every object has a set magnitude; the faintest the naked eye can see is about magnitude six. Flux is a relative unit that is that is linearly scaled, with one representing maximum brightness. Much information can be extracted by making observations in different filters, or bandpasses of the electromagnetic spectrum. The most commonly used optical system at the present is the Johnson/Cousins UBVRI system, consisting of five filters in the Ultraviolet (U), Blue (B), Visual (V), Red (R), and Infrared (I) portions of the spectrum. A plot of each filter's transmission rate versus wavelength is shown in Figure 1-3.

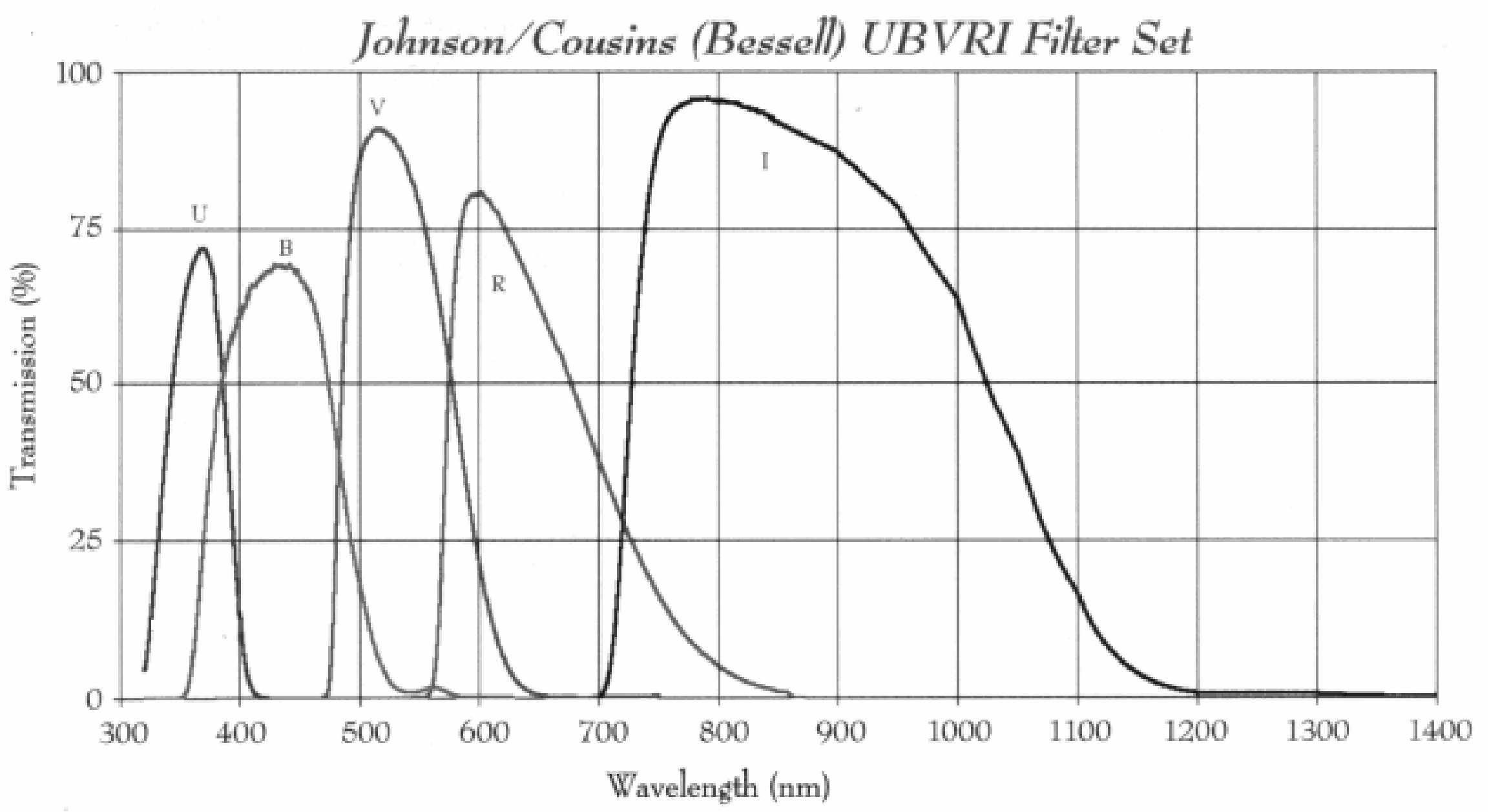

Fig. 1-3: Transmission rate vs. wavelength for the Johnson/Cousins filter set.

In Astronomy, time is conventionally marked in Heliocentric Julian Days (HJD). The Julian Date (JD) is the amount of time that has passed, in days, from noon Universal Time (UT), the time in Greenwich, England, on January 1$^{st}$, 4713 BC. This has the convenience of being a continuously running clock, unobstructed by any local time changes such as leap years or daylight savings. However, as the Earth revolves about the Sun, the distance from Earth with respect to a star system can vary up to two Astronomical Units (AU), the average distance between the Earth and Sun, as shown in Figure 1-4. This corresponds to about sixteen light-minutes, the distance light travels in a minute, and thus due to the light-time effect a yearly variation up to sixteen minutes is induced when observing a system. To correct for this error, the Sun is used as a stable reference point, and thus when the extra light travel time to Earth is added or subtracted JD is converted to Heliocentric Julian Date, or HJD.

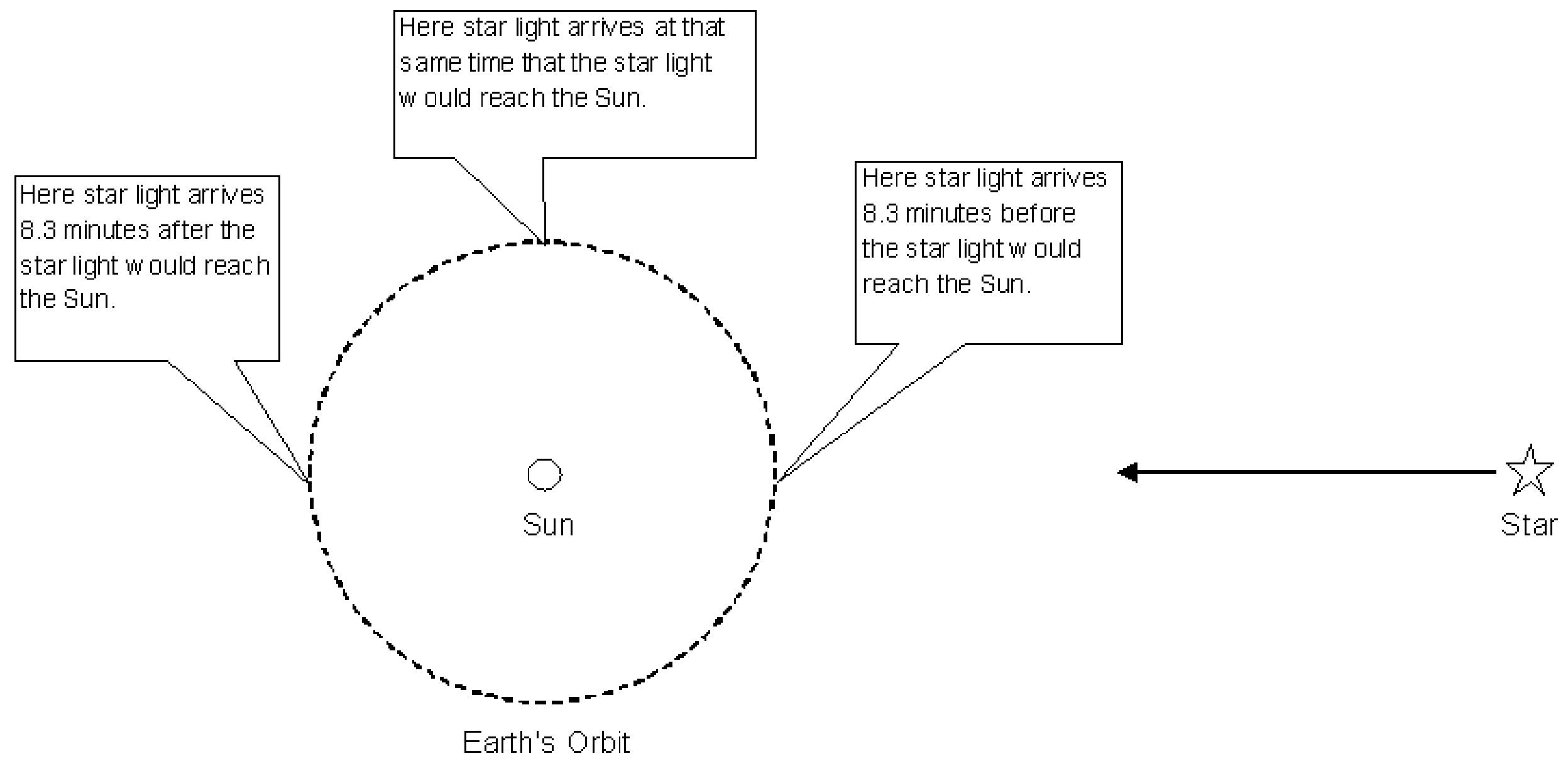

Fig 1-4: Illustration of the induced light travel time as Earth revolves about the Sun.

When dealing with light curves, units of phase are used, with zero to one representing a complete orbit containing two eclipses. When the primary star, which is the more luminous, and thus usually more massive, hotter, and larger star, is behind or eclipsed by the secondary star, the system is defined to be at a phase of zero. This is also referred to as the primary eclipse, and for non-eccentric, or perfectly circular systems, the secondary eclipse, when the primary star is in front, occurs a phase of 0.5. If the system has some eccentricity, thus following elliptical orbits, the secondary eclipse will be shifted by an amount proportional to the eccentricity. The parts of the light curve around phases 0.25 and 0.75, when the system is out of eclipse, are referred to as the shoulders. Often for graphical clarity a light curve will be drawn from zero to two. In this case the data from phase zero to one is simply repeated from one to two. If the period of a system is known, along with a time in HJD when the system was at primary eclipse, referred to as an ephemeris, data in HJD taken over a long baseline can be phased to form a light curve.

### 1.5 Stellar Spectra and Radial Velocity Curves

When the light from a star is dispersed through a prism or grating such that the variance of its intensity with respect to wavelength is plotted, it is referred to as a spectrum, and contains much information about the star. By looking at what wavelength peak emission occurs at surface temperature can be determined via Wien’s law. The excitement of certain spectral lines or presence of absorption bands indicates the chemical composition of the stars’ atmosphere, as well as further specifics about the temperature. The width of spectral lines gives an indication of the density of the atmosphere, and thus surface gravity, which is key in distinguishing between normal hydrogen-burning, main-sequence stars, helium-burning supergiants, and collapsed stellar remnants such as white dwarfs. Since both stars are viewed at once the spectrum of a binary system consists of two superimposed spectra, one from each component.

While light curves mostly contain information about the system's temperature distribution and size, radial velocity curves are needed to determine the basic parameters of component masses and thus the scale of the system. Radial velocity, the motion of a star towards or away from the observer, is measured via the Doppler Shift, which causes the spectrum of a star to be shifted towards shorter or longer wavelengths. By measuring exactly how much certain spectral line features are shifted with respect to their rest wavelengths over one orbital period the velocity of one or both stars in the system can be quantified into units such as kilometers per second. (Since the two stars in a system will always be moving in opposite directions, it is possible to separate out which spectral lines belong to which star as they shift in opposite directions.) If the absolute velocity of each star is known, along with the period of the system, each star's individual mass can be directly computed via orbital mechanics. In cases where the primary component is significantly more luminous than the secondary, the spectral lines of the secondary are overwhelmed by the primary, and not visible, and thus only the radial velocity of the primary star can be known. In these cases, a mass ratio must be first be found via light curve modeling in order to compute each individual component mass. If no spectra or radial velocity curves exist for a system, one may guess its scale by using the photometric color of the system to estimate a temperature, and then derive a mass based on that temperature using a mass-temperature relationship.

### 1.6 Telescopes and CCD Cameras

Telescopes are the main instrument in collecting observational data, and can be basically thought of as light buckets. Greater apertures allow for greater light collection areas and thus higher signal-to-noise for fainter objects. Most modern telescopes are Cassegrains, wherein light enters the telescope and reflects off a primary mirror that causes individual rays to converge toward a secondary mirror back at the top of the telescope. The secondary mirror reflects the light back down through a hole in the center of the primary mirror, where it focuses at either an eyepiece or camera. An illustration of this set-up is shown in Figure 1-5.

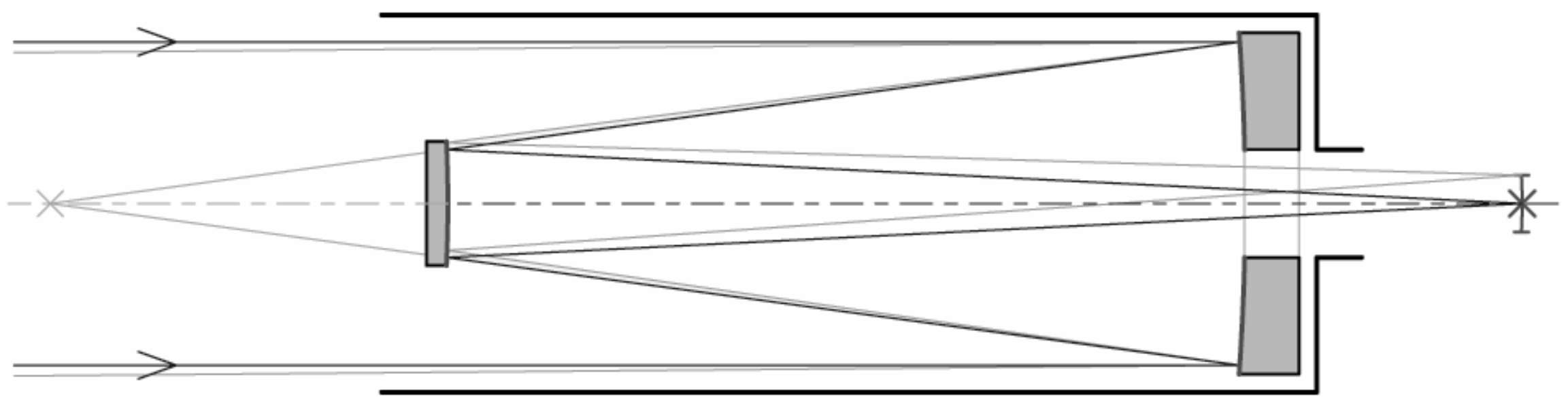

Fig. 1-5: Configuration and light paths of a Cassegrain telescope.

Prior to the digital revolution of the early 1990's, measurements were made by attaching a photomultiplier tube to the telescope, referred to as photoelectric observations. The modern instrument for accurately measuring changes in brightness is the Charge-Coupled Device (CCD) camera, whose photoactive region is usually a semi-conducting layer of silicon. When an incoming photon strikes the surface of a CCD camera, an electron is released via the photoelectric effect, and contained to a local region, a pixel, via voltage gates. Usually over 70% of incident photons cause the release of an electron, making CCD's very sensitive, especially compared to conversion rates of 2% for photographic film and even less for the human eye. During an exposure, electrons accumulate in each pixel, and are then read out and converted to intensity to form an image. The Signal-to-Noise, or S/N, is measured as the ratio of the received photons from a star to that produced by noise.

There are three principal sources of noise introduced into CCD cameras that must be accounted for by correcting the images, called science frames, with calibration frames. When starting a new exposure, not every pixel will be wiped completely clean to start from zero, and thus the camera has some measure of offset, called a Bias. To correct for this, an exposure of zero seconds is taken, called a Bias frame, and is subtracted from the science frames. Although CCD cameras are typically cooled to temperatures of -30°C, there are still some thermal photons emitted from the camera itself, referred to as Dark Current. This current accumulates during the length of the exposure, and to compensate Dark Frames are subtracted, which are exposures of equal length taken with the shutter closed. The third source of error is the fact that not every pixel will respond to incoming photons at exactly the same rate. Thus science images are divided by a Flat Frame, which is an exposure of an evenly illuminated source, such as a blank screen or the sky just above the horizon at twilight. This also has the added effect of compensating for any dust or debris in the optical path of the system.

## 1.7 Photometry

When taking an image of a star field, the light from a star will be spread out over many pixels. This is due in part to any imperfections in the focus of a telescope, but mainly due to scintillation, the distortion of a star's light as it travels through the Earth's atmosphere. Thus, the intensity count of all pixels containing signal from a star are summed up, and the background of light from the sky, determined from an average of pixels away from the star, is subtracted. As a star rises and sets, its light will be traveling through varying distances of the Earth's atmosphere, called its air mass, and thus its light will be dimmed accordingly. As well, high altitude clouds or changes in the atmosphere over time will distort the intensity readings of a single star. To compensate for this, the process of differential photometry is invoked. In each CCD image, the variable star is compared to a reference star that is assumed to be constant. Thus, as atmospheric disturbances over such a small region of the sky are constant, and since it is always the difference of the object with respect to the reference that is taken, atmospheric effects become negligible.

Traditionally, when taking differential photometric measurements, a single reference star is used to measure the magnitude of both the variable star, referred to as the object, and a third star called the check. The check star should also be theoretically constant, and is used to verify the constancy of the reference star, as well as to provide a measure of the variability inherent in the system. If the check star has a certain standard deviation from its average, one would expect the object to have a similar deviation from its true value. However, all the measurements performed in this paper deviate slightly from this traditional approach.

The program used to extract photometric measurements from images, MaximDL, has the ability to specify multiple reference stars. The object is measured with respect to each reference star, and the resulting average taken as the final object measurement. This has the effect of compensating for any systematic errors introduced via image calibration or non-linear atmospheric effects. Any random errors in individual reference stars are statistically cancelled out, and the deviation of reference star measurements with respect to each other can be used to fulfill the traditional role of the check star in ensuring reference constancy. Also, corrections for atmospheric extinction effects arising if the object and reference having vastly different colors can be neglected as a multitude of reference stars ensure a range of colors. This new method can provide measurement accuracy exceeding the S/N measurements of individual stars. The technique has been found to approach the milli-magnitude level in precision for reasonably faint stars on sub-meter class telescopes, which has previously only been demonstrated on larger meter-class telescopes.

# II. MINIMUM TIMINGS AND O-C DIAGRAMS

## 2.1 Time of Minima and Resulting Science

A time of minimum for an eclipsing binary system is defined as the moment at which the system reaches minimum light either during primary or secondary eclipse. The elapsed time between two successive times of primary or secondary minima is the orbital period for the system at that point in time. By accurately measuring many times of minima the system's period can be calculated to within fractions of a second; an accuracy around $10^{-7}$ compared to a typical orbital period of about a day. Times of minima measured over a significantly long baseline in comparison to the orbital period can reveal changes in the orbital period at the same level of fractions of a second. A variety of mechanisms are known to induce observed period changes, such as mass transfer between components, the presence of spots, angular momentum transport via magnetic braking, and light time effects induced by a third body orbiting the binary pair. Thus by studying period changes one can both study the physics of component interaction, magnetic cycles, and as well deduce the presence of low-mass objects such as brown dwarfs and even planets.

## 2.2 The Kwee-van Worden Method

Although various minima fitting techniques exist, such as visual estimation or curve fitting, by far the most widely used method is the Kwee-van Woerden method (Kwee and van Woerden, 1956). Not only does it provide an extremely accurate measurement of the time of minima, but additionally that of the standard error, which is tantamount in performing period studies. The basic idea is that one visually chooses a time of minima that is roughly close to the true time of minima. The eclipse, which should be symmetric, is folded about this first guess, and the resulting asymmetry measured via the difference in magnitudes at equal times before and after the initial time of minima. The asymmetry of the initial guess is corrected for, and one is left with the true time of minima, with the number of employed data points and their scatter used to measure the standard error.

## 2.3 O-C Diagrams

The most common form of tracking period changes is by applying an ephemeris to a set of time of minima and taking the difference between that which was observed and calculated, called the O-C. In this way changes in the period are evident as the trend in minimum timings begin to deviate from zero. Traditionally the O-C values are plotted on the y-axis as a function of epoch on the x-axis, or the number of orbits that have elapsed since the ephemeris time of primary minimum. Also, JD can be used for the x-axis, which has the advantage of comparing O-C diagrams compiled by different authors using different ephemerides. It should be noted that even though some individual minimum timings can have large errors, such as those performed visually, they can still show important details due to the extremely long baselines of some O-C diagrams, on the order of several decades, and thus should not be neglected.

# III. LIGHT CURVE MODELING AND PARAMATER DETERMINATION

## 3.1 Basic Concepts

The basic assumption one makes when modeling eclipsing binary systems is that there exists a unique combination of physical parameters that will produce the observed light and radial velocity curves. This works well in practice, with any standard errors in parameter measurement resulting from observational errors. Usually one observes and compiles a light curve for a given system in multiple filters or bandpasses. The difference in morphology between light curves in different filters contains much information about temperature distribution in the system. Thus, if a certain model adequately matches all of the observed light and radial velocity curves, then one can be confident that the employed parameters are those actually possessed by the system.

## 3.2 The Physics of Eclipsing Binary Models

In general, an eclipsing binary model operates by dividing each component of the system into a number of separate elements, referred to as the grid, and based on the parameters it assigns a certain luminosity to each component. The geometry of the system is determined by the components' radii and separation, and then the system is rotated in N steps through 360 degrees, and the resulting flux as would be seen by an observer recorded at each point to form a theoretical light curve. The main element affecting the luminosity of a component is its surface temperature, due to the quadratic dependence of luminosity on temperature. However, temperature will not be uniform over the surface of a star. Due to rotation, stars are oblate, with their equatorial radius greater than their polar radius. As a result, the poles have a higher surface gravity, resulting in a higher temperature and thus luminosity. Another effect, limb darkening, is the apparent dimming of light seen from the edges, or limbs, of stars as opposed to their centers. Light emitted from stellar limbs have to travel though a greater amount of the star's atmosphere as seen by the observer, resulting in greater scattering. Also, the two stars in a system will mutually irradiate each other, so that the interior facing sides of each star will have a slightly higher surface temperature. This is referred to as the reflection effect, as it is modeled by computing the amount of light from one star that reflects off the other. Finally, star spots are modeled by assigning a circular region on the star's surface a lower temperature than the rest of the stellar surface. While in nature they are never perfect circles, but rather concentrated groups of smaller roughly circular spots, the data is never precise enough to discern that level of detail, and thus the general spot model works well in practice.

To derive luminosity as a function of wavelength from a given surface temperature, one may assume a general blackbody distribution. However, this is usually a gross oversimplification, as stellar spectra are dominated by absorption and emission lines. Thus, in modern programs, luminosities are extracted from model stellar spectra, providing an accurate measure of luminosity for a given bandpass. This allows for very precise determination of surface temperature, although it is of vast importance to correctly specify the scale of the system, as the model atmospheres are sensitive to local values of surface gravity.

## 3.3 Solution Schemes in Multi-Parameter Space

When solving for the parameters of an eclipsing binary system, there can be anywhere from two to twenty unknown parameters, each with a wide range of possible values. The accuracy of a particular model is evaluated by measuring the sum of the absolute deviations with respect to the actual data points. As one wants this value as low as possible, one encounters the mathematical problem of finding the minimum of an equation with a large number of variables. The multi-parameter solution space is almost never smooth, but rather quite chaotic and filled with a large number of local minima.

The simplest method, referred to as differential corrections, is to start with some initial value for each parameter, and to vary each parameter one at a time, fixing a new value when the local minimum of the absolute deviation is found. If the solution space is large, the computing time can be prohibitively long, and thus one can take the derivative for each parameter as it is varied in order to predict the best-fit value for that parameter. However, this method will only find the local minima in the solution space, and thus is highly dependant on the assumed initial conditions, as well as the step size used in adjusting each parameter.

A more advanced solution method is the genetic algorithm developed by P. Charbonneau (1995). Light curves for an initial population of solutions, generated with random parameters constrained by user-specified minimum and maximum values, are computed and compared to the observational light curve. Their corresponding value of absolute deviation is used as a measure of fitness for natural selection, with parameters from good fits being passed on to a second generation of solutions, and parameters from bad fits being eliminated. After being subject to random mutations, to maintain parameter diversity, this second generation is compared to the observational data, and bred into a third generation of solutions. The process continues for N generations, until a satisfactorily accurate solution is found. Thus, a large population ensures a complete exploration of the parameter space, and a large number of generations ensures that the global minimum absolute deviation, or absolute best fit, is found. This method converges on the global best solution rather quickly, thus providing the fastest, most-accurate, user unbiased method known.

## 3.4 Currently Employed Codes

The first, and most widely-known, eclipsing light curve modeling code is the Wilson-Devinney (WD) code (Wilson and Devinney, 1971). It has been gradually expanded and updated since the 70's as new physics have been uncovered and modeling techniques invented. Today it can accurately model almost any eclipsing system known to exist. However, its solution-finding method is limited to differential corrections, and thus contains the inherent limitations of that method previously enumerated. A more recent code, the Eclipsing Light Curve (ELC) code, was developed in 2000, and contains all the abilities of the WD code, except for the ability model over-contact binaries (Orosz and Hauschildt, 2000). In addition, it incorporates the previously described genetic algorithm, as well as other techniques, making it more powerful than the WD code. Also worthy of mention, Binary Maker is a user-friendly, student-oriented, Windows program that employs the WD code, but does not have any solving ability. It does have a good graphics interface however that is useful for producing visual models, and was used to produce the visual models in this thesis.

As ELC is primarily used in this thesis, a definition of its parameters will now be given. ELC defines star 1 as the star that is closest to the observer at primary, or phase zero, and thus star 2 is eclipsed at that time. Star 2 is then usually the primary component. $T_1$ and $T_2$ are the effective temperatures of star 1 and star 2 respectively. The mass ratio, Q, is the mass of star 2 divided by the mass of star 1. The orbital separation (Sep.), is defined as the distance between the two components' centers of mass. The inclination of the system is designated by *i*, with 90 degrees being an exactly edge-on orbit. The effective radii, $Reff_1$ and $Reff_2$, are the radius of star 1 and star 2 respectively divided by the orbital separation. Each component's size can also be given as the fractional measure of the amount that each star fills its respective Roche Lobe, called the fill factor, designated by $f_1$ and $f_2$. For the spot parameters, $TF_1$ and $TF_2$ are the temperature factors, or ratio of the spot temperature to the underlying temperature, of spots 1 and 2. $Rad_1$ and $Rad_2$ are the radii of the spots, where 90 degrees covers exactly half the star. $Lat_1$, $Lat_2$, $Lon_1$, and $Lon_2$ are the respective latitudes and longitudes of spots 1 and 2. The north pole is at a latitude of 0 degrees, the equator at 90 degrees, and the south pole at 180 degrees. The longitude is zero degrees at the inner Lagrangian point, 90 degrees on the leading side, 180 degrees at the back end, and 270 degrees on the trailing side.

# IV. RT ANDROMEDAE

## 4.1 Background

The system, RT And, [RA: 23 11 10, Dec: +53 01 33, $V_{mag} \approx 9$], consists of two stars of about $1.1M_\odot$ and $0.8M_\odot$. It is a RS CVn type binary system, which are known for their high level of spot activity and flaring due to increased magnetic effects resulting from the rapid system rotation. As such its light curve is not constant, but varies over a wide range of time scales from several hours to years as the location and intensity of spots on its surface varies.

Due to its relatively bright magnitude, there have been a large number of observations and papers published on RT And, with the most recent being Prilluba et al. in 2000, Kjurkchieva, Marchev, and Ogloza, in 2001, and Erdem, Demircan, and Güre, also in 2001. However, a number of key parameters of the system remain in question. Prilluba derives $M_1 = 1.10M_\odot$, $M_2 = 0.83M_\odot$, and i = 87.6 degrees, and Kjurkchieva derives $M_1 = 1.23M_\odot$, $M_2 = 0.91M_\odot$, and i = 82 degrees. The difference in component masses is directly a result of the difference in inclination, as the observed radial velocity curves only give each mass as a function of inclination. Thus, if the inclination can be definitively solved through modeling of the light curves, then the component masses can be definitively determined. Erdem studied the period changes of RT And and suggests either a third body with an orbital period of 105 years, or a cyclic magnetic activity modulation of 65 years. Additionally, although almost all of the studies on RT And have resulted in measurements of spot positions, there have been little in the way of analyzing trends in spot movement or position, especially on short timescales.

Since all of the most recent papers were published before the development of model atmospheres, ELC, and implementation of the genetic algorithm, a modern re-observation and study of RT And is much needed.

## 4.2 Observations

Observations of RT And were taken with Emory Observatory's 24" telescope and an Apogee 47 CCD camera cooled to -30°C on the nights of October 15th and November 8th and 9th in 2004 by Horace Dale, and on September 12th, 13th, 14th, and 30th, October 3rd, 27th, and 28th, and November 29th and 30th in 2005, and October 13th and 14th in 2006 by the author, in U, B, V, R, and I filters. Differential photometry was performed via MaximDL with respect to GSC 3998-1794, 2256, and 2415 in all filters, and additionally GSC 3998-983 and 2231 in V, R, and I. GSC 3998-983 and 3998-2231 were not used in U and B due to S/N <100 in those filters. GSC 3998-2415 was photometrically calibrated by Heckert (1995) by observations of Landolt standards, and other reference stars were calibrated differentially. All times were corrected to HJD.

## 4.3 Minimum Timings and O-C Diagram

Williamon (1974) was the first to thoroughly study major period changes exhibited by RT And by collecting all available times up to the time of publication, JD 2442000. He found two major period changes around JD 2430000 and 2438000, each a decrease by 2.2 and 0.8 seconds respectively, and attributed it to instantaneous mass transfer. Williamon's O-C diagram is reproduced in Figure 4-1.

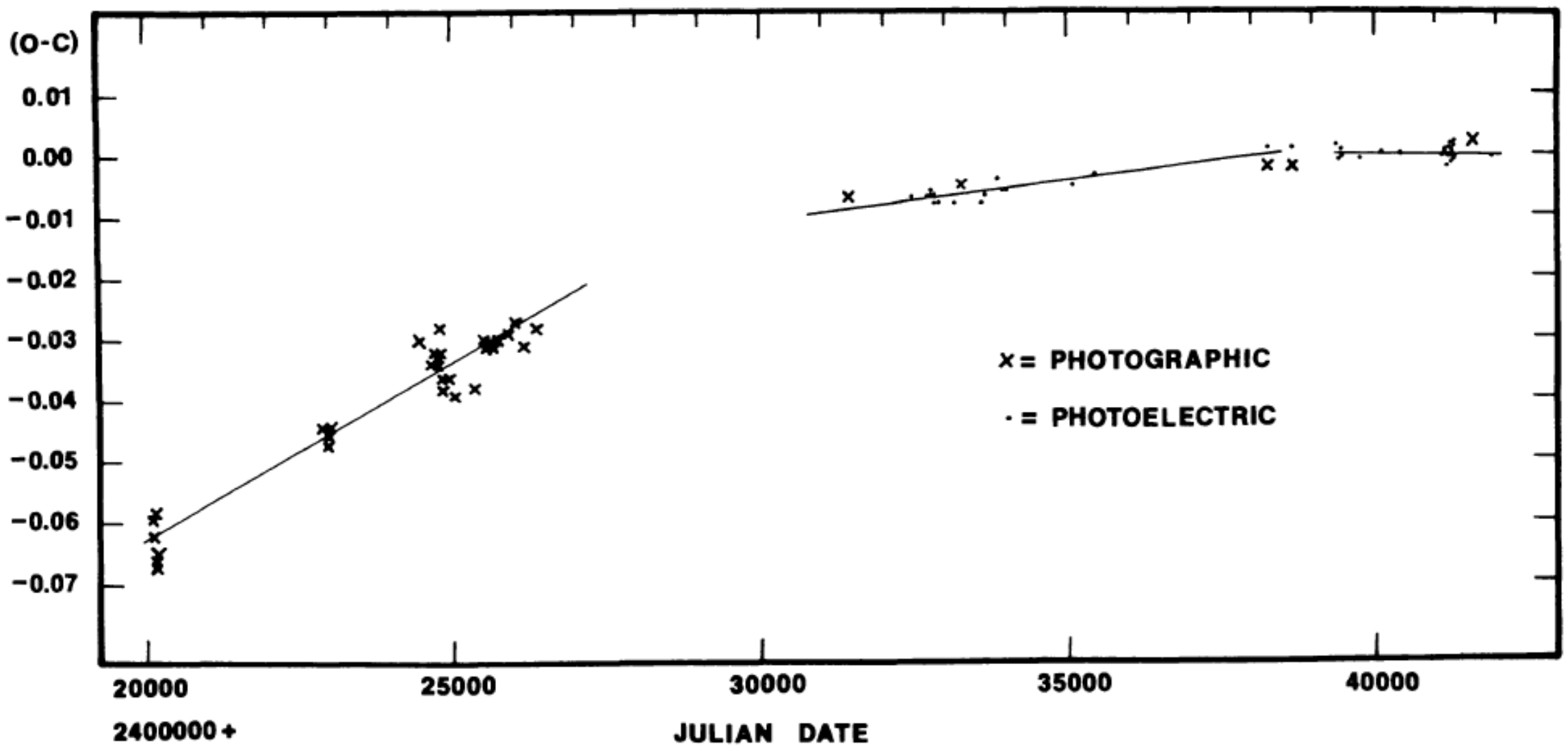

Figure 4-1: O-C diagram from Williamon (1974) showing two major period changes of the RT And system.

With respect to this thesis' observations, all observed times of minimum were determined via the method of Kwee-van Woerden, and shown in Table 4-1 with errors, employed filter, and type (primary or secondary eclipse). All previously published times of minima available after JD 2437000, right before the second period shift in Williamon's data, were compiled and assigned a weight that was inversely proportional to its error. In cases where no error was given, a value of ±.005 days was assumed. A linear least-squared fit to the data was then performed and a new ephemeris calculated to be

$T_{pri}$ (HJD) = 2437349.45168980 + 0.62892898•E, where E is the epoch. An O-C diagram of the data is shown in Figure 4-2.

Table 4-1: Observed times of minima for RT And.

| $T_{min}$ (HJD) | Error (±) | Filter | Type |
|---|---|---|---|
| 2453294.687483 | 0.000064 | U | Pri |
| 2453294.687563 | 0.000058 | R | Pri |
| 2453294.687598 | 0.000057 | I | Pri |
| 2453294.687640 | 0.000022 | B | Pri |
| 2453294.687751 | 0.000038 | V | Pri |
| 2453318.586764 | 0.000048 | R | Pri |
| 2453318.586775 | 0.000054 | V | Pri |
| 2453318.586836 | 0.000074 | I | Pri |
| 2453318.587003 | 0.000029 | B | Pri |
| 2453318.587019 | 0.000066 | U | Pri |
| 2453626.762604 | 0.000107 | I | Pri |
| 2453626.762702 | 0.000098 | R | Pri |
| 2453626.762729 | 0.000110 | B | Pri |
| 2453626.762795 | 0.000084 | V | Pri |
| 2453626.762988 | 0.000157 | U | Pri |
| 2453628.649328 | 0.000126 | B | Pri |
| 2453628.649330 | 0.000071 | R | Pri |
| 2453628.649363 | 0.000081 | V | Pri |
| 2453628.649431 | 0.000063 | I | Pri |
| 2453628.649613 | 0.000093 | U | Pri |
| 2453672.672519 | 0.000103 | B | Pri |
| 2453672.672626 | 0.000081 | V | Pri |
| 2453672.672733 | 0.000245 | R | Pri |
| 2453672.672870 | 0.000108 | I | Pri |
| 2454023.615307 | 0.000095 | U | Pri |
| 2454023.615364 | 0.000060 | B | Pri |
| 2454023.615420 | 0.000063 | R | Pri |
| 2454023.615457 | 0.000071 | V | Pri |
| 2454023.615667 | 0.000099 | I | Pri |
| 2453319.528425 | 0.000394 | V | Sec |
| 2453319.529685 | 0.000373 | R | Sec |
| 2453319.530005 | 0.000196 | B | Sec |
| 2453319.531680 | 0.000163 | I | Sec |
| 2453319.532366 | 0.000549 | U | Sec |
| 2453627.704716 | 0.000268 | B | Sec |
| 2453627.704945 | 0.000346 | V | Sec |
| 2453627.705161 | 0.000261 | R | Sec |
| 2453627.705267 | 0.000205 | I | Sec |
| 2453627.706754 | 0.000765 | U | Sec |
| 2453644.686696 | 0.000567 | V | Sec |
| 2453644.686819 | 0.000434 | I | Sec |
| 2453644.686876 | 0.000479 | R | Sec |
| 2453644.687470 | 0.000429 | B | Sec |

Table 4-1 (Cont.)

| $T_{min}$ (HJD) | Error (±) | Filter | Type |
|---|---|---|---|
| 2453644.688541 | 0.000568 | U | Sec |
| 2453671.730192 | 0.000199 | I | Sec |
| 2453671.731033 | 0.000223 | R | Sec |
| 2453671.731633 | 0.000362 | V | Sec |
| 2453671.731742 | 0.000867 | U | Sec |
| 2453671.731916 | 0.001188 | B | Sec |
| 2453705.687104 | 0.001361 | B | Sec |
| 2453705.693063 | 0.001648 | R | Sec |
| 2453705.693096 | 0.000559 | V | Sec |
| 2453705.697002 | 0.000490 | I | Sec |
| 2453705.697421 | 0.002187 | U | Sec |
| 2454022.670767 | 0.000340 | I | Sec |
| 2454022.672038 | 0.000135 | U | Sec |
| 2454022.672115 | 0.000130 | B | Sec |
| 2454022.672396 | 0.000157 | V | Sec |
| 2454022.672410 | 0.000304 | R | Sec |

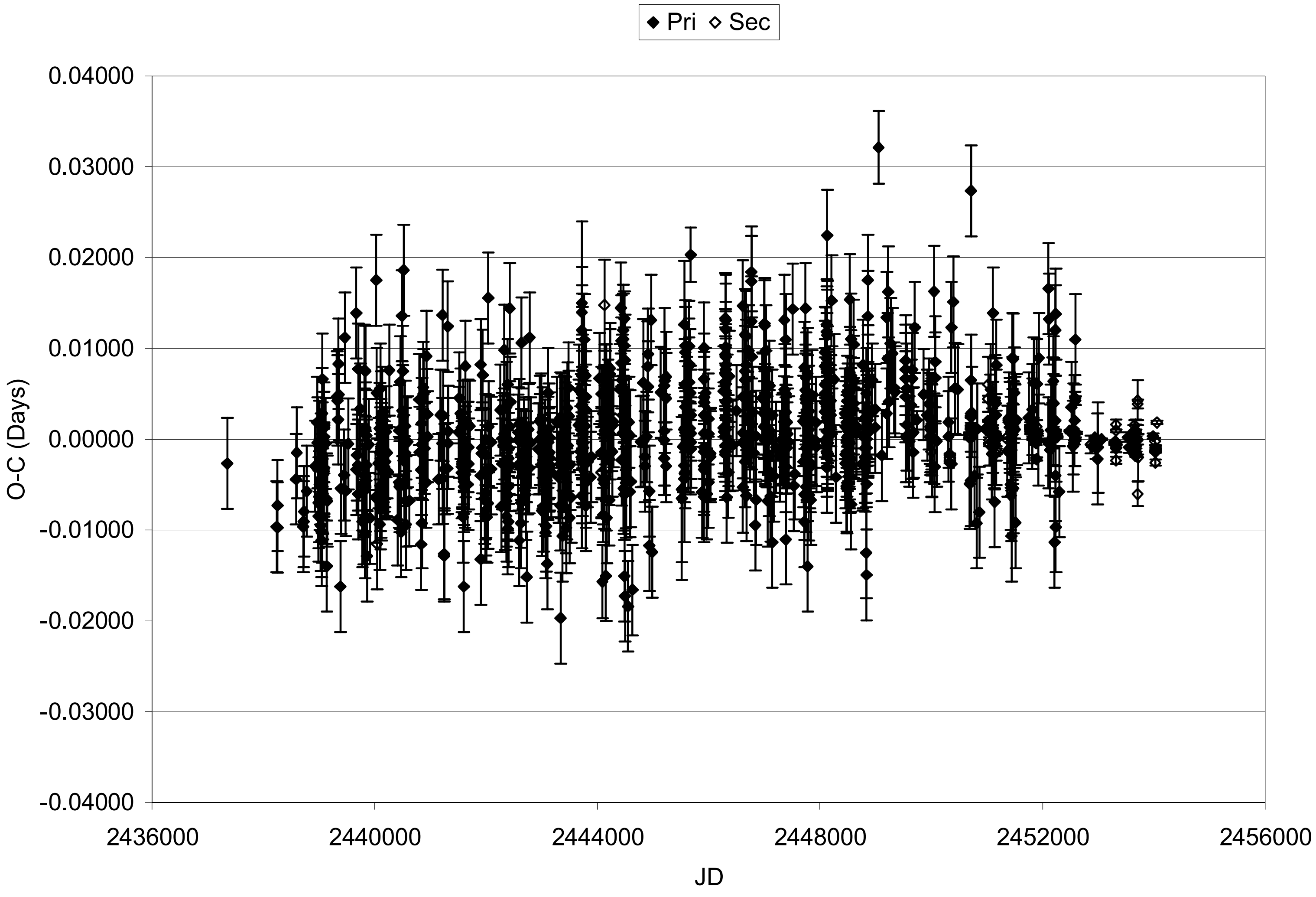

Figure 4-2: O-C diagram for RT And

The second period change identified by Williamon can just be seen to the far left in Figure 4-2. However, it is interesting that afterwards the period appears to be relatively stable, with no period shifts on the order of those discovered by Williamon. While there does not appear to be any noticeable variation on the timescales suggested by Erdem (2001), there does appear to be a quasi-sinusoidal oscillation just above the noise level with an apparent period of about 10.5 years and an amplitude of 0.004 days. The presence of spots in a system will offset the photometric minima from the true geometric minima. Although Pribulla et. al (2000) argues for a more or less random distribution of starspots, they do claim to have observed a 6.8 year cycle in the local position of a particular starspot repeatedly observed at a longitude of 270 degrees on the primary component. As well, Pribulla et al. (2000) calculates that spots of those typically found in the RT And system would displace minima up to 0.0035 days. Keeping this in mind, the O-C variations of RT And seem likely to be due by a spot cycle on the order of 10.5 years.

### 4.4 Light Curves and Modeling

All data collected from Emory observatory was compiled into composite U, B, V, R, and I light curves, and first tested against the parameters of Kjurkchieva (2001) and Prilluba (2000). The parameters of inclination, mass ratio, temperature, and radius derived from each study were inputted into the ELC program, and the resulting model light curves are plotted against this study's observational data in Figures 4-3 and 4-4 for Kjurkchieva (2001) and Prilluba (2000).

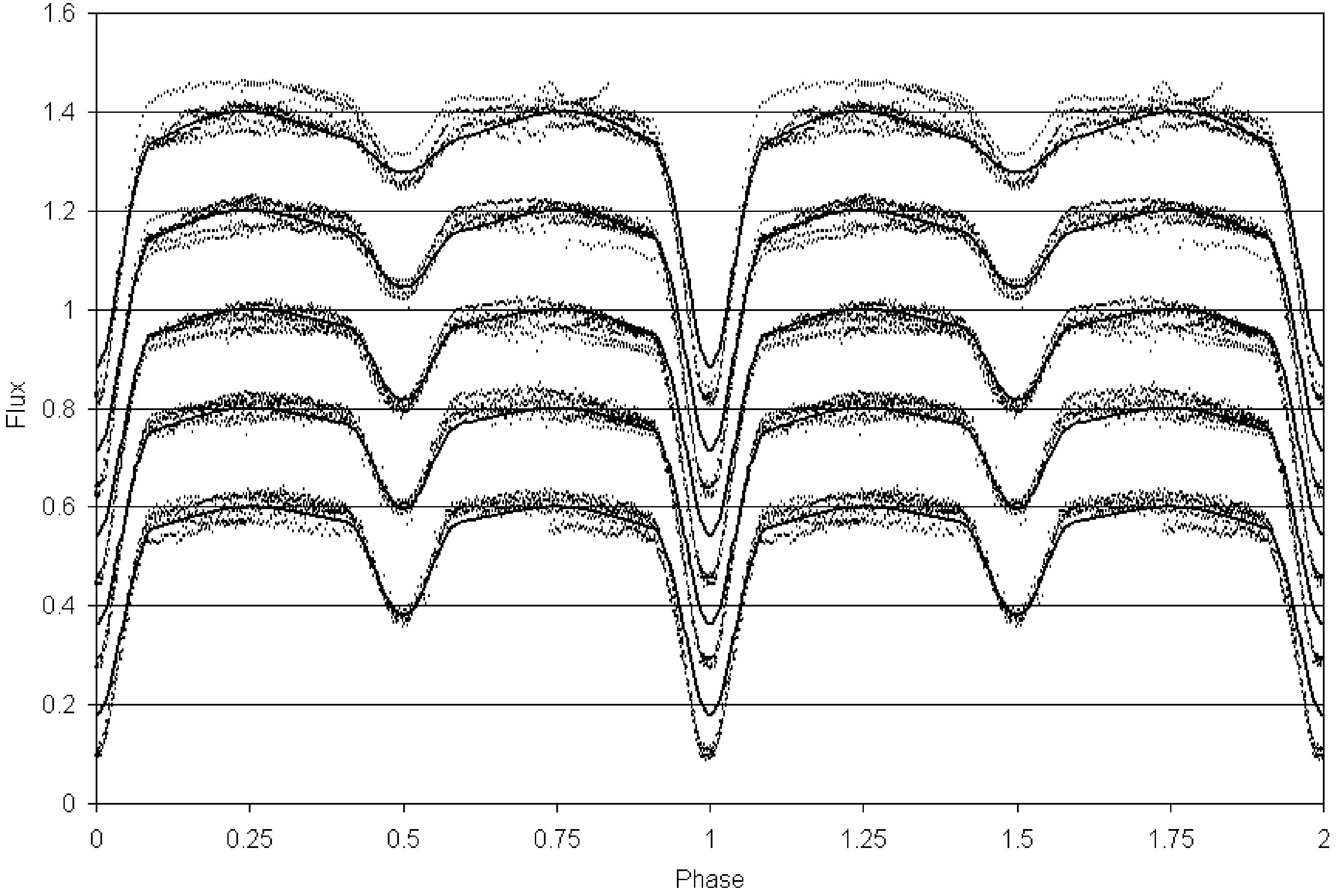

Figure 4-3: Top to bottom: Observed U, B, V, R, and I curves (points) for RT And with Kjurkchieva’s model (solid lines).

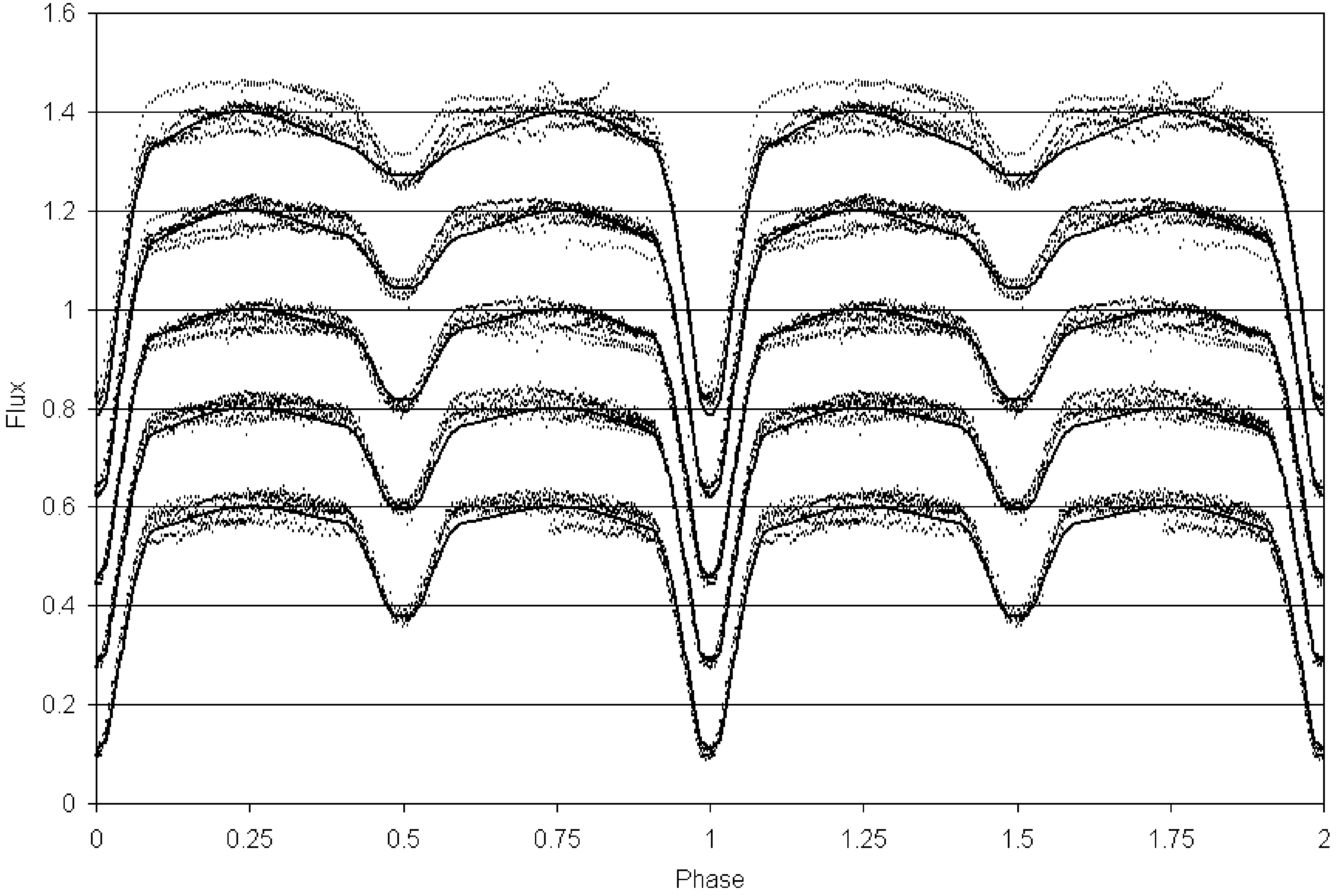

Figure 4-4: Top to bottom: Observed U, B, V, R, and I curves (points) for RT And with Prilluba's model (solid lines).

As can be seen from Figures 4-3 and 4-4, the model of Prilluba et al. (2000) provides a much better fit to the data, especially with respect to the primary eclipse, and signifies that the high orbital inclination is the correct solution. To confirm this finding, several attempts were made via the genetic mode of ELC to find an acceptable solution by confining 80° < i < 85°, and allowing the mass ratio, temperatures, and radii to vary. No possible combination of parameters with i < 85° were found to yield a satisfactory fit. Attempts to find a better solution than presented by Prilluba et al. (2000) with the composite data proved fruitless, and thus that solution was taken as the correct orbital solution.

The data was subsequently divided into 7 separate light curves. These were the nights of October 15$^{th}$ 2004, November 8$^{th}$ and 9$^{th}$ 2004, September 12$^{th}$, 13$^{th}$, and 14$^{th}$ 2005, September 30$^{th}$ and October 3$^{rd}$ 2005, October 27$^{th}$ and 28$^{th}$ 2005, November 29$^{th}$ and 30$^{th}$ 2005, and October 13$^{th}$ and 14$^{th}$ 2006. Since most of these groups are comprised of two nights in a row, they are mostly all complete light curves free from spot movement during the course of observation. Data taken by Williamon in 1974 and digitized recently by Sowell, (Sowell, 2007), were added as an eighth individual curve. Each curve was independently modeled with the genetic code of ELC with the orbital parameters fixed, and the free parameters being the temperature ratio, angular size, and location of two spots on the primary component and one spot on the secondary. The eight individual curves with model fits are shown in Figures 4-5 to 4-12.

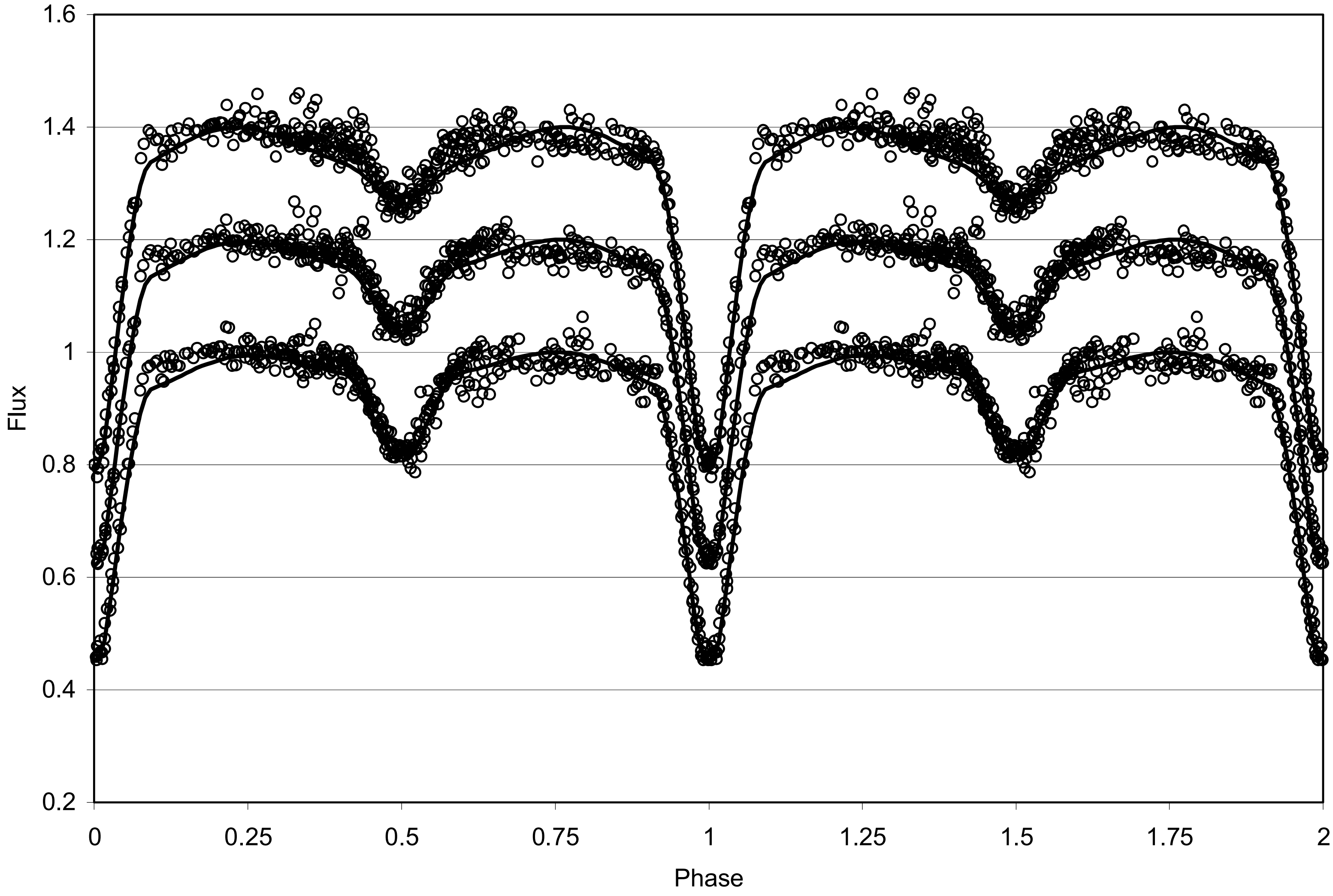

Figure 4-5: Top to bottom: Observed U, B, V, R, and I curves (circles) for 1974 with model fits (solid lines).

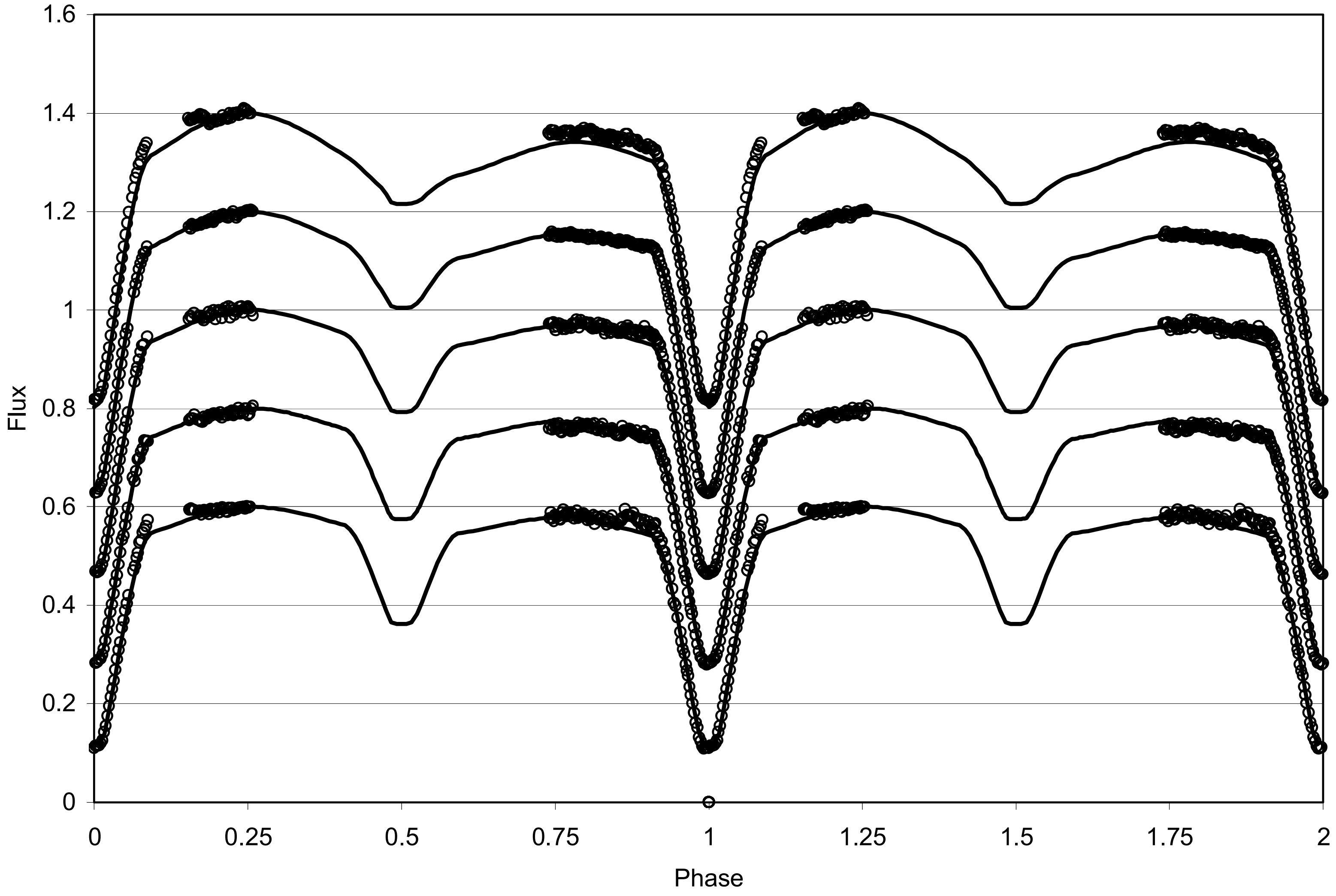

Figure 4-6: Top to bottom: Observed U, B, V, R, and I curves (circles) for Oct. 15th 2004 with model fits (solid lines).

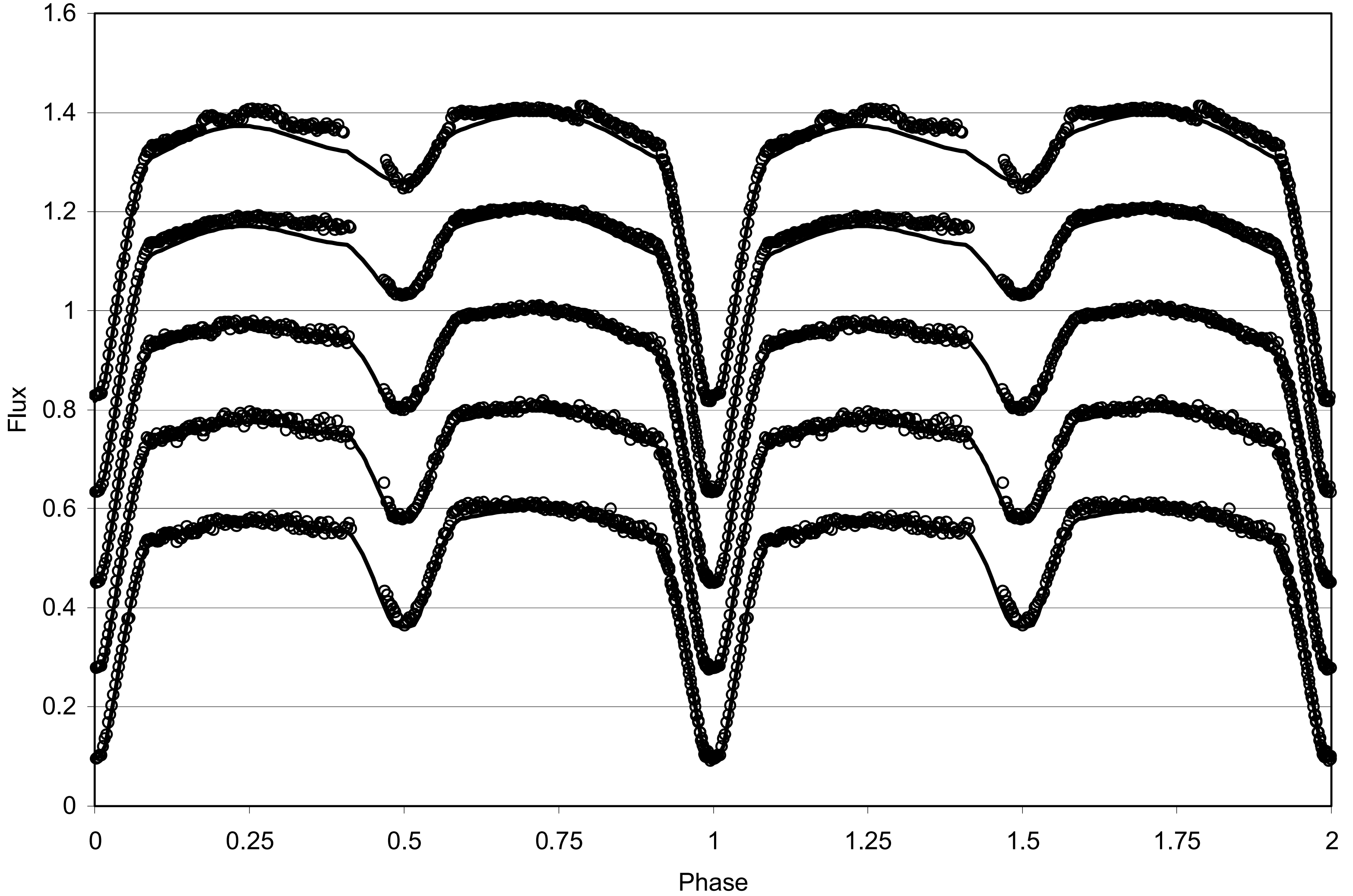

Figure 4-7: Top to bottom: Observed U, B, V, R, and I curves (circles) for Nov. 8th-9th, 2004 with model fits (solid lines).

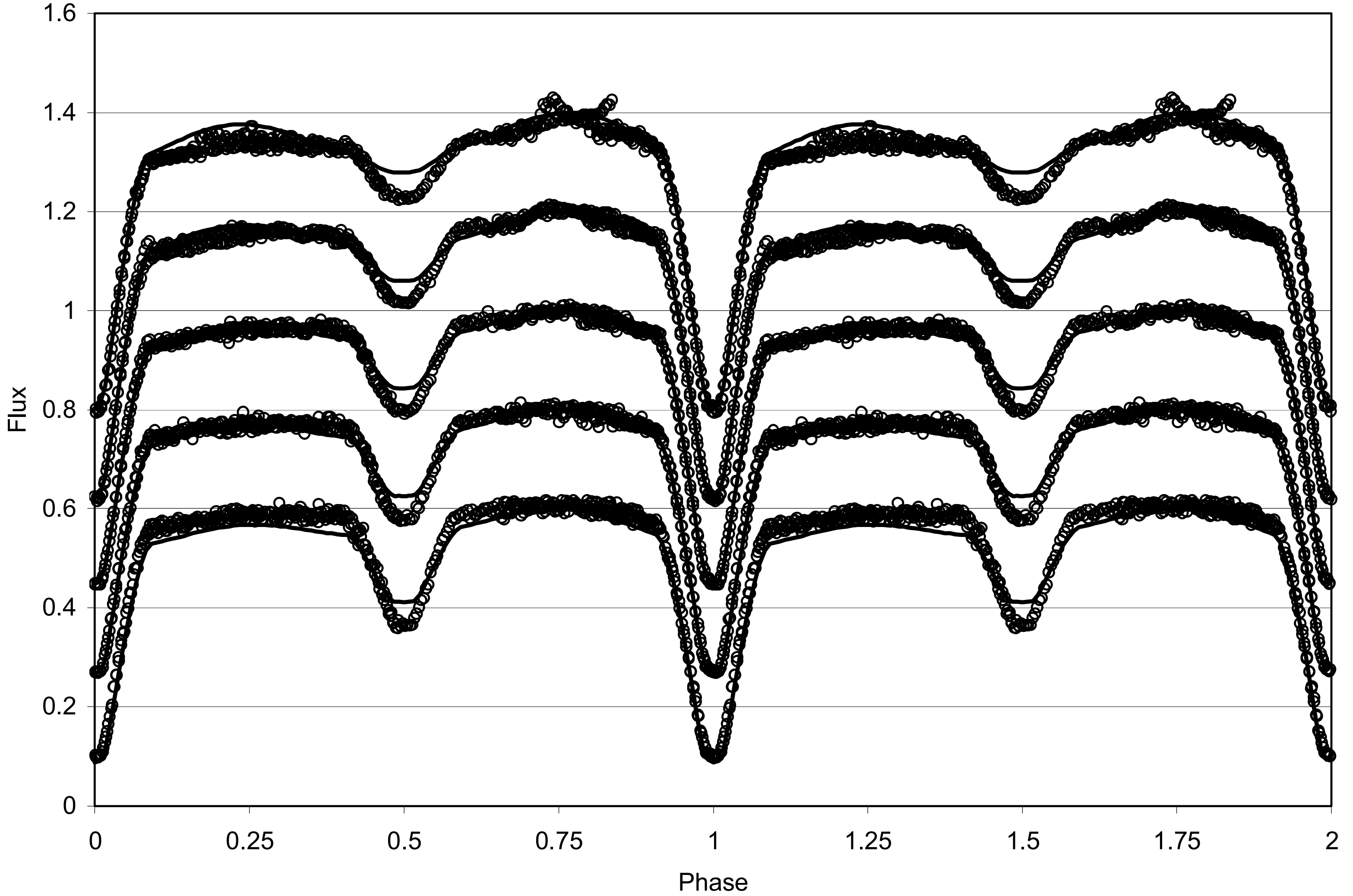

Figure 4-8: Top to bottom: Observed U, B, V, R, and I curves (circles) for Sept. 12$^{th}$-14$^{th}$, 2005 with model fits (solid lines).

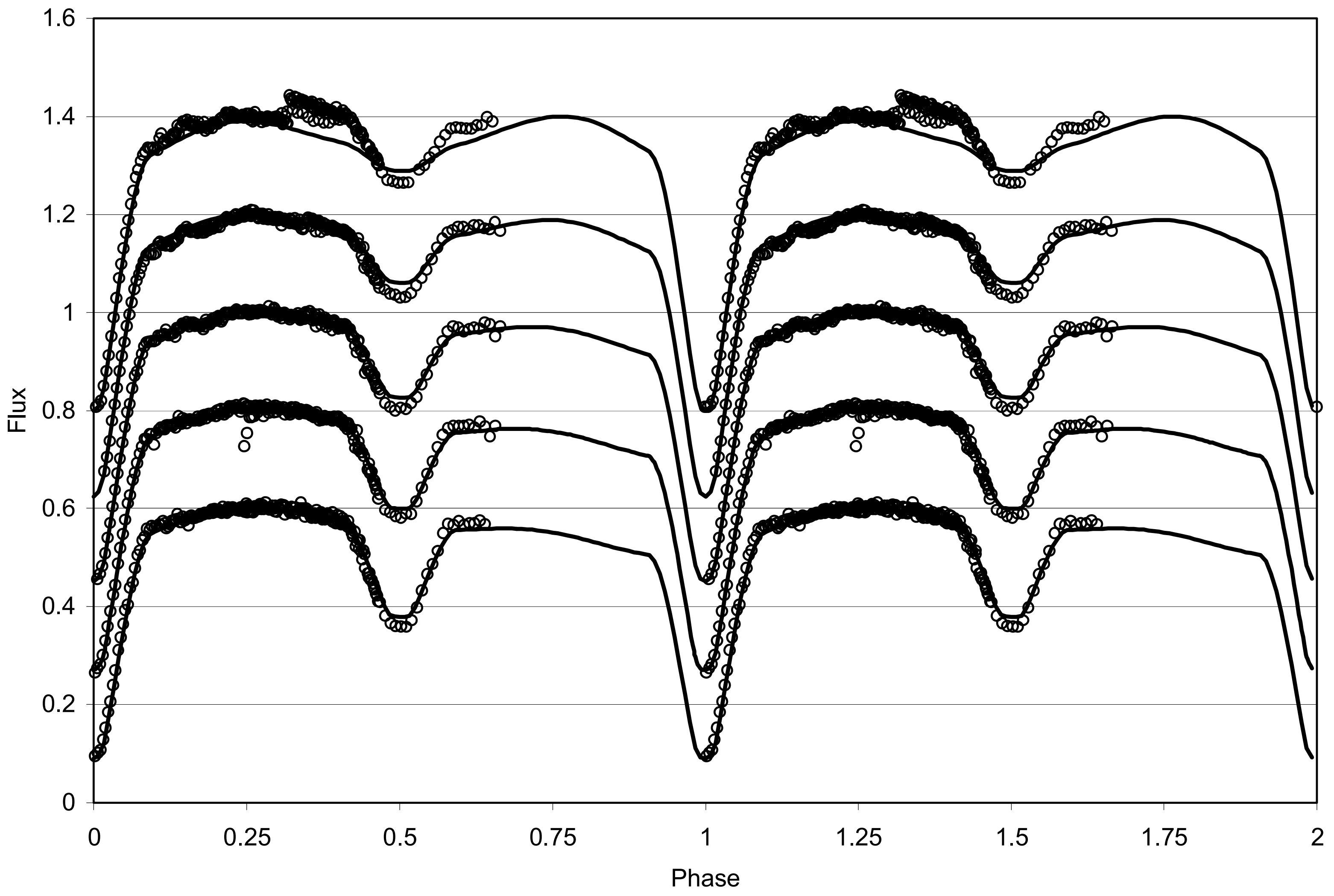

Figure 4-9: Top to bottom: Observed U, B, V, R, and I curves (circles) for 9/30-10/3, 2005 with model fits (solid lines).

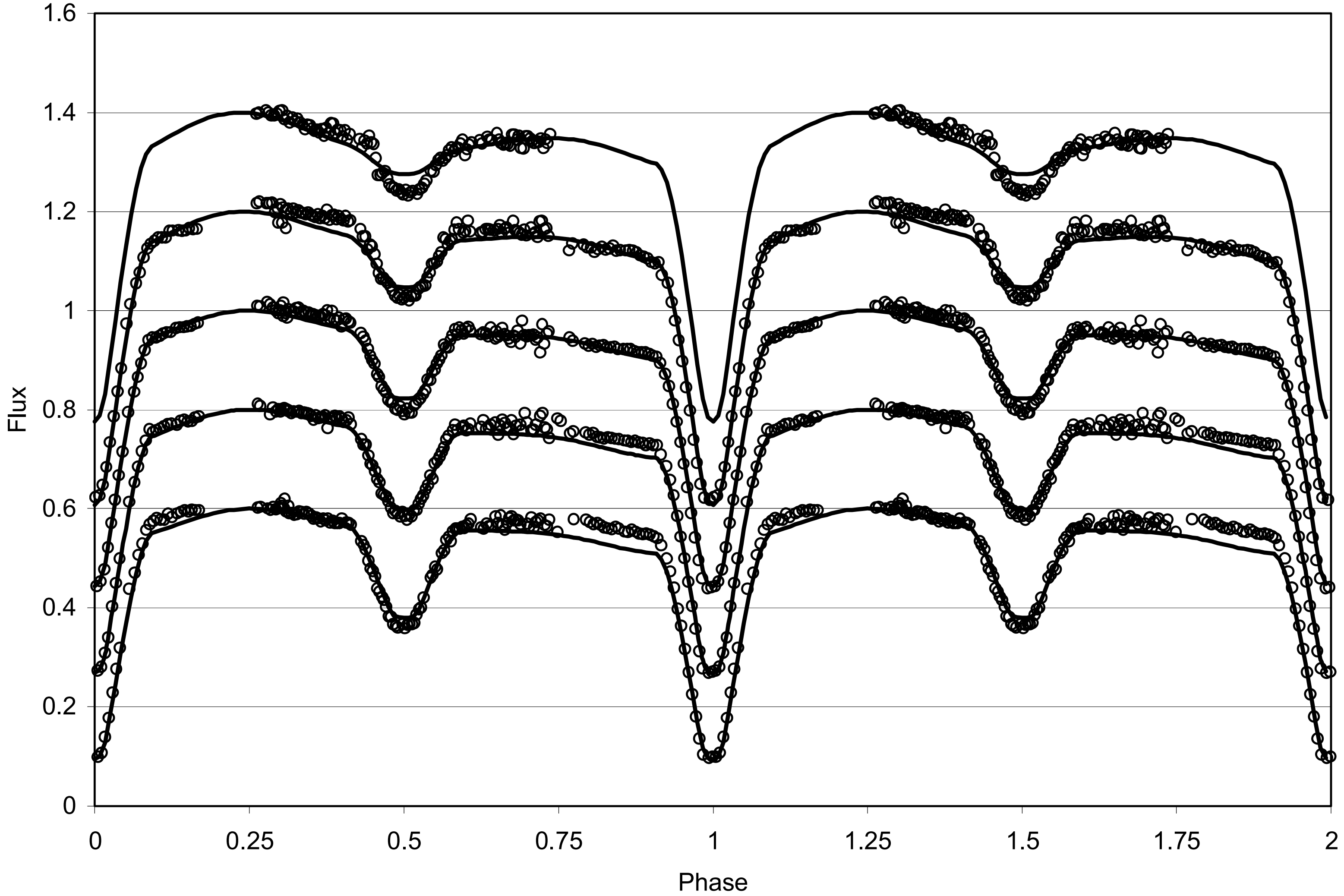

Figure 4-10: Top to bottom: Observed U, B, V, R, and I curves (circles) for Oct. 27th-28th, 2005 with model fits (solid lines).

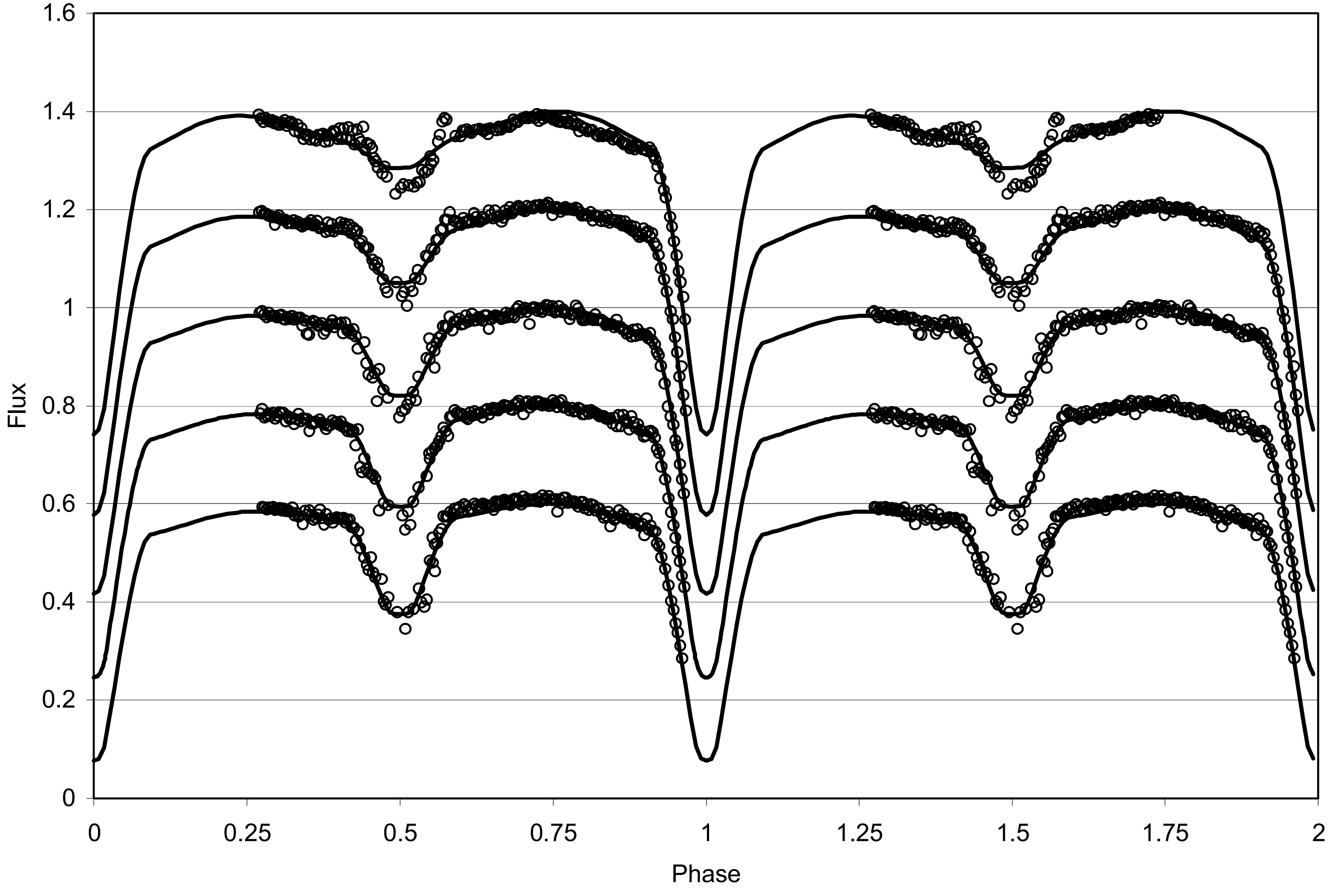

Figure 4-11: Top to bottom: Observed U, B, V, R, and I curves (circles) for Nov. 29th-30th, 2005 with model fits (solid lines).

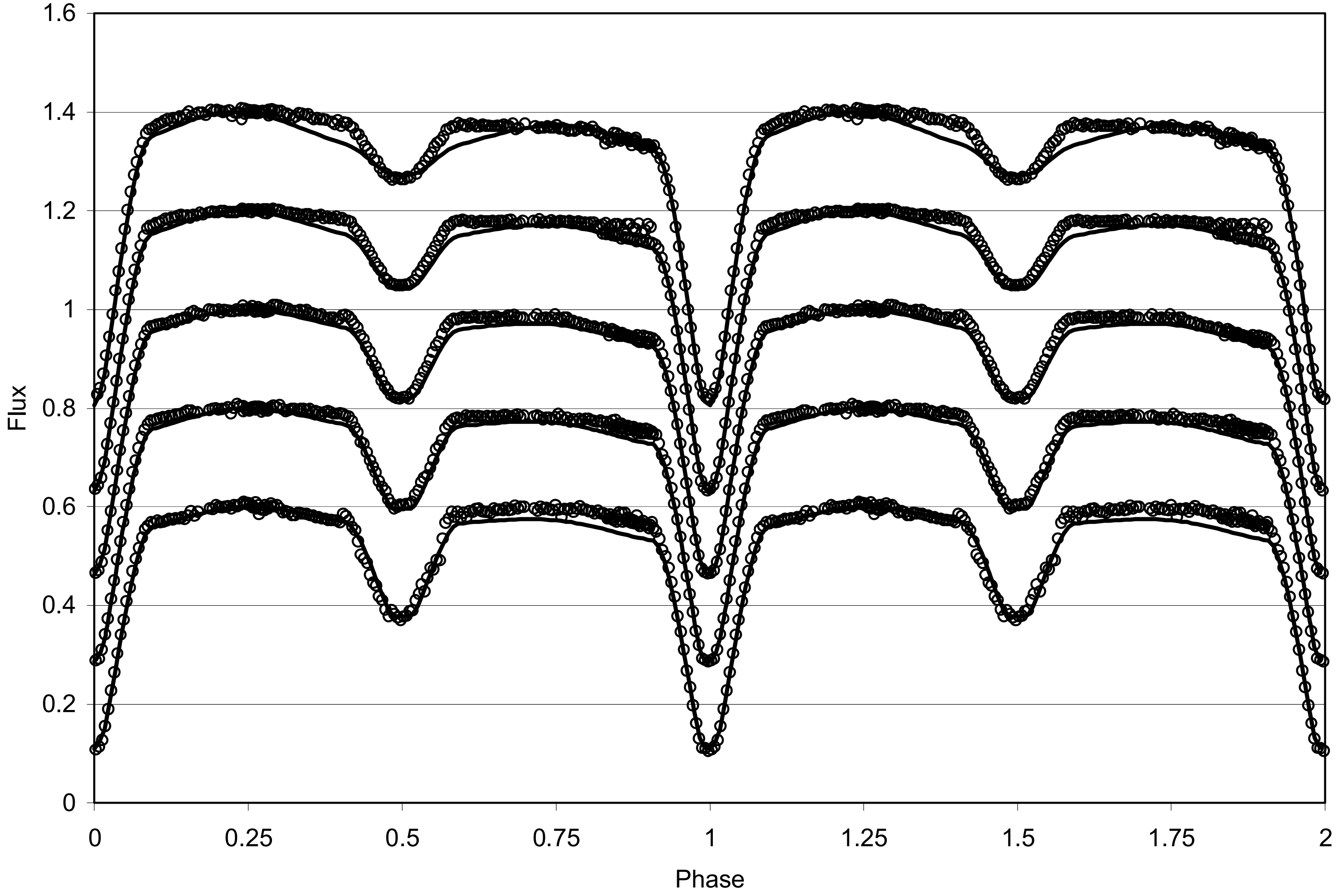

Figure 4-12: Top to bottom: Observed U, B, V, R, and I curves (circles) for Oct. 13th-14th, 2006 with model fits (solid lines).

## 4.5 Results

The results for the orbital parameters of RT And by modeling the composite light curves is shown in Table 4-2, and a geometrical model of the system shown in Figure 4-13. Spot parameters for each of the eight separated light curves is shown in Table 4-3, a plot of the spatial starspot distribution is shown in Figure 4-14.

Table 4-2: Orbital Solutions for RT And

| **Parameter** | **Value** |
|---|---|
| Inclination | 87.6° |
| Mass Ratio ($M_{pri}/M_{sec}$) | 1.36 |
| Primary Temperature | 5900K |
| Secondary Temperature | 4651K |
| Fill Factor of Primary | 0.635 |
| Fill Factor of Secondary | 0.490 |
| Fractional Radius of Primary | 0.320 |
| Fractional Radius of Secondary | 0.224 |

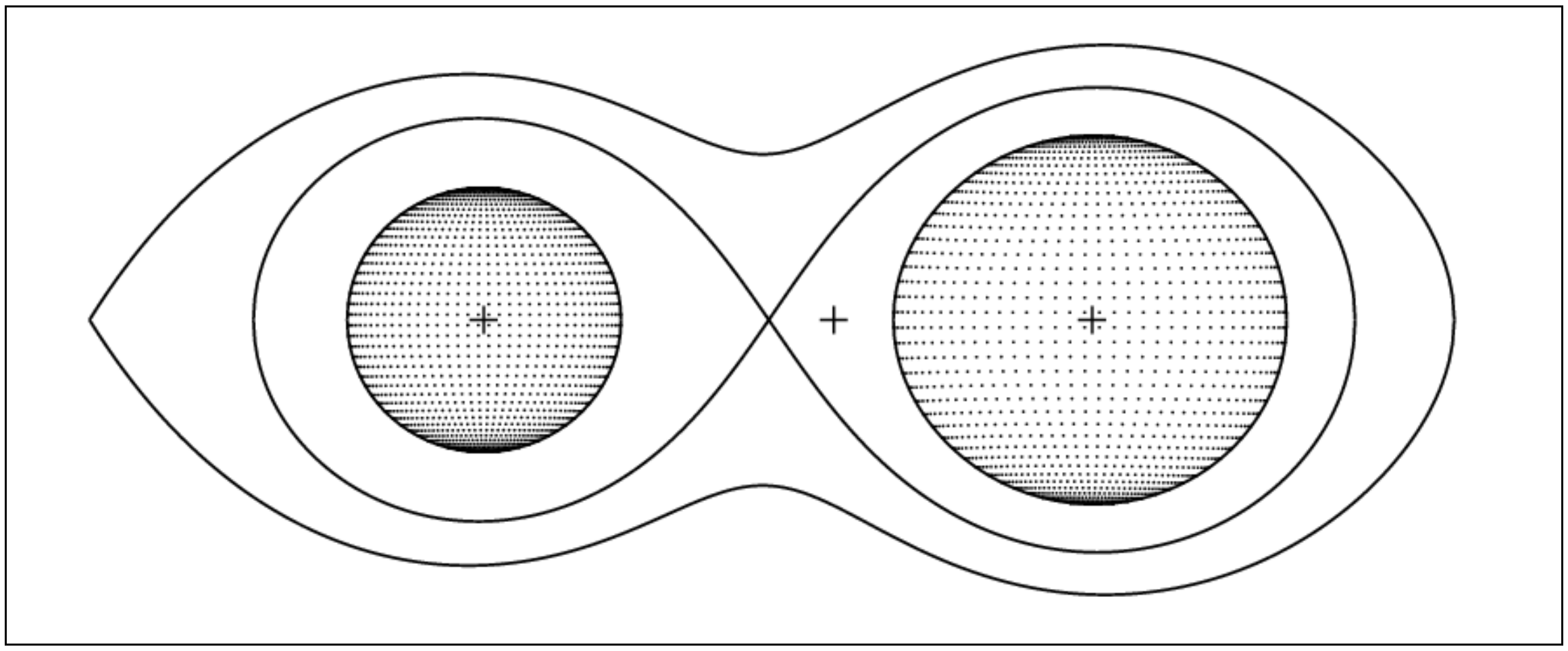

Figure 4-13: A geometrical model of the RT And system. The crosses represent the center of each star and the barycenter of the system. The inner solid line represents the Roche Lobe for each star. The primary, hotter star is on the right.

Table 4-3: Spot Parameters for the Eight Separated Light Curves of RT And

| Set | Pri. 1 TF | Pri. 1 Size | Pri. 1 Lat. | Pri. 1 Long. | Pri. 2 TF | Pri. 2 Size | Pri. 2 Lat. | Pri. 2 Long. | Sec. TF | Sec Size | Sec Lat. | Sec Long. |
|---|---|---|---|---|---|---|---|---|---|---|---|---|
| 1974 | 0.55 | 2.7 | 55.1 | 278.6 | 1.12 | 14.7 | 164.2 | 339.3 | 0.71 | 29.1 | 127.9 | 159.6 |
| Oct. 15, 2004 | 1.28 | 23.0 | 168.5 | 336.8 | 0.91 | 19.6 | 53.9 | 313.4 | 1.44 | 25.2 | 23.3 | 89.3 |
| Nov. 8-9, 2004 | 0.56 | 46.0 | 0.0 | 294.8 | 1.17 | 56.8 | 0.4 | 17.4 | 1.18 | 25.1 | 95.6 | 319.6 |
| Sept. 12-14, 2005 | 0.57 | 2.4 | 143.2 | 135.8 | 1.23 | 29.9 | 178.6 | 98.0 | 0.53 | 71.6 | 9.9 | 107.3 |
| Sept.30-Oct.3, 2005 | 0.59 | 8.4 | 94.6 | 131.0 | 1.35 | 22.8 | 4.3 | 133.7 | 0.55 | 34.7 | 92.5 | 232.4 |
| Oct. 27-28, 2005 | 0.82 | 5.2 | 0.0 | 2.4 | 0.58 | 20.5 | 145.8 | 73.2 | 0.82 | 39.9 | 148.0 | 225.3 |
| Nov. 29-30, 2005 | 0.67 | 43.8 | 0.0 | 164.4 | 1.04 | 9.0 | 63.4 | 160.0 | 0.93 | 57.2 | 146.7 | 129.0 |
| Oct. 13-14, 2006 | 0.65 | 14.7 | 123.3 | 56.1 | 1.27 | 18.2 | 165.8 | 26.7 | 1.00 | 53.5 | 86.5 | 39.8 |

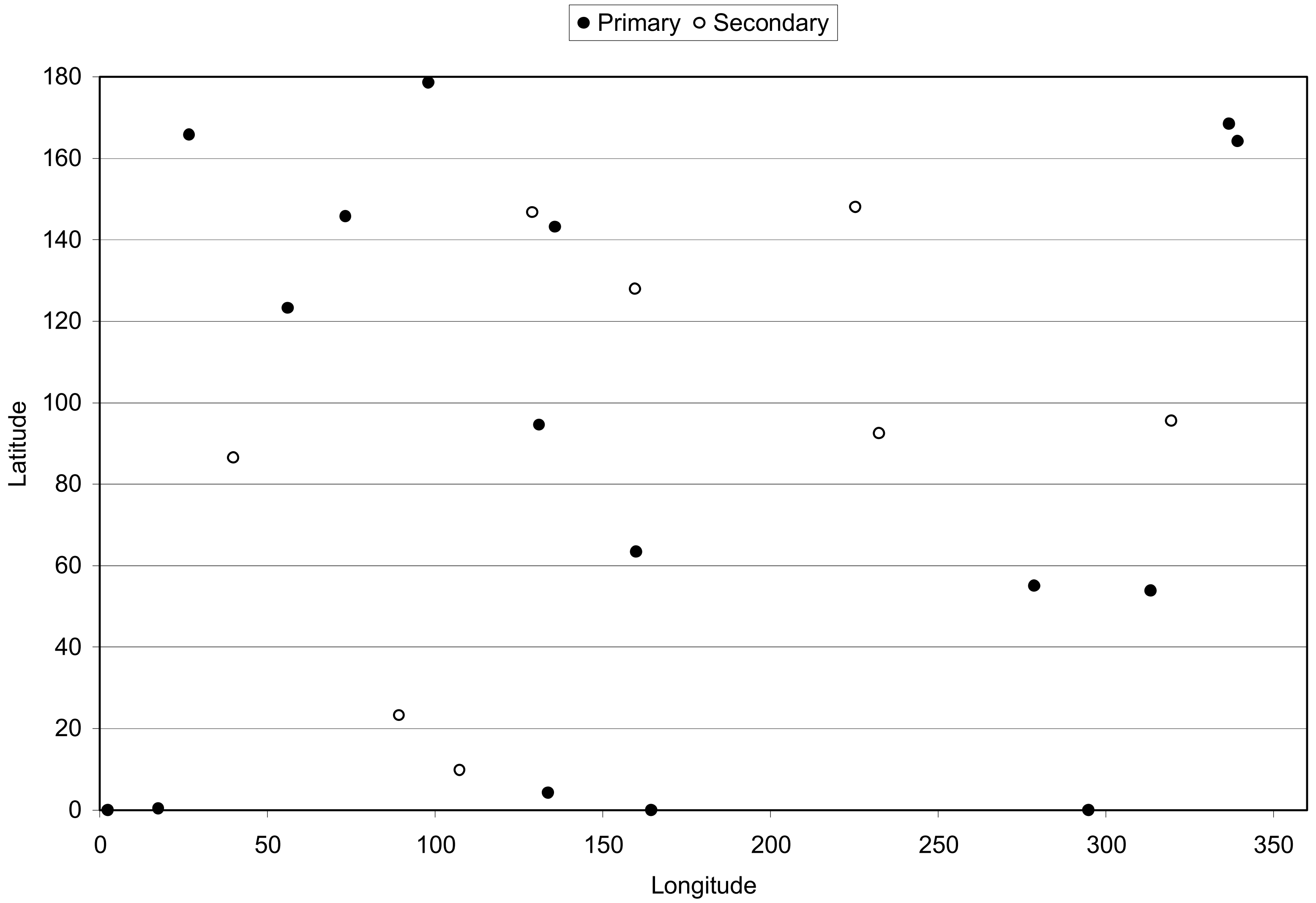

Figure 4-14: Spatial Spot Distribution for spots on the primary (filled circles) and secondary (open circles) components.

## 4.6 Discussion

The ease of confirming the high inclination solution as found by Pribulla et al. (2000) over that of Kjurkchieva (2000) is most likely due to the use of model atmospheres. While it was possible in the past to have a range of valid solutions when using blackbody functions for intensity, the model atmospheres ensure a uniqueness of solution. The confirmation of the inclination as 87.6° definitively sets the masses of the components of the RT And system at $1.10M_{\odot}$ and $0.83M_{\odot}$.

There does not appear to be any preferential location for spots on either component as shown by Figure 4-14. A cursory search for any sort of trends in spot location, intensity, or size with respect to time did result in the apparent tracking of a spot on the secondary component during the 2005 season. The spot appears to move from low to high latitudes at a near constant longitude. Thus, while the starspot distribution in the RT And system is essentially random over long time scales, it is possible to track spots on the time scales observed in this work. With respect to the model fits of the composite and individual light curves, further work should be conducted in order to obtain better fits, especially in the U filter where the secondary is not properly fitted. This is most likely a result of either incorrect surface temperatures or possibly incorrect values for limb darkening or other physical constants. It is possible that by when the U filter is properly fitted, the spot solutions will be more accurate. In addition, further observation, as well as re-analysis of published spot solutions and/or light curves may result in definite spot trends for the system and a deeper understanding of RS CVn systems.

# V. TU BOOTIS

## 5.1 Background

TU Boo, [RA: 14 04 59, Dec: +30 00 00, $V_{mag} \approx 12$], is a short period system, ($P \approx 0.324$ days), and classified as a W Uma type eclipsing binary. It has only been subjected to a single published analysis, that of Niarchos, Hoffmann, and Duerbeck (1996), who modeled the system in 1995 based on only B and V light curves obtained photoelectrically in 1982, and obtained a slightly over-contact solution for the system. They derived $T_{pri}$ = 5800K, $T_{sec}$ = 5787K, i = 88°, q ($m_{pri}/m_{sec}$) = 2.01, $r_{pri}$ = 0.510, and $r_{sec}$ = 0.396. They also identified several interesting aspects of the system, namely that although its physical characteristics, such as P < 0.4 days and T < 6000K, would classify it as a W-type W Uma system, the fact that its secondary eclipse is total, resulting from significantly different component masses, would normally classify it as an A-type W Uma system, which are usually hotter and have longer periods. Thus, the system must truly be at a unique place in its evolution and deserves a detailed modern analysis.

## 5.2 Observations

Observations in U, B, V, R, and I were taken with Lowell Observatory's 42" Hall Telescope and FLI SITe 2048x2048 CCD camera, cooled by liquid nitrogen to -133°C, on the nights of April $20^{th}$, $21^{st}$, $22^{nd}$, $23^{rd}$, and $24^{th}$ in 2006. Differential photometry was performed via MaximDL with respect to GSC 2012-878, 2545-811, 2545-1000, 2012-479, 2012-831, which were photometrically calibrated by observations of Landolt field standards. All times were corrected to HJD.

## 5.3 Minimum Timings and O-C Diagram

All observed times of minimum were determined via the method of Kwee-van Woerden, and shown in Table 5-1 with errors, employed filter, and type (primary or secondary eclipse). All previously published times of minima available were compiled and assigned a weight that was inversely proportional to its error. In cases where no error was given, a value of ±.005 days was assumed. A linear least-squared fit to the data was then performed for data after JD 2450000 and a new ephemeris calculated to be $T_{pri}$ (HJD) = 2424609.51631688 + 0.32428343•E, where E is the epoch. An O-C diagram of the data is shown in Figure 5-1.

Table 5-1: Observed Times of Minimum for TU Boo

| $T_{min}$ (HJD) | Error (±) | Filter | Type |
|---|---|---|---|
| 2453845.936530 | 0.003593 | V | Pri |
| 2453845.936557 | 0.000638 | B | Pri |
| 2453845.936603 | 0.000877 | R | Pri |
| 2453845.937036 | 0.001312 | I | Pri |
| 2453847.881942 | 0.000149 | U | Pri |
| 2453847.882126 | 0.000729 | B | Pri |
| 2453847.882361 | 0.000591 | V | Pri |
| 2453847.882465 | 0.000760 | R | Pri |
| 2453847.882470 | 0.001102 | I | Pri |
| 2453848.855010 | 0.000266 | U | Pri |
| 2453848.855100 | 0.000917 | B | Pri |
| 2453848.855258 | 0.000145 | V | Pri |
| 2453848.855379 | 0.000506 | I | Pri |
| 2453848.855407 | 0.000301 | R | Pri |
| 2453849.827944 | 0.000761 | B | Pri |
| 2453849.828036 | 0.000175 | U | Pri |
| 2453849.828067 | 0.000341 | V | Pri |

Table 5-1 (Cont.)

| $T_{min}$ (HJD) | Error (±) | Filter | Type |
|---|---|---|---|
| 2453849.828159 | 0.000143 | R | Pri |
| 2453849.828247 | 0.000089 | I | Pri |
| 2453845.774816 | 0.001091 | I | Sec |
| 2453845.775159 | 0.003841 | B | Sec |
| 2453845.775320 | 0.001533 | V | Sec |
| 2453846.747825 | 0.000674 | I | Sec |
| 2453846.747921 | 0.001766 | B | Sec |
| 2453846.747934 | 0.000485 | V | Sec |
| 2453846.748038 | 0.000146 | R | Sec |
| 2453846.748191 | 0.000839 | U | Sec |
| 2453847.720438 | 0.000608 | U | Sec |
| 2453847.720559 | 0.001036 | R | Sec |
| 2453847.720962 | 0.001083 | I | Sec |
| 2453847.721135 | 0.000169 | B | Sec |
| 2453847.721415 | 0.001085 | V | Sec |
| 2453848.693499 | 0.000758 | I | Sec |
| 2453848.693662 | 0.000845 | B | Sec |
| 2453848.693705 | 0.000596 | R | Sec |
| 2453848.693718 | 0.000375 | V | Sec |
| 2453848.693779 | 0.000365 | U | Sec |
| 2453849.666273 | 0.000357 | V | Sec |
| 2453849.666320 | 0.001623 | I | Sec |
| 2453849.667217 | 0.001572 | U | Sec |

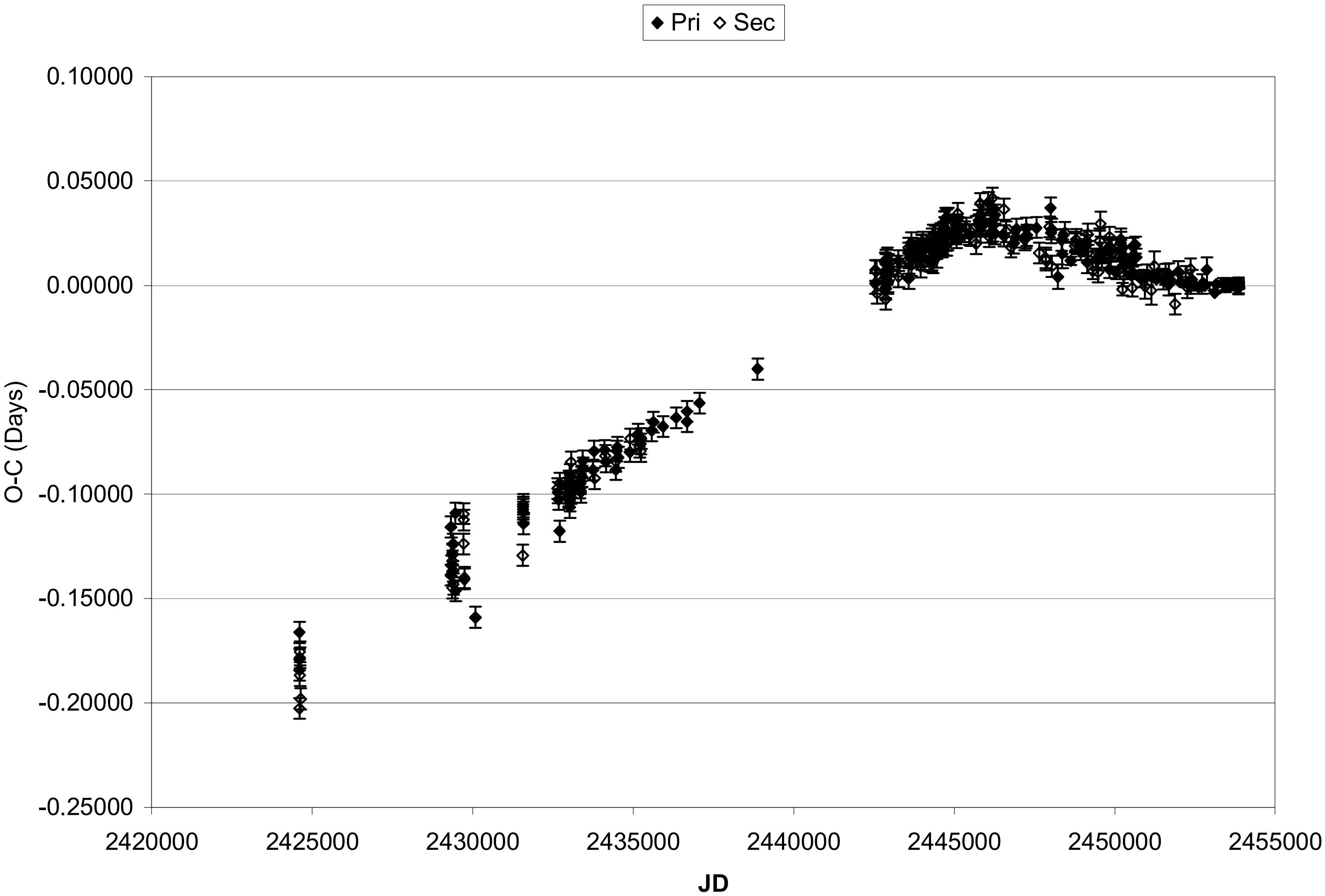

Figure 5-1: O-C Diagram for TU Boo

The major period change just after JD 2445000 was noted by Niarchos (1996), who calculated it to be a period shortening of 0.413 seconds, and noted that it occurred just after their 1982 observations. Figure 5-1 shows evidence for minor but continuous period changes afterwards, with the most noticeable abrupt change around JD 2452000. Performing weighted least squares fits to the data before and after each major change yields period decreases of 0.304 and 0.02 seconds at JD 2445000 and 2452000 respectively. These are most likely due to rapid mass transfers between the two components, and could in theory be used to calculate the amount of mass transferred in each case.

## 5.4 Light Curves and Modeling

All data was compiled into U, B, V, R, and I light curves and simultaneously solved using the ELC program. Since no radial velocity curves are available, in order to set the scale of the system to an appropriate value, which affects the model atmospheres, the mass of the secondary was set to be 0.5 $M_{\odot}$. The values of mass ratio, inclination, each component's fill factor, each components temperature, as well as the temperature factor, size, and location of two spots on the primary and one spot on the secondary, were allowed to vary. The final solutions allowed for a grid resolution of 1600 points for each component and 360 points per model light curve. The observed light curves with model fits are shown in Figure 5-2.

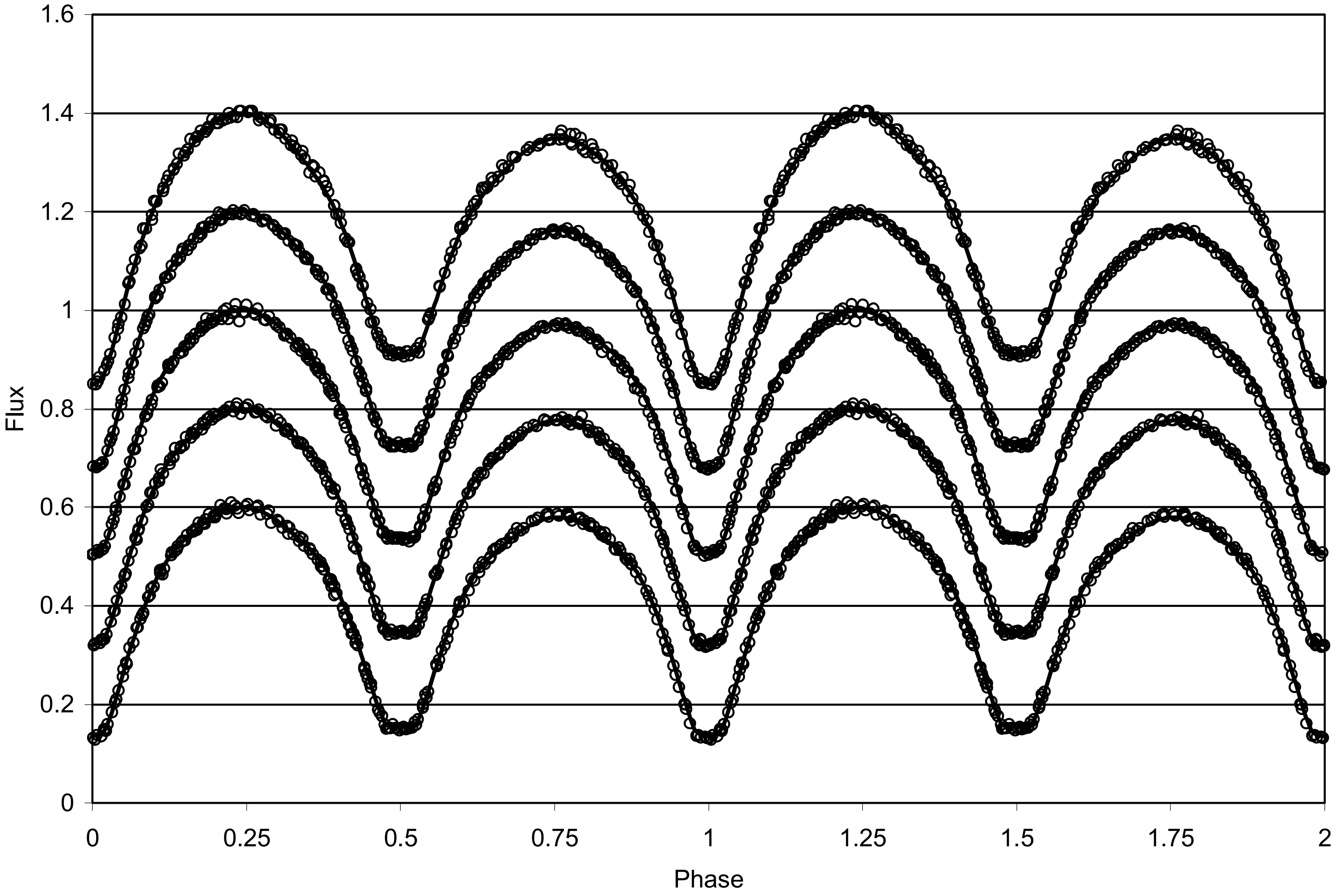

Figure 5-2: Top to bottom: Observed U, B, V, R, and I light curves (open circles) for TU Boo with model fits (solid lines).

## 5.5 Results

The results for the orbital and spot parameters of TU Boo by modeling the composite light curves is shown in Table 5-2, a geometrical model of the system is shown in Figure 5-3, and a 3D model of the system showing the spot distribution is shown in Figure 5-4.

Table 5-2: Orbital and Spot Solutions for TU Boo

| Parameter | Value |
|---|---|
| Inclination | 88.32° |
| Mass Ratio ($M_{pri}/M_{sec}$) | 2.08 |
| Primary Temperature | 5821K |
| Secondary Temperature | 5691K |
| Fill Factor of Primary | 0.998 |
| Fill Factor of Secondary | 0.989 |
| Fractional Radius of Primary | 0.446 |
| Fractional Radius of Secondary | 0.317 |
| Temperature Factor of Spot 1 on Primary | 0.704 |
| Size of Spot 1 on Primary | 12.4° |
| Latitude of Spot 1 on Primary | 76.6° |
| Longitude of Spot 1 on Primary | 180.9° |
| Temperature Factor of Spot 2 on Primary | 0.960 |
| Size of Spot 2 on Primary | 44.6° |
| Latitude of Spot 2 on Primary | 160.0° |
| Longitude of Spot 2 on Primary | 104.8° |
| Temperature Factor of Spot 1 on Secondary | 0.940 |
| Size of Spot 1 on Secondary | 28.1° |
| Latitude of Spot 1 on Secondary | 45.9° |
| Longitude of Spot 1 on Secondary | 248.9° |

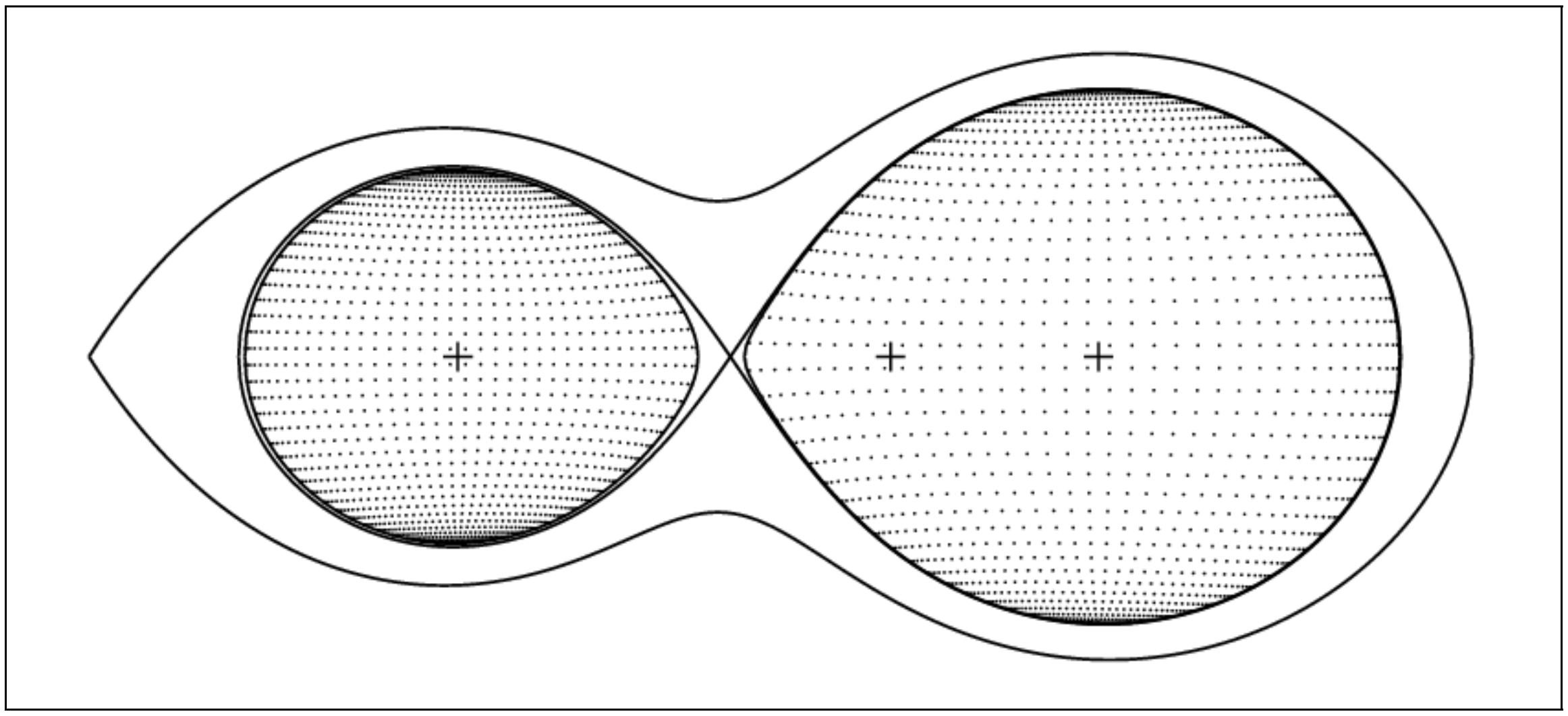

Figure 5-3: A geometrical model of the TU Boo system. The crosses represent the center of each star and the barycenter of the system. The inner solid line represents the Roche Lobe for each star. The primary, hotter star is on the right.

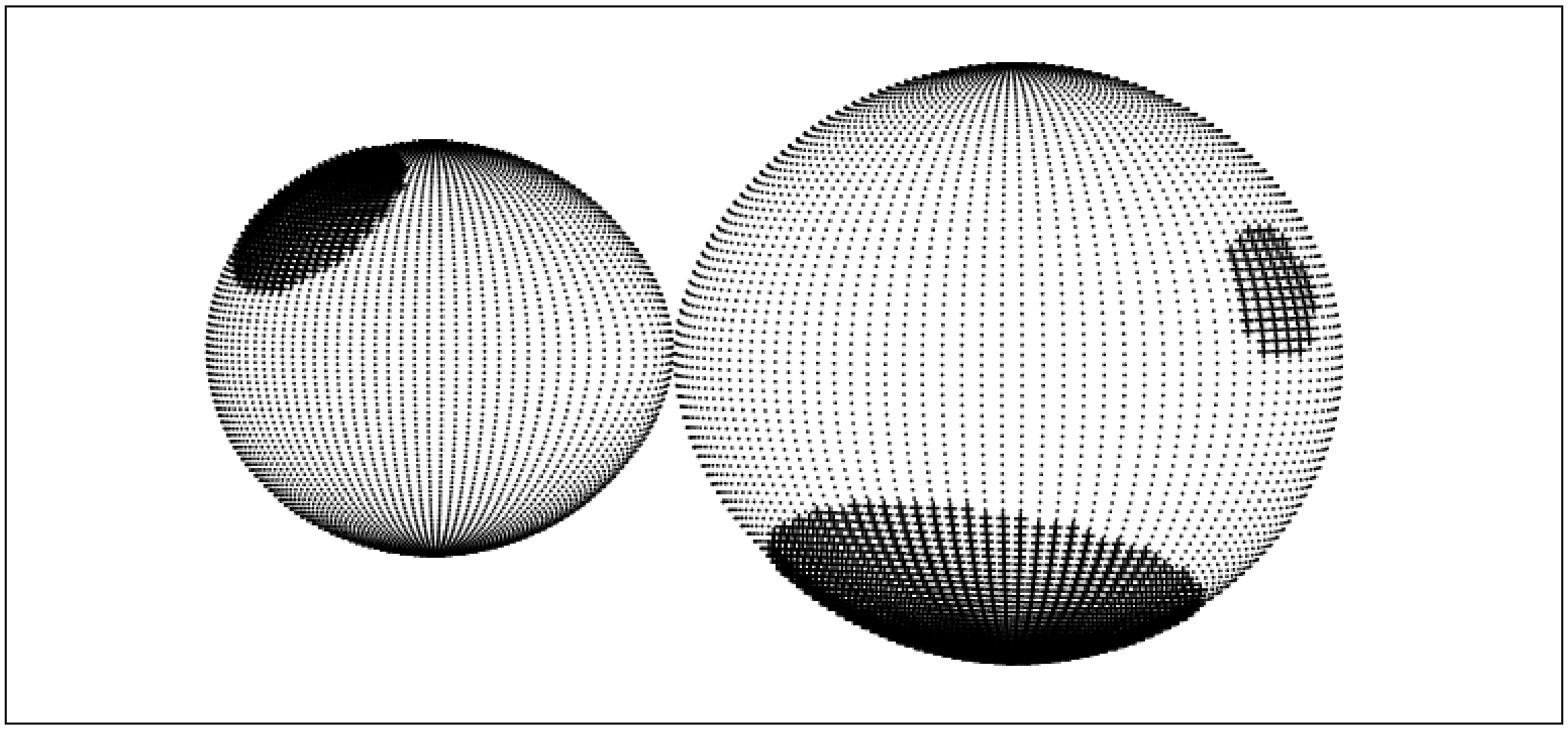

Figure 5-4: A 3D model of TU Boo showing the spot distribution. Note that the smallest spot has a temperature factor ≈ 0.70 while the larger spots are ≈ 0.95.

### 5.6 Discussion

The modeling results for TU Boo seem to be somewhat dependant on the employed spot model. Without spots, the ELC program converges to fill factors of 1.0 for both stars while trying to fit the light curve. However, with the above solution including spots, the primary star 99.8% fills its Roche Lobe while the secondary fills slightly less, leaving the system detached. The use of spots in the model is justified by the inequality of the shoulders in the light curve, with phase 0.25 being brighter than phase 0.75. The only way to explain this asymmetry is the presence of spots. The exact number of spots that need be invoked to fit the light curve while still having a valid solution for the orbital parameters however is unknown, but the derived spot parameters appear to be physically and logically valid. Radial velocity curves might ultimately be needed to definitely distinguish between the possible models.

The re-classification of TU Boo as a near-contact or barely semi-detached system would help in explaining the anomalies reported by Niarchos (1996) that the system appears to be W-type W Uma but has some A-type aspects. The system would in fact still be in the process of evolving into a W-type W Uma system, with mass transfer still occurring from the primary component to the secondary component. The high mass ratio then, normally only exhibited by A-type W Uma systems, is a result of the incomplete evolution. I would like to present the theory that A-type W Uma systems evolve into W-type W Uma systems, in which case TU Boo would be a missing evolutionary link. It is also possible that TU Boo oscillates between a contact and non-contact system due to an inability to achieve thermal equilibrium as generally proposed for W Uma systems by Lucy (1975). Further modeling should be employed to probe if this configuration is truly accurate, as it would be a keystone in understanding the formation scenario of W-type W Uma type systems.

# VI. KV GEMINORUM

## 6.1 Background

The star, KV Gem, [RA: 06 47 13, Dec: +15 43 34, Vmag ≈ 12.5], was first designated as a variable star by Kukarkin et. al (1968), and was classified as a $RR_c$ Lyrae type, a type of short-period pulsating star, by the General Catalogue of Variable Stars, or GCVS (Kholopov, 1985, 1987). The period, according to the GCVS, was 0.2185467 days, which would have made it the shortest period RR Lyrae type star known, and thus was rightfully under suspicion as being misclassified. This could happen if it was a near-contact eclipsing binary with nearly equal minima that were not distinguishable from each other. If plotted with half its true period, it would mimic the light curve of a pulsating star. In 1991, Schmidt reclassified the system as an eclipsing binary based on new photoelectric observations that revealed details of the light curve morphology and color, and thus doubled the period to 0.43713 days. However, the period was based on limited minimum timings, and the photoelectric observations themselves had a large amount of scatter and the exact type of eclipsing binary was unable to be determined. Schmidt (1991) estimates that as many as four or five-hundred stars are misclassified as RR Lyrae type in the GCVS, and thus modern observation and modeling of KV Gem would help to understand the nature of these systems.

## 6.2 Observations

Observations of KV Gem were taken with Emory Observatory's 24" telescope and an Apogee 47 CCD camera cooled to -30°C on the nights of February 21st, 26th, and 27th of 2007 in B, V, R, and I filters. Differential photometry was performed via MaximDL with respect to GSC 1330-0101, 1330-1460, 1330-0119, 1330-0741, and another star lacking a designation. All reference stars were photometrically calibrated with respect to Landolt standards. All times were corrected to HJD.

## 6.3 Minimum Timings and O-C Diagram

All observed times of minimum were determined via the method of Kwee-van Woerden, and are shown in Table 6-1 with errors, employed filter, and type (primary or secondary eclipse). All previously published times of minima available were compiled and assigned a weight that was inversely proportional to their error. In cases where no error was given, a value of ±.005 days was assumed. A linear least-squared fit to the data was then performed and the new ephemeris calculated to be $T_{pri}$ (HJD) = 2450839.32621582 + 0.35852290•E, where E is the epoch. An O-C diagram of the data is shown in Figure 6-1. The quasi-sinusoidal trend in the O-C diagram, if real, would most likely be due to spots in the system.

Table 6-1: Observed Times of Minimum for KV Gem

| $T_{min}$ (HJD) | Error (±) | Filter | Type |
|---|---|---|---|
| 2454158.533387 | 0.000875 | R | Pri |
| 2454158.534536 | 0.000992 | B | Pri |
| 2454159.605693 | 0.000245 | I | Pri |
| 2454159.605849 | 0.000408 | B | Pri |
| 2454159.605975 | 0.000296 | R | Pri |

Table 6-1 (Cont.)

| | | | |
|---|---|---|---|
| 2454159.606242 | 0.000170 | V | Pri |
| 2454158.710547 | 0.000237 | I | Sec |
| 2454158.711410 | 0.000162 | V | Sec |
| 2454158.711736 | 0.000241 | B | Sec |
| 2454158.712346 | 0.000407 | R | Sec |
| 2454159.786679 | 0.000368 | B | Sec |
| 2454159.787231 | 0.000543 | I | Sec |
| 2454159.787925 | 0.001062 | V | Sec |
| 2454159.788062 | 0.001052 | R | Sec |

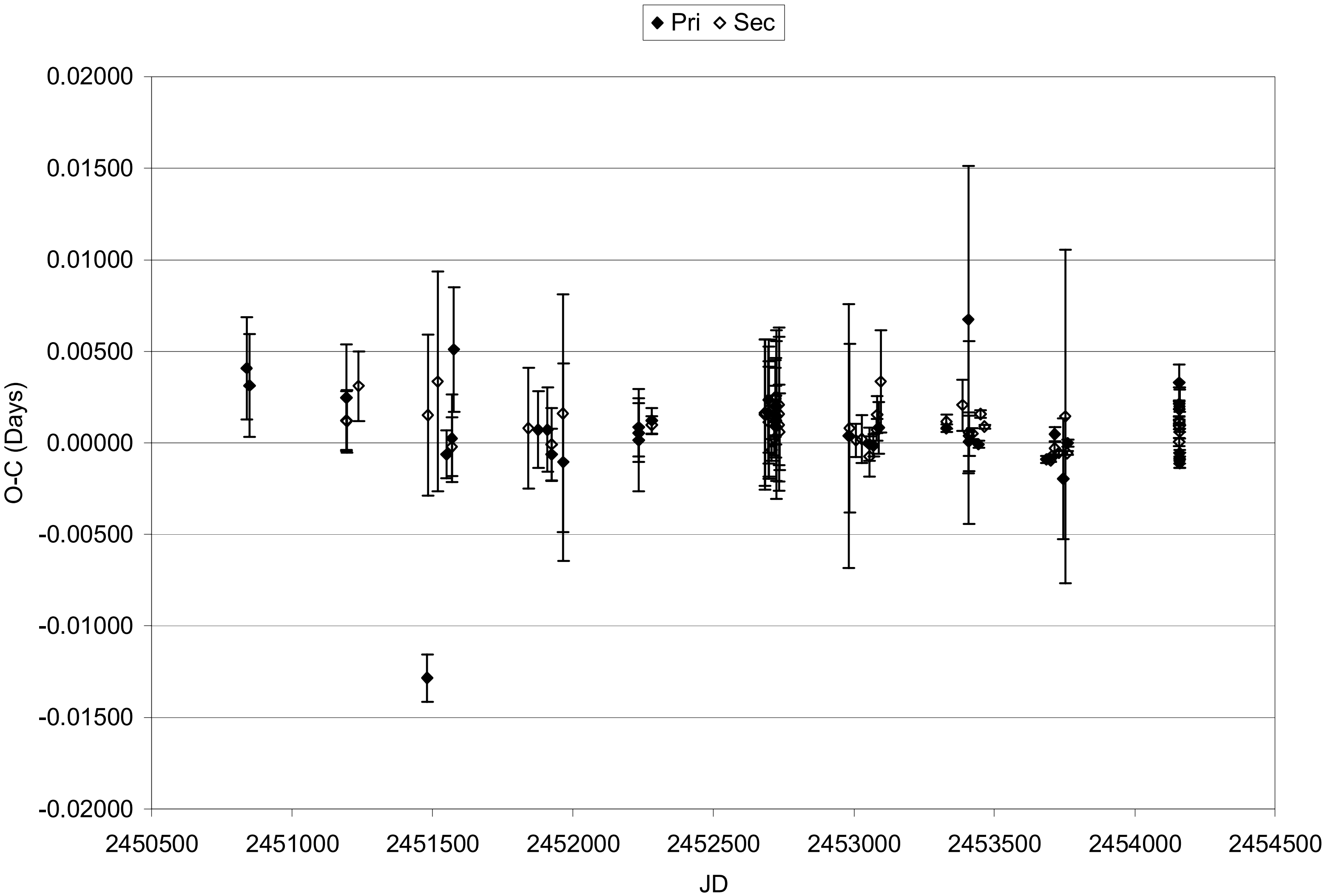

Figure 6-1. O-C Diagram for KV Gem

### 6.4 Light Curves and Modeling

All data was compiled into B, V, R, and I light curves and simultaneously solved using the ELC program. Since no radial velocity curves are available, in order to set the scale of the system to an appropriate value, which affects the model atmospheres, the separation of the two components was set to $2.25R_\odot$, based on assumed masses for the initial derivations of temperature. The values of mass ratio, inclination, each component's fill factor, each component's temperature, as well as the temperature factor, size, and location of one spot on the primary and one spot on the secondary, were allowed to vary. The presence of spots in the system is supported by the unequal height of the shoulders around phase 0.25 and 0.75. The final solutions allowed for a grid resolution of 1600 points for each component and 360 points per model light curve. The observed light curves with model fits are shown in Figure 6-2.

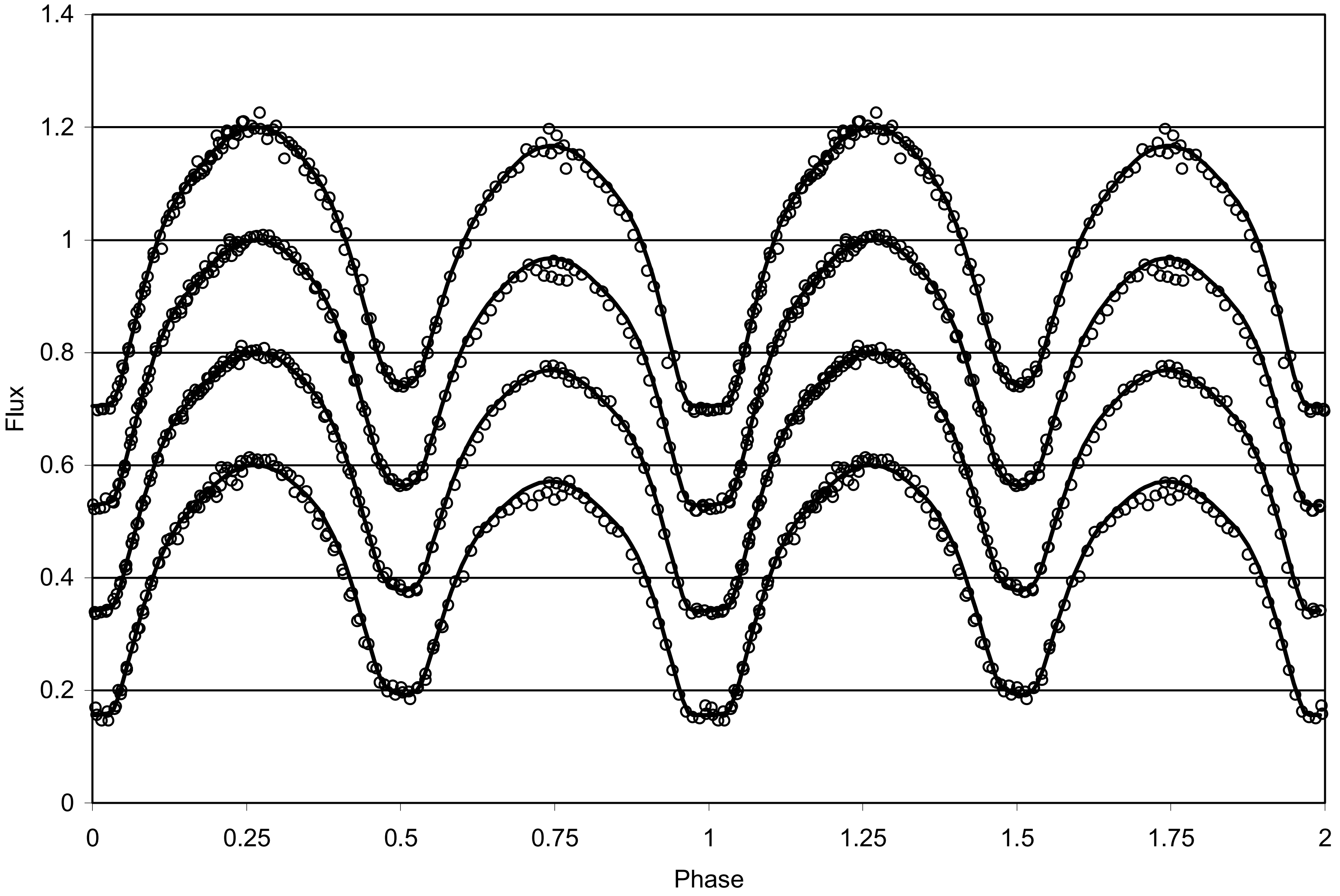

Figure 6-2: Top to bottom: Observed B, V, R, and I light curves (open circles) for KV Gem with model fits (solid lines).

## 6.5 Results

The results for the orbital and spot parameters of KV Gem by modeling the composite light curves are shown in Table 6-2, a geometrical model of the system is shown in Figure 6-3, and a 3D model of the system showing the spot distribution is shown in Figure 6-4.

Table 6-2: Orbital and Spot Solutions for KV Gem

| Parameter | Value |
|---|---|
| Inclination | 89.30° |
| Mass Ratio ($M_{pri}/M_{sec}$) | 0.348 |
| Primary Temperature | 6000K |
| Secondary Temperature | 5799K |
| Fill Factor of Primary | 0.979 |
| Fill Factor of Secondary | 0.990 |
| Fractional Radius of Primary | 0.291 |
| Fractional Radius of Secondary | 0.474 |
| Temperature Factor of Spot 1 on Primary | 0.678 |
| Size of Spot 1 on Primary | 28.1° |
| Latitude of Spot 1 on Primary | 124.2° |
| Longitude of Spot 1 on Primary | 152.7° |
| Temperature Factor of Spot 1 on Secondary | 0.636 |
| Size of Spot 1 on Secondary | 37.4° |
| Latitude of Spot 1 on Secondary | 155.8° |
| Longitude of Spot 1 on Secondary | 187.0° |

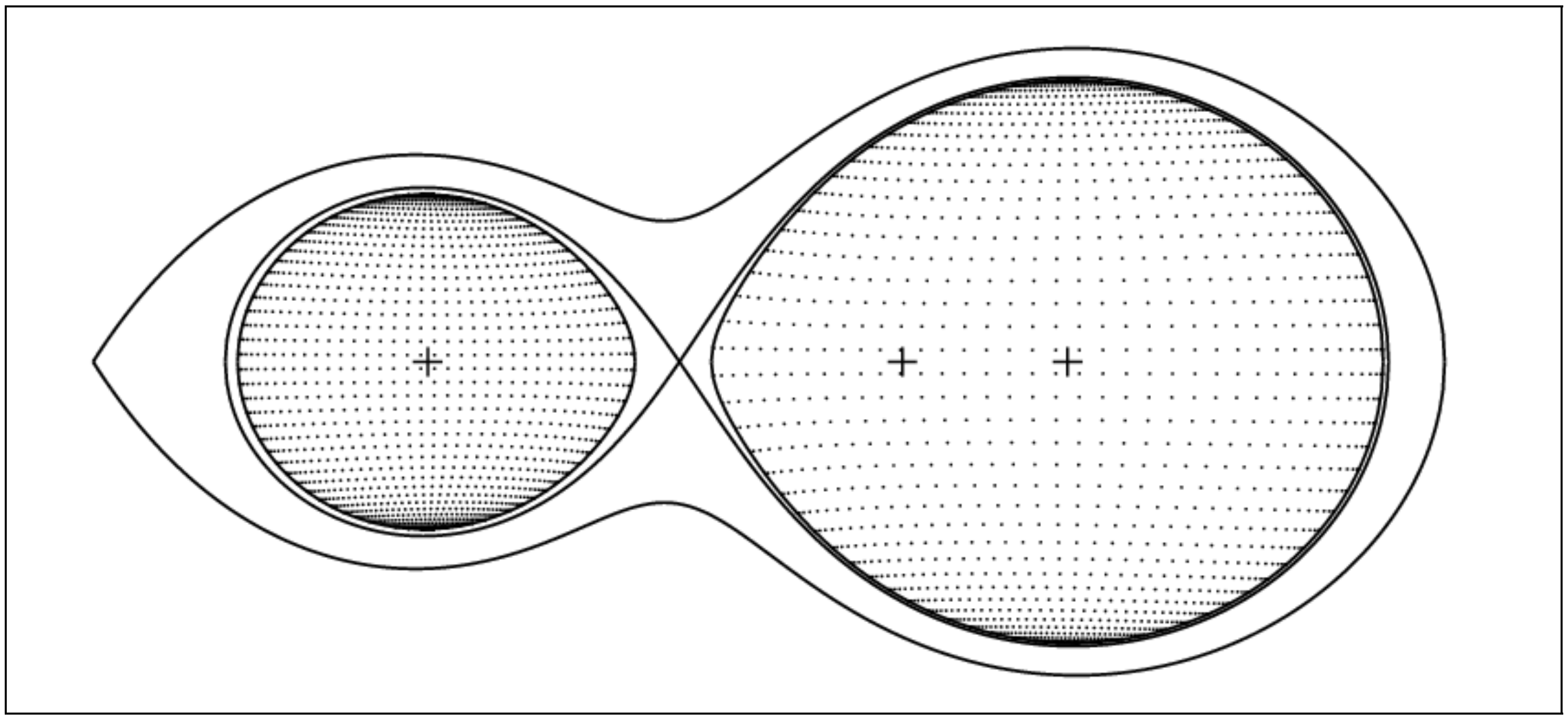

Figure 6-3: A geometrical model of the KV Gem system. The crosses represent the center of each star and the barycenter of the system. The inner solid line represents the Roche Lobe for each star. The primary, hotter star is on the left.

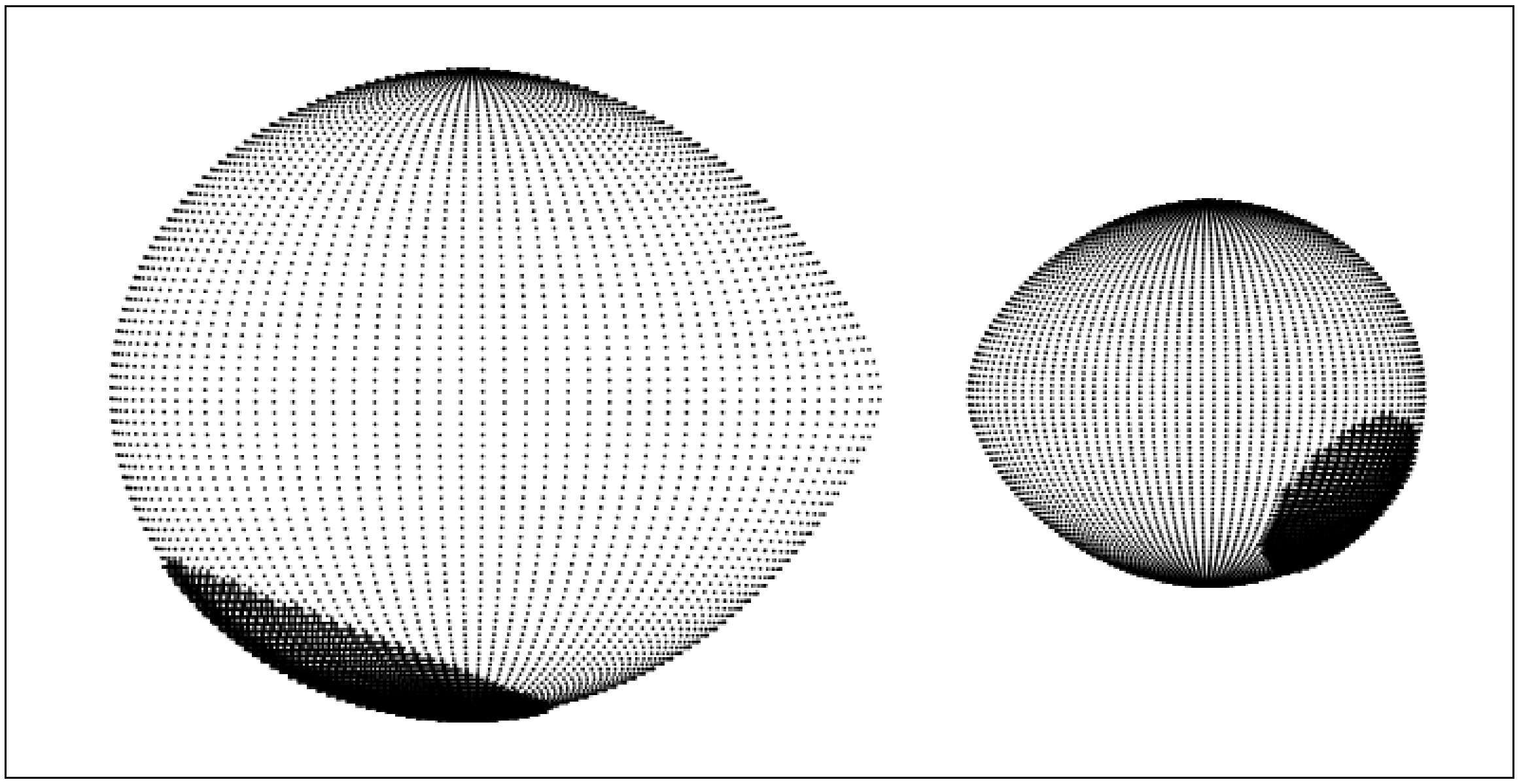

Figure 6-4: A 3D model of KV Gem showing the spot distribution. The spots each have temperature a temperature factor ≈ 0.65.

## 6.6 Discussion

The eclipsing binary nature of KV Gem has been definitively confirmed and the first accurate multi-color light curve obtained. The period has been refined to a value of ≈ 0.358 days, proving the GCVS data extremely unreliable for this system.

The derived orbital and spot models appear to fit the observed light curves extremely well. The most striking feature of this system is the inverse mass ratio; the hotter, primary star is less massive component by almost a factor of three. The fact that the secondary is almost filling its Roche Lobe should also be noted of importance. It is highly unlikely that the system is at its first period of interaction, due to the only 200K difference in surface temperatures, and the inverse mass ratio. A plausible scenario is that the primary used to be more the more massive component, and evolved first, filling its Roche Lobe and transferring mass to the secondary in an A-type W Uma configuration. Mass transfer continued until the secondary became the more massive component, and the system detached. The extra mass speeded up the evolution of the secondary, which is now about to transfer mass back onto the primary, most likely causing the system to come into contact as a W-type W Uma system. Thus, KV Gem, and other misclassified RR Lyrae stars, might be important in understanding stellar evolution in binaries and formation scenarios of W Uma type systems.

# VII. UU LYNCIS

## 7.1 Background

UU Lyn, [RA: 09 15 31, Dec: +42 42 12, Vmag ≈ 11.5], is a barely detached Algol type system consisting of components with roughly 1.4 and 0.6 $M_{\odot}$. The system has been subjected to only one detailed analysis by Yamasaki, Okazaki, and Kitamura in 1983 based on photoelectric observations obtained in 1981. Their findings that each component is critically close to filling their respective Roche lobes put UU Lyn in an interesting place evolutionarily speaking, possibly as a pre-cursor to W Uma type, contact systems.

## 7.2 Observations

Observations of UU Lyn were taken with Emory Observatory's 24" telescope and an Apogee 47 CCD camera cooled to -30°C on the nights of January 30$^{th}$, and February 14$^{th}$, 15$^{th}$, and 18$^{th}$ of 2007 in B, V, R, and I filters. Differential photometry was performed via MaximDL with respect to GSC 2990-0253, 2990-0321, 2990-0347, and 2990-0461. All reference stars were photometrically calibrated with respect to Landolt standards. All times were corrected to HJD.

## 7.3 Minimum Timings and O-C Diagram

All observed times of minimum were determined via the method of Kwee-van Woerden, and shown in Table 7-1 with errors, employed filter, and type (primary or secondary eclipse). All previously published times of minima available were compiled and assigned a weight that was inversely proportional to its error.

In cases where no error was given, a value of ±.005 days was assumed. A linear least-squared fit to the data was then performed and the new ephemeris calculated to be $T_{pri}$ (HJD) = 2425687.35655483 + 0.468460086 •E, where E is the epoch. An O-C diagram of the data is shown in Figure 7-1.

Table 7-1: Observed Times of Minimum for UU Lyn

| $T_{min}$ (HJD) | Error (±) | Filter | Type |
|---|---|---|---|
| 2454131.784342 | 0.000392 | B | Pri |
| 2454131.784484 | 0.000472 | V | Pri |
| 2454131.784640 | 0.000209 | R | Pri |
| 2454131.785299 | 0.000358 | I | Pri |
| 2454146.775324 | 0.000562 | B | Pri |
| 2454146.774578 | 0.000430 | V | Pri |
| 2454146.775688 | 0.000291 | R | Pri |
| 2454146.775642 | 0.000450 | I | Pri |
| 2454147.711858 | 0.000140 | B | Pri |
| 2454147.712613 | 0.000196 | V | Pri |
| 2454147.711866 | 0.000332 | R | Pri |
| 2454147.711300 | 0.000571 | I | Pri |
| 2454147.945851 | 0.000734 | R | Sec |
| 2454147.948746 | 0.000530 | I | Sec |
| 2454150.757671 | 0.002121 | R | Sec |
| 2454150.761036 | 0.001926 | I | Sec |

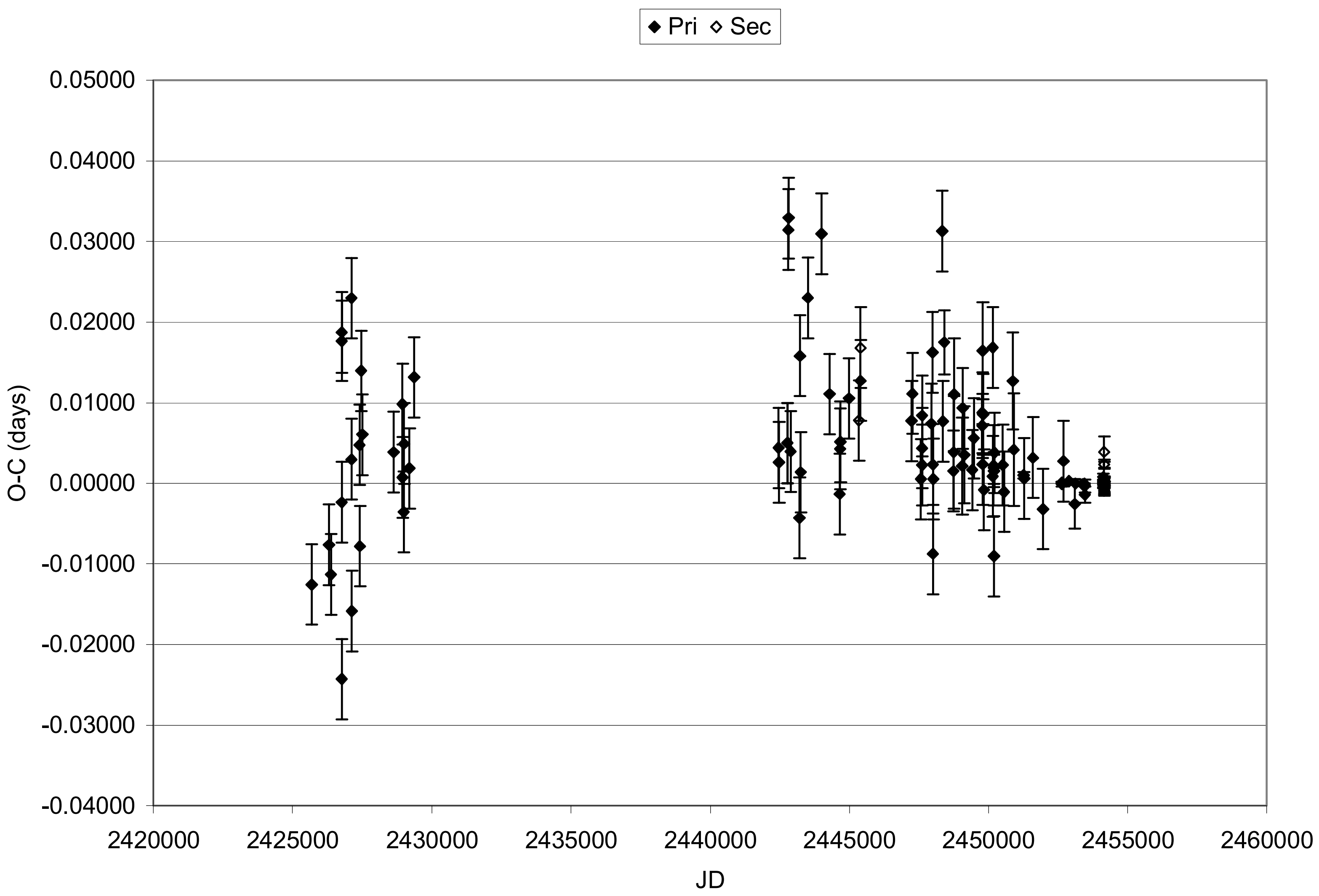

Figure 7-1: O-C Diagram for UU Lyn

## 7.4 Light Curves and Modeling

All data was compiled into B, V, R, and I light curves and simultaneously solved using the ELC program. In order to set the scale of the system to an appropriate value, the separation of $3.2R_{\odot}$ derived by Yamasaki (1981) was fixed. The values of mass ratio, inclination, each component's fill factor, each components temperature were allowed to vary. The final solutions allowed for a grid resolution of 1600 points for each component and 360 points per model light curve. The observed light curves with model fits are shown in Figure 7-2.

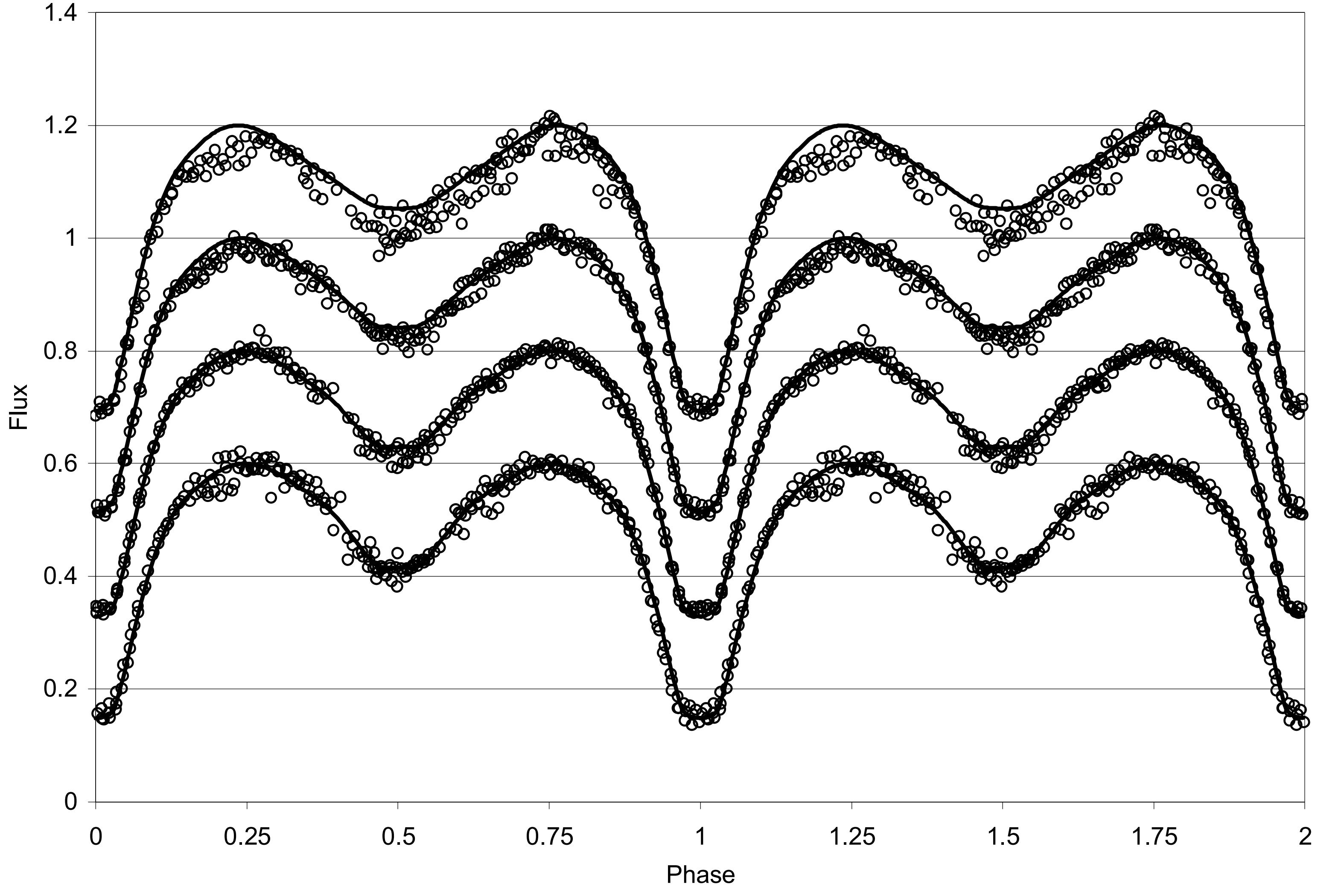

Figure 7-2: Top to bottom: Observed B, V, R, and I light curves (open circles) for UU Lyn with model fits (solid line).

### 7.5 Results

The results for the orbital parameters of UU Lyn by modeling the composite light curves are shown in Table 7-2, and a geometrical model of the system is shown in Figure 7-3.

Table 7-2: Orbital Parameter Solutions for UU Lyn

| Parameter | Value |
|---|---|
| Inclination | 89.31° |
| Mass Ratio ($M_{pri}/M_{sec}$) | 2.81 |
| Primary Temperature | 6795K |
| Secondary Temperature | 4453K |
| Fill Factor of Primary | 1.000 |
| Fill Factor of Secondary | 0.999 |
| Fractional Radius of Primary | 0.499 |
| Fractional Radius of Secondary | 0.317 |

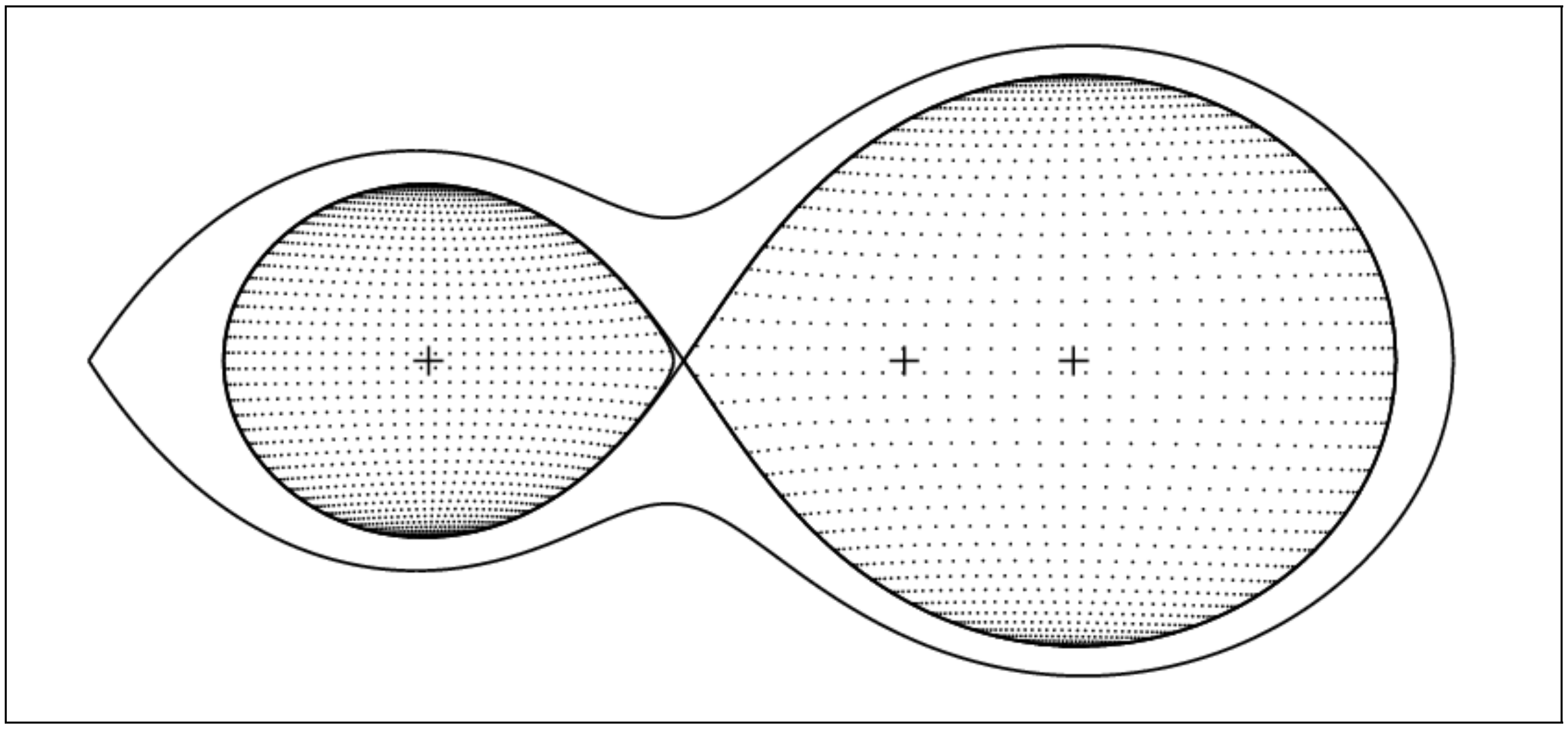

Figure 7-3: A geometrical model of the UU Lyn system. The crosses represent the center of each star and the barycenter of the system. The inner solid line represents the Roche Lobe for each star. The primary, hotter star is on the right.

## 7.6 Discussion

The derived values for the inclination, mass ratio, and surface temperatures appear to match fairly well those of Yamasaki (1983) who derived i = 88.5°, q ($m_{pri}/m_{sec}$) = 2.44, and $T_{pri} - T_{sec} \approx$ 2000-3000K. Yamasaki's (1983) values for the fractional radii translate to fill factors of ≈ 0.98 for each component, and thus classify the system as barely detached. The values found in this work indicate that the system is at least semi-detached and most likely actually in contact. If so, it must be explained how this system's components have not reached thermal equilibrium while the previously presented systems TU Boo and KV Gem have. An explanation first proposed for near-contact systems in general by Shaw (1994), is that UU Lyn is in the process of transitioning from a close but fully detached, non-interacting system to an A-type W Uma system, which again are classified according to P > 0.4 days, $T_{pri}$ > 6000K, and a high mass ratio.

As proposed in sections 5.6 and 6.6, it is possible that while systems are in the A-type phase, as mass is transferred the surface temperatures reach equilibrium, and finally enough mass is transferred from the primary to the secondary to the point where the system detaches. We would then be left with a KV Gem or TU Boo like system, during which the secondary begins to evolve and the system again comes into contact as a W-type W Uma system. In order to definitely prove this theory a large number of systems would likely need to be observed for their astrophysical parameters, and most importantly an indication of their evolutionary status, as we would expect the W-types to be more evolved according to this proposed scenario.

# VIII. MY CYGNI

## 8.1 Background

MY Cyg, [RA: 20 20 03, Dec: +33 56 35, Vmag ≈ 9], is a well-detached Algol type system consisting of two nearly identical stars with about 1.8 $M_{\odot}$, and a period of about four days. The interesting aspect of the system is its slightly eccentric orbit, which results in the displacement of the secondary minima from phase 0.50 in the light curve. Eccentric orbits also experience apsidal motion, the gradual displacement of its longitude of periastron, which is due to both the classical effects of gravitational tidal forces as well as general relativistic effects. Thus, MY Cyg provides a means to observationally test the theoretical predictions of classical gravitational effects and general relativity.

In the first detailed analysis of the system, Williamon (1975) found that when the primary minimum was centered directly at phase 0.0, secondary minimum was centered at phase 0.5022, and derived the values of the eccentricity and longitude of periastron to be $e = 0.010$ and $\omega = 69.6°$. He also found evidence that the secondary had been further displaced from phase 0.50 over the past 45 years prior to the study, and speculated that it was due to apsidal motion. Although the data from Williamon (1975) has been subjected to a modern re-analysis with the Wilson-Devinney code by Tucker, Sowell, and Williamon (2006) for the orbital parameters, a modern light curve is not available in order to measure any changes in $e$ or $\omega$ from 1975 to the present, and hence the motivation for the current study.

## 8.2 Observations

Observations of MY Cyg were taken with Emory Observatory's 24" telescope and a SBIG8 CCD camera cooled to -30°C on the nights of October 20th, 22nd, and 23rd of 2006 in B, V, R, and I filters. The observation dates were chosen so as to obtain data for both eclipses and a shoulder of the light curve. Differential photometry was performed via MaximDL with respect to GSC 2680-0641, 2680-1534, and 2680-731 .Reference stars were not photometrically calibrated, so all measurements are in differential magnitudes. All times were corrected to HJD.

## 8.3 Minimum Timings and O-C Diagram

All observed times of minimum were determined via the method of Kwee-van Woerden, and shown in Table 8-1 with errors, employed filter, and type (primary or secondary eclipse). All previously published times of minima available were compiled and assigned a weight that was inversely proportional to its error. In cases where no error was given, a value of ±.005 days was assumed. A linear least-squared fit to all data was performed and a value of 4.00519019 days obtained for the period. Recent minima from this work and that of Caton (2005) were then used to calculate an initial time of minimum for the primary and secondary separately. The new ephemerides were thus calculated to be

$T_{pri}$ (HJD) = 2426001.42509328 + 4.00519019·•E, and

$T_{sec}$ (HJD) = 2426001.43775391 + 4.00519019•E, where E is the epoch.

An O-C diagram created with the primary ephemeris is shown in Figure 8-1.

Taking the times of minimum from Williamon (1975) as the most accurate in that time period, the difference of the averaged O-C values between primary and secondary minima is calculated to be 0.00919 days. Taking the times of minima presented in this work, as well as those of Caton (2005), the difference of the averaged O-C values between primary and secondary minima is calculated to be 0.01237 days. Dividing by the orbital period yields projected deviations of the secondary from phase 0.50 by 0.00229 at the era of Williamon's study and 0.00309 in this one. This is in good agreement with Williamon's (1975) published deviation of 0.0022. As well, taking this work's phased light curve with primary placed exactly at phase 0.0, the Kwee-van Woerden method was used to determine the phase occurrence of secondary in each filter, and the resulting weighted average yielded a value of 0.50314, a displacement of 0.00314 from phase 0.50, well in agreement with the value of 0.00309 from O-C timings.

Table 8-1: Minimum Timings for MY Cyg

| $T_{min}$ (HJD) | Error (±) | Filter | Type |
|---|---|---|---|
| 2454029.746978 | 0.000327 | B | Pri |
| 2454029.744578 | 0.000360 | V | Pri |
| 2454029.744448 | 0.000598 | R | Pri |
| 2454029.745125 | 0.000264 | I | Pri |
| 2454031.761635 | 0.000889 | I | Sec |
| 2454031.761581 | 0.000254 | V | Sec |
| 2454031.759089 | 0.001559 | R | Sec |
| 2454031.759525 | 0.001007 | B | Sec |

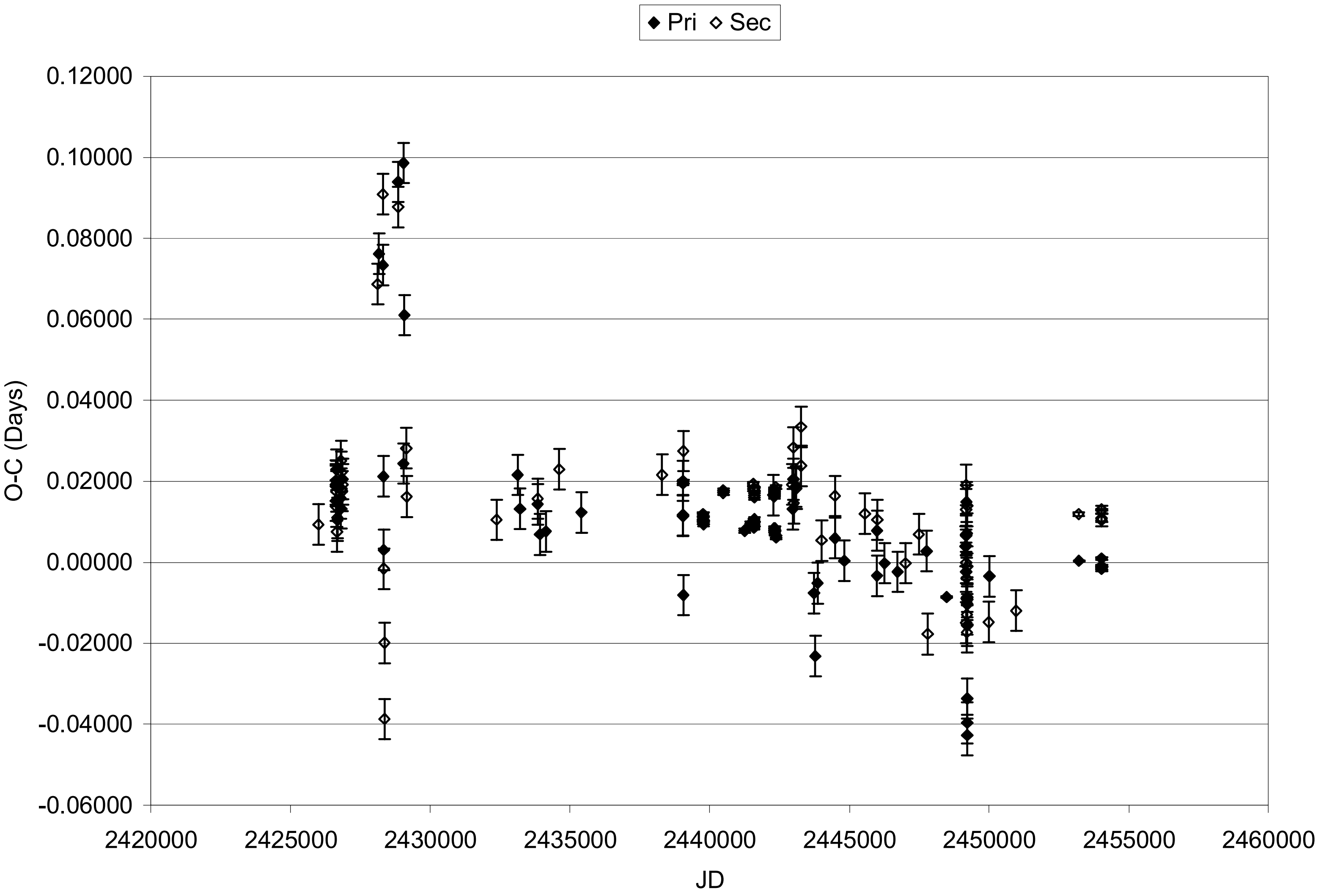

Figure 8-1. O-C Diagram for MY Cyg according to the updated primary ephemeris

Equation 8-1 can be used to calculate ω, the longitude of periastron, given D, the displacement of the secondary in days, P, the period of the system in days, and e, the eccentricity (Sterne, 1939).

$$D = \frac{P}{\pi}\left[\tan^{-1}\left(\frac{e\cos\omega}{(1-e^2)^{1/2}}\right) + \frac{e\cos\omega}{1-e^2\sin^2\omega}(1-e^2)^{1/2}\right] \tag{8-1}$$

Using P = 4.00519019 days, and the same value for e as derived by Williamon (1975), e = 0.010, the longitude of periastron is calculated to be 69.8° based on Williamon's (1975) value for the displacement from his light curves, 68.9° based on the O-C deviation from Williamon's (1975) time of minima, 60.4° based on the value for the displacement from this work's light curves, and 61.0° based on this work's time of minima. Given the possibility of period changes and greater inherent error in using the O-C values to measure the secondary displacement in comparison to compiled light curves, the values for the longitude of periastron of 69.8° at the time of Williamon's (1975) observations and 60.4° in the present study will be assumed as the most accurate. Given an elapsed time of 11,709 days, or 32.058 years between the average of Williamon's (1975) secondary eclipse measurements and this study's, a value of 0.293°/yr for the regression of the longitude of periastron is calculated.

Theoretically, the total apsidal motion is due to two components, the classical which arises from tidal interactions on the oblate stars, and the relativistic, which is a direct consequence of Einstein's Theory of General Relativity. The classical part may be calculated by equation 8-2 (Sterne, 1939), where $f_2(e) = (1 + (3/2)e^2)(1-e^2)^{-5}$, P is the period in days, $M_1$ and $M_2$ are the masses in solar masses, $k_{2,1}$ and $k_{2,2}$ are known as the apsidal motion constants of each component, $r_1$ and $r_2$ are the fractional radii of each component, $\omega_{r,1}$ and $\omega_{r,2}$ are the star's angular rotation speeds, and $\omega_k$ is the mean angular Keplerian velocity, equal to 2π/P.

(8-2)

$$\begin{aligned}\dot{\omega}_{CL}^{theo}(\frac{deg}{yr}) &= 365.25(\frac{360}{P})\{k_{2,1}{r_1}^5[15f_2(e)(\frac{M_2}{M_1}) \\ &+ (\frac{\tilde{\omega}_{r,1}}{\tilde{\omega}_k})^2(\frac{1+M_2/M_1}{(1-e^2)^2})] \\ &+ k_{2,2}{r_2}^5[15f_2(e)(\frac{M_1}{M_2}) \\ &+ (\frac{\tilde{\omega}_{r,2}}{\tilde{\omega}_k})^2(\frac{1+M_1/M_2}{(1-e^2)^2})]\},\end{aligned}$$

Table 1 of Jeffery (1984) gives values of $k_2$ based on computations of main-sequence stellar interiors. Using the derived value of log g = 4.1 by Popper (1971), the value of $k_{2,1} = k_{2,2} = 0.0045$ is chosen. Popper (1971) gives no mention of any line broadening or rotational velocities, and thus we may assume that the stars are tidally locked so that $\omega_{r,1} / \omega_k = \omega_{r,2} / \omega_k = 1$. Popper (1971) and Williamon (1975) give $M_1 = 1.81\ M_\odot$ and $M_2 = 1.78\ M_\odot$, and Williamon (1975) gives $r_1 = 0.141$ and $r_2 = 0.136$. Using e = 0.010 and P = 4.00519019 days as before, a value of 0.257°/yr is obtained.

The equation for the relativistic contribution is ironically much simpler, and given in equation 8-3, where $M_1$ and $M_2$ are in $M_\odot$, and P is in days (Kopal, 1953).

(8-3)

$$\dot{\omega}_{GR}^{theo}(\frac{deg}{yr}) = 9.2872 \times 10^{-3}\frac{(M_1+M_2)^{2/3}}{(P/2\pi)^{5/3}(1-e^2)}$$

Using the values given above, the theoretical relativistic contribution is calculated to be 0.0461°/yr. Adding together the classical and relativistic contributions yields a total theoretical rate of change of the longitude of periastron of 0.303°/yr. This is in excellent agreement with the observed rate of 0.293°/yr.

### 8.4 Light Curves and Modeling

The observed compiled light curves for MY Cyg are shown in Figure 8-2. Modeling of the system based on the new data was not performed due to incompleteness of the curve.

### 8.5 Results

Since modeling of the system was not performed, there are no results to report other than those reported in section 8.3.

### 8.6 Discussion

The observational and theoretical determinations of the rate of change of the longitude of periastron explain the long-term shifts in the O-C noticed by Williamon (1975). As well the very close match between observation and theory validates the classical derivation of Sterne (1939), General Relativity, and the models for internal stellar structure as derived by Jeffery (1984).

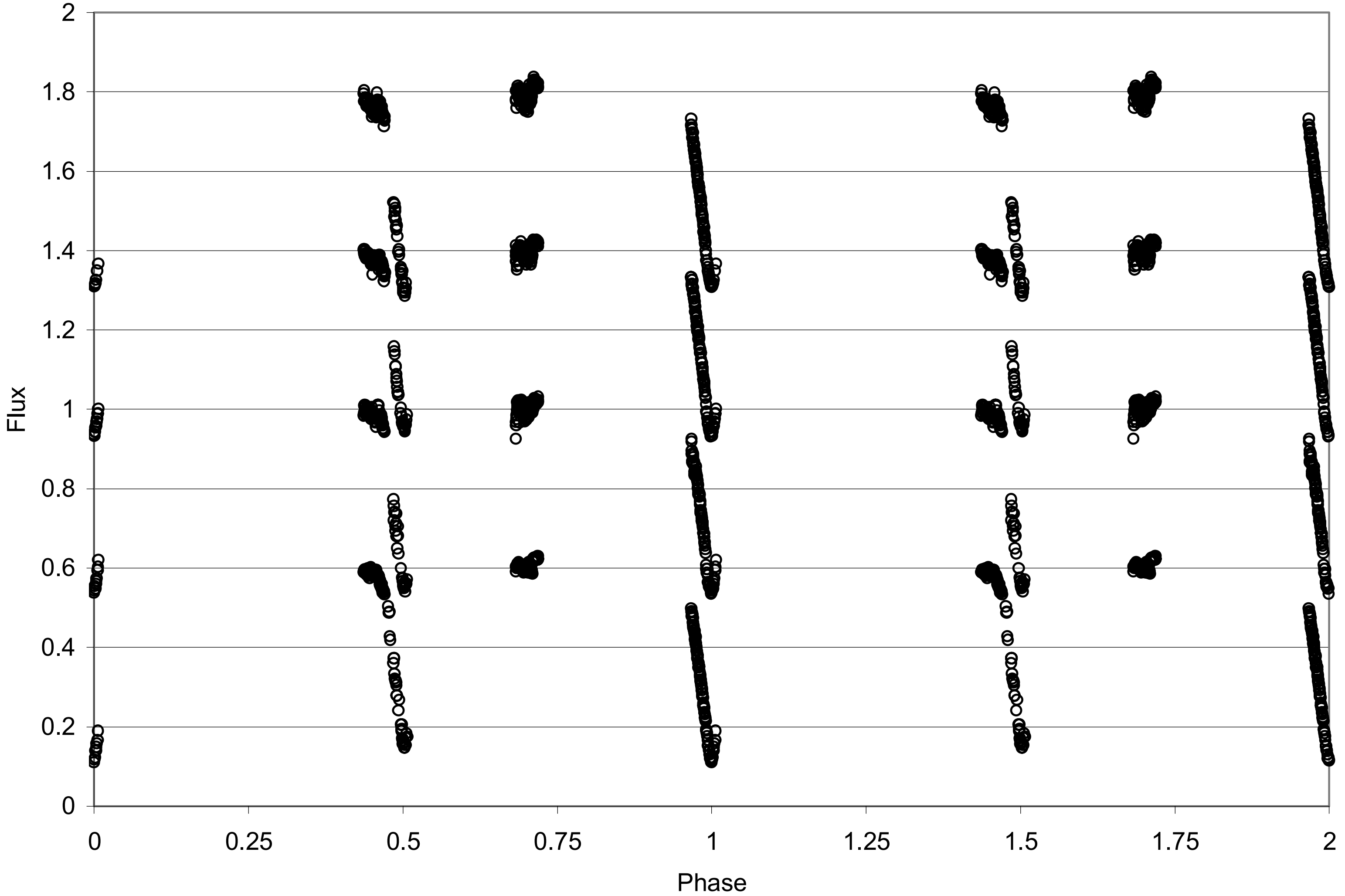

Figure 8-2: Top to bottom: Observed B, V, R, and I light curves (open circles) for MY Cyg.

# IX. KR PERSEI

## 9.1 Background

KR Per, [RA: 04 37 09, Dec: +44 12 40, Vmag ≈ 10.5], is a detached system consisting of nearly equal components of spectral type F5V. The period of the system is almost exactly one day, and thus it is almost impossible to obtain a complete light curve from one location, as one would see the exact same portion of the light curve night after night. This has limited it to only one detailed study by Chen et al. (1985) who combined efforts at the Fernbank Science Center Observatory in Atlanta, GA, USA and Yunnan Observatory in Kunming, Yunnan, China during the 1982-1983 observing season. Aside from the basic geometric parameters they found a small eccentricity to the orbit, and give $e = 0.009$ and $\omega = 169°$. As the system has been mostly neglected since, a re-observation and compilation of available data was needed.

## 9.2 Observations

Observations of KR Per were taken with Emory Observatory's 24" telescope and an Apogee 47 CCD camera cooled to -30°C on the nights of February 7$^{th}$ and 21$^{st}$, March 6$^{th}$, April 17$^{th}$, and October 22$^{nd}$ of 2006 in U, B, V, R, and I filters. Differential photometry was performed via MaximDL with respect to GSC 2892-1153 and 2892-0516. Reference stars were not photometrically calibrated, so all measurements are in differential magnitudes. All times were corrected to HJD.

## 9.3 Minimum Timings and O-C Diagram

All observed times of minimum were determined via the method of Kwee-van Woerden, and shown in Table 9-1 with errors, employed filter, and type (primary or secondary eclipse). All previously published times of minima available were compiled and assigned a weight that was inversely proportional to its error. In cases where no error was given, a value of ±.005 days was assumed. A linear least-squared fit to all data was performed and a value of 0.99607801 days obtained for the period. The new ephemeris was thus calculated to be

$T_{pri}$ (HJD) = 2429491.00857035+ 0.99607801·•E, where E is the epoch.

An O-C diagram created with the new ephemeris is shown in Figure 9-1.

Table 9-1: Observed Times of Minima for KR Per

| $T_{min}$ (HJD) | Error (±) | Filter | Type |
|---|---|---|---|
| 2454031.879195 | 0.000147 | I | Sec |
| 2454031.879297 | 0.000248 | B | Sec |
| 2454031.879396 | 0.000316 | R | Sec |
| 2454031.879642 | 0.000208 | V | Sec |
| 2454031.879742 | 0.000397 | U | Sec |

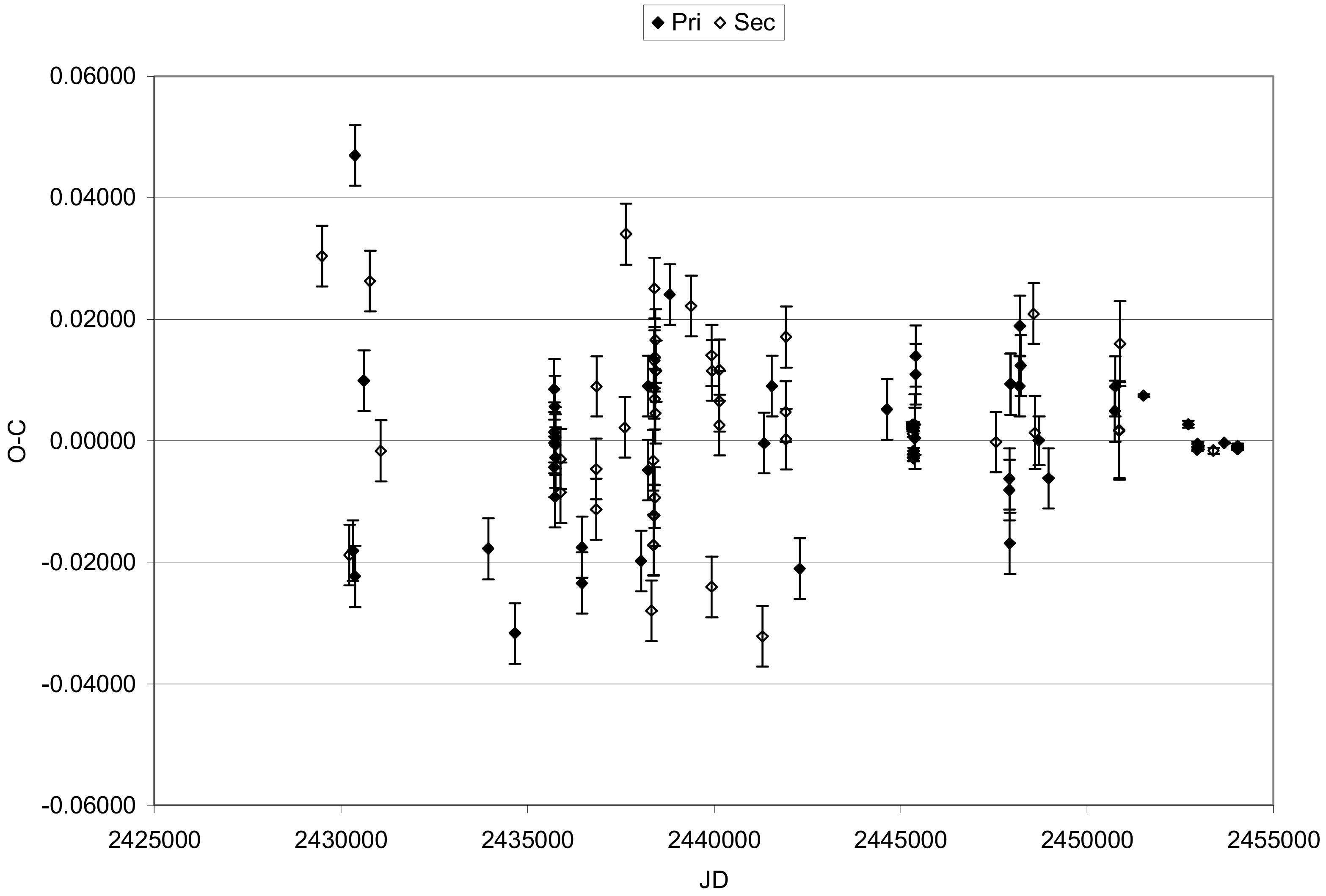

Figure 9-1: O-C Diagram for KR Per

## 9.4 Light Curves and Modeling

The observed light curves for KR Per are shown in Figure 9-2. However, not enough of the curve is available for modeling, and the data used in Chen et al. (1985) was not published for a modern analysis.

## 9.5 Results

A lack of an adequate number of recent accurate minimum timings precludes the possibility of attaining an accurate measurement of the current longitude of periastron. As well, without spectroscopic radial velocity data, a large number of assumptions would have to be made in order to calculate a theoretical value for the yearly change in the longitude of periastron.

## 9.6 Discussion

The fact that eccentricity exists in such a short period system certainly warrants further study via both spectroscopic and photometric data. Efforts should be made to obtain a radial velocity curve and observe primary and secondary minima by observing the system at the very beginning and end of each observing season.

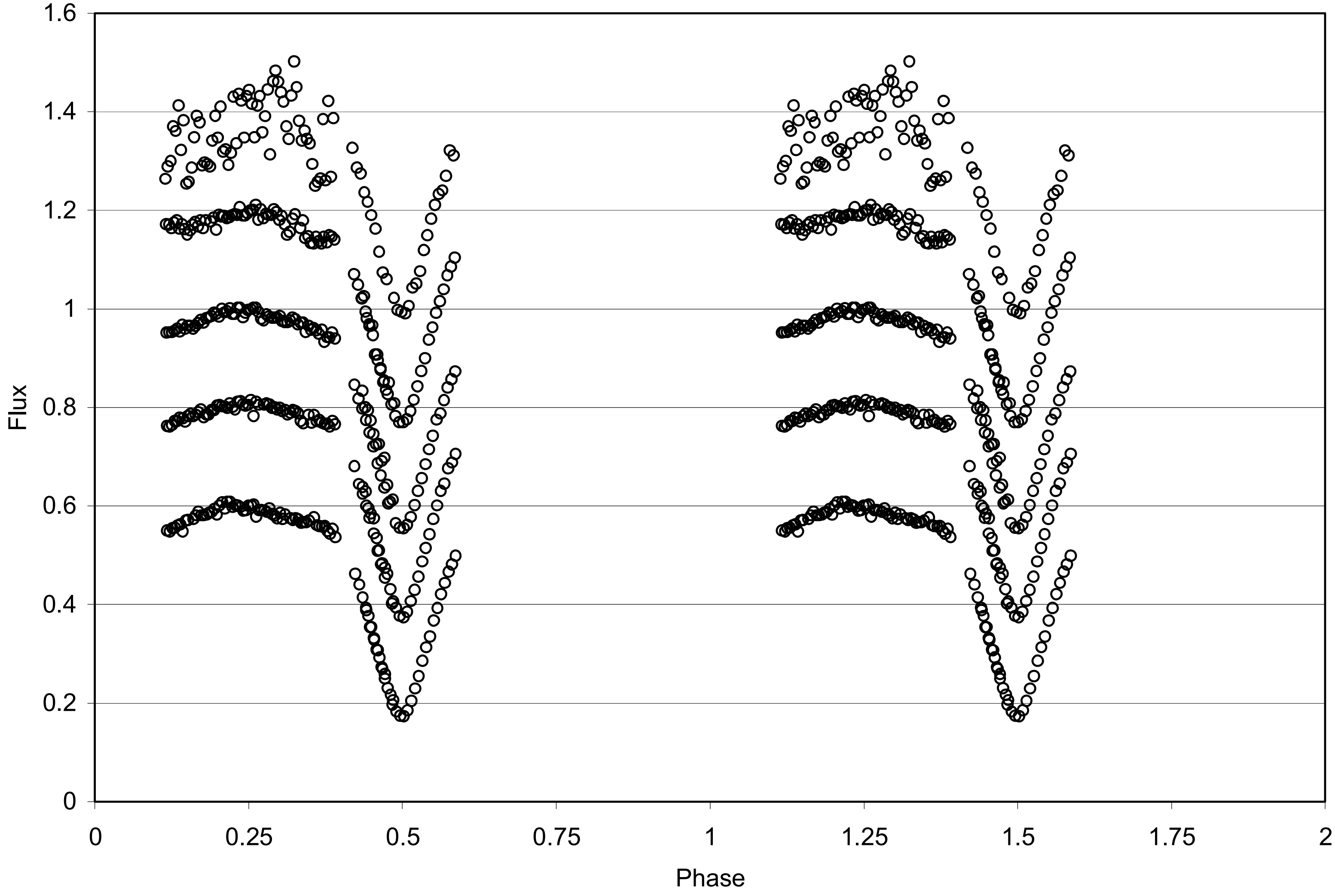

Figure 9-2: Top to bottom: Observed U, B, V, R, and I light curves (open circles) for KR Per.

# X. RU ERIDANIS

## 10.1 Background

RU Eri, [RA: 03 54 44, Dec: -14 56 07, Vmag ≈ 10], is a barely detached Algol-like system very similar to UU Lyn, with a period of about 0.63 days. There have been three studies of the system by Sarma and Sanwal in 1981, Russo in 1982, and Nakamura, Yamasaki, and Kitamura in 1984. Nakamura's (1984) solutions, which was the first to employ radial velocities and the Wilson-Devinney code, found $M_{pri} = 1.4M_{\odot}$, $M_{sec} = 0.76M_{\odot}$, $R_{pri} = 1.6R_{\odot}$, $R_{sec} = 1.2R_{\odot}$, and $i = 76.5°$, based on photoelectric B and V light curves. However, there is no derivation of surface temperatures, which would be helpful in studying the evolution of thermally decoupled, barely detached systems, to thermally coupled contact systems.

## 10.2 Observations

Observations of RU Eri were taken with Emory Observatory's 24" telescope and an SBIG8 CCD camera cooled to -30°C on the nights of October 23rd and November 19th of 2006 in B, V, R, and I filters. Differential photometry was performed via MaximDL with respect to GSC 5882-1173. The reference star was not photometrically calibrated, so all measurements are in differential magnitudes. All times were corrected to HJD.

## 10.3 Minimum Timings and O-C Diagram

All observed times of minimum were determined via the method of Kwee-van Woerden, and shown in Table 10-1 with errors, employed filter, and type (primary or secondary eclipse). All previously published times of minima available were compiled and assigned a weight that was inversely proportional to its error. In cases where no error was given, a value of ±.005 days was assumed. A linear least-squared fit to all data was performed and a value of 0.63219840 days obtained for the period. The new ephemeris was thus calculated to be

$T_{pri}$ (HJD) = 2440188.37208868 + 0.63219840•E, where E is the epoch.

An O-C diagram created with the new ephemeris is shown in Figure 10-1.

Table 10-1: Observed Times of Minima for RU Eri

| $T_{min}$ (HJD) | Error (±) | Filter | Type |
|---|---|---|---|
| 2454032.885072 | 0.000095 | B | Pri |
| 2454032.885521 | 0.000145 | V | Pri |
| 2454032.885000 | 0.000157 | R | Pri |
| 2454032.884375 | 0.000225 | I | Pri |
| 2454059.755276 | 0.000571 | B | Sec |
| 2454059.752313 | 0.001852 | V | Sec |
| 2454059.751609 | 0.000939 | R | Sec |
| 2454059.749803 | 0.000776 | I | Sec |

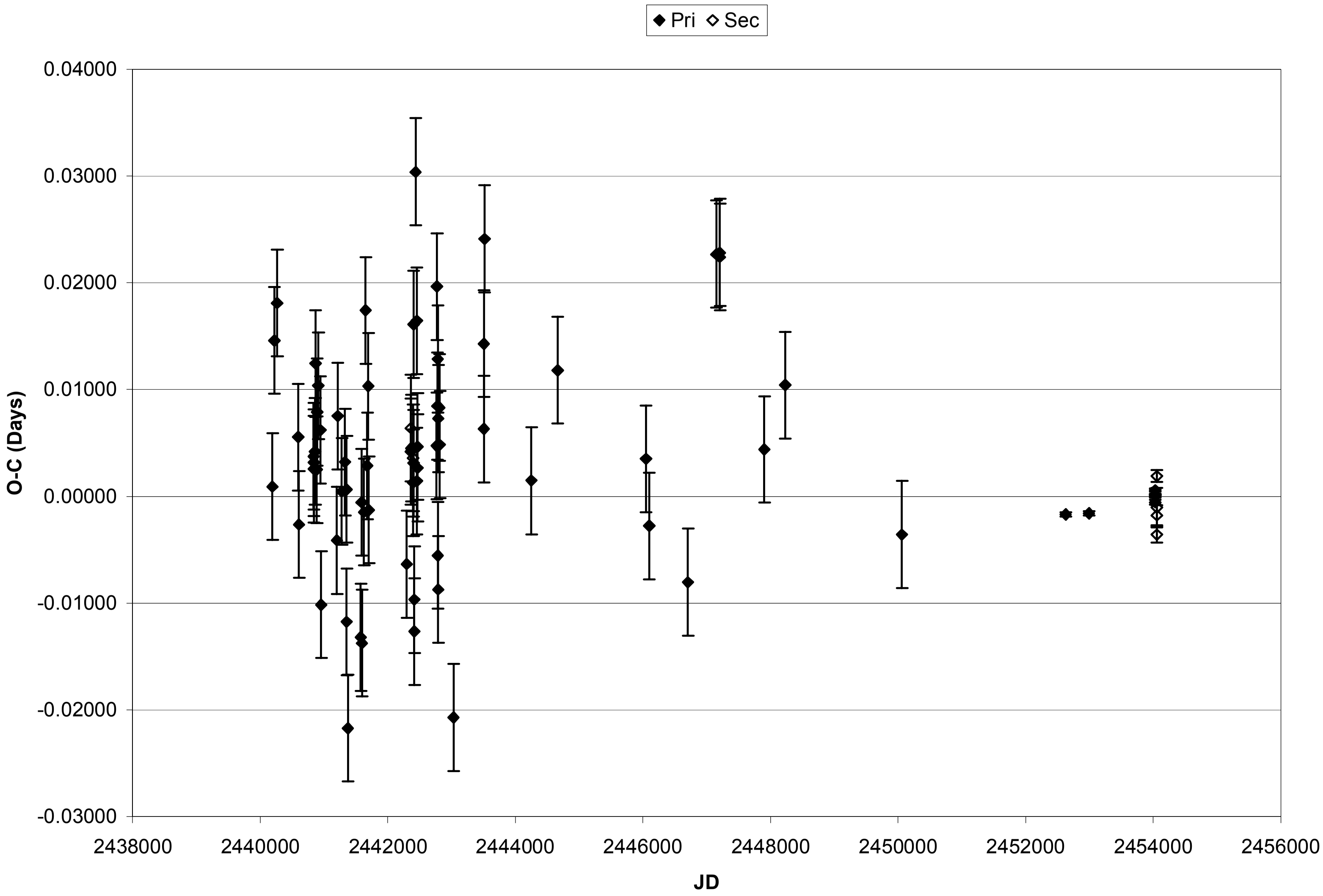

Figure 10-1: O-C Diagram for RU Eri

### 10.4 Light Curves and Modeling

All data was compiled into U, B, V, R, and I light curves and simultaneously solved using the ELC program. Although only about 80% of the curve was obtained, since only parts of the shoulders are missing an accurate model is able to be obtained. In order to set the scale of the system to an appropriate value, the separation was fixed to a value of $3.8R_{\odot}$ as found by Nakamura (1984). The values of mass ratio, inclination, each component's fill factor, and each components temperature were allowed to vary. The final solutions allowed for a grid resolution of 1600 points for each component and 360 points per model light curve. The observed light curves with model fits are shown in Figure 10-2.

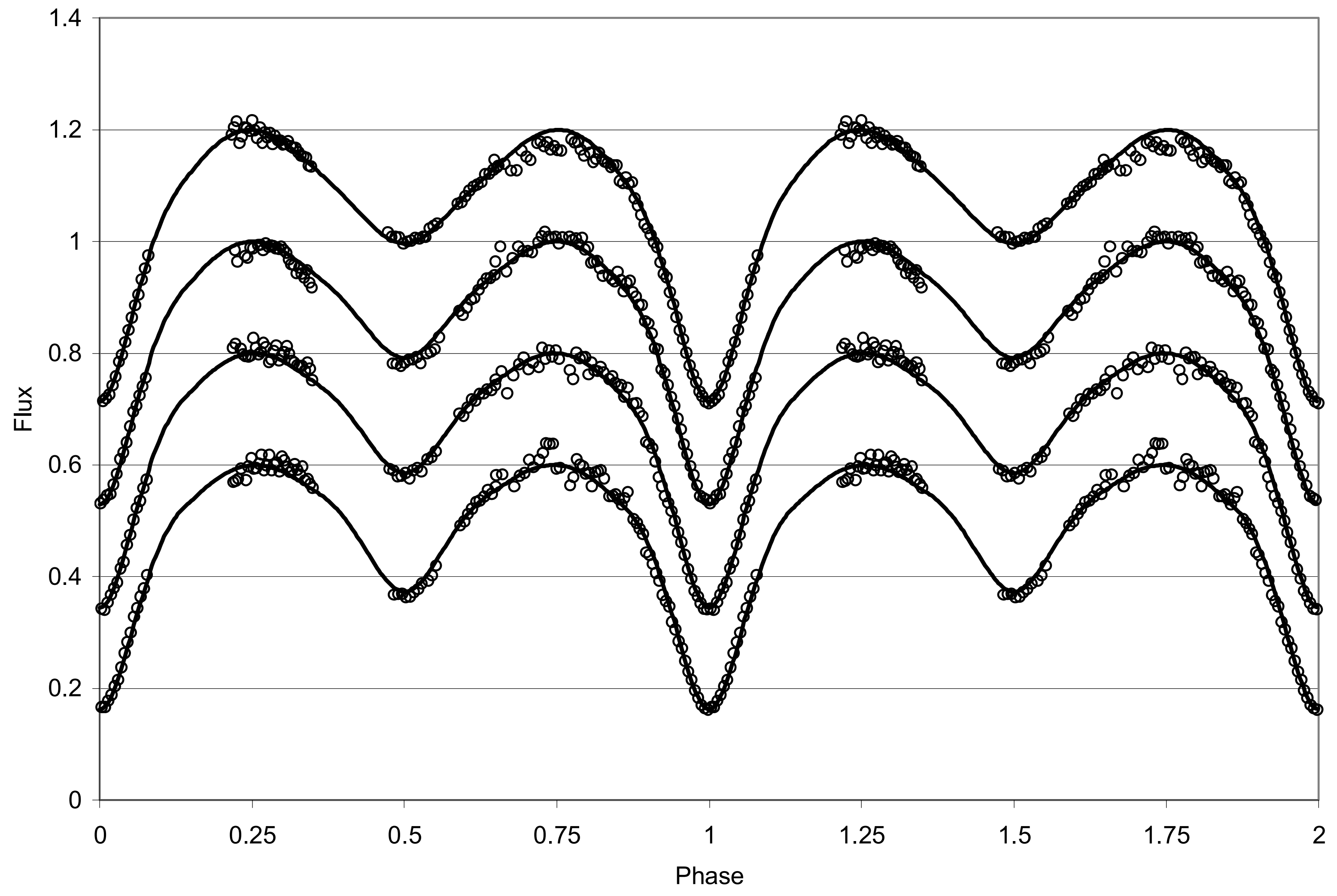

Figure 10-2: Top to bottom: Observed B, V, R, and I light curves (open circles) for RU Eri with model fits (solid lines).

### 10.5 Results

The results for the orbital parameters of RU Eri by modeling the composite light curves is shown in Table 10-2, and a geometrical model of the system shown in Figure 10-3.

Table 10-2: Orbital Parameter Solutions for RU Eri

| Parameter | Value |
|---|---|
| Inclination | 75.48° |
| Mass Ratio ($M_{pri}/M_{sec}$) | 1.51 |
| Primary Temperature | 8540K |
| Secondary Temperature | 5551K |
| Fill Factor of Primary | 0.906 |
| Fill Factor of Secondary | 0.845 |
| Fractional Radius of Primary | 0.415 |
| Fractional Radius of Secondary | 0.344 |

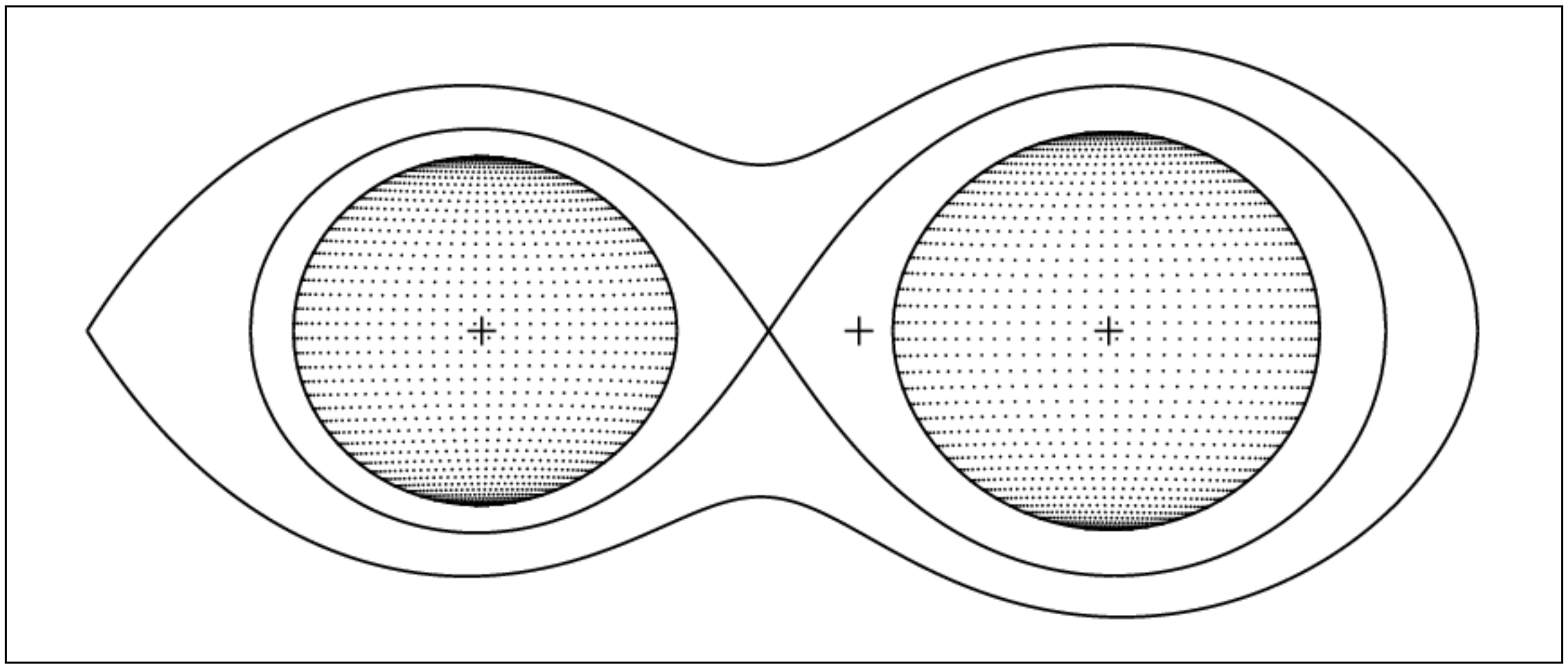

Figure 10-3: A geometrical model of the RU Eri system. The crosses represent the center of each star and the barycenter of the system. The inner solid line represents the Roche Lobe for each star. The primary, hotter star is on the right.

## 10.6 Discussion

Nakamura's (1984) solutions of i = 76.5°, q ($m_{pri}/m_{sec}$) = 1.84, $r_{pri}$ = 0.421, and $r_{sec}$ = 0.316 agree fairly well with this work's photometric analysis, and the derived temperatures are what would be expected for stars of the given masses. With respect to the difference in the derived mass ratios, as Nakamura (1984) only obtained a radial velocity curve for the primary component, there is no way to distinguish between which mass ratio is the more accurate value. A radial velocity curve of the secondary component, which should be obtainable with modern equipment, as well a completion of the light curve, is needed for a definitive analysis. With respect to the evolutionary scenario proposed in sections 5.6, 6.6, and 7.6, RU Eri would not yet have had any interaction between its components, but as the primary evolved would eventually head towards a UU Lyn type state on the way to an A-type W Uma system.

# XI. YY CETI

## 11.1 Background

YY Cet, [RA: 02 00 12, Dec: -18 12 28, $V_{mag} \approx 10$], is a semi-detached beta Lyrae type system with a period of approximately 0.79 days. It has had only one detailed analysis by McFarlane, King, and Hilditch in 1986. Based on a photoelectric V light curve and radial velocities for both components, McFarlane et al. (1986) derived $M_{pri} = 1.84M_{\odot}$, $M_{sec} = 0.94M_{\odot}$, $R_{pri} = 2.09R_{\odot}$, $R_{sec} = 1.63R_{\odot}$, $i = 87°$, $T_{pri} = 7500K$, and $T_{sec} = 5314K$. However, with only one color light curve there must be an appreciable measure of uncertainty in these measurements, and thus a multiple-color solution would be critical in confirming the above parameters.

## 11.2 Observations

Observations of YY Cet were taken with Emory Observatory's 24" telescope and a SBIG8 CCD camera cooled to -30°C on the nights of November $12^{th}$, $13^{th}$, and December $4^{th}$ of 2006 in B, V, R, and I filters. Differential photometry was performed via MaximDL with respect to GSC 5856-564. The reference star was not photometrically calibrated, so all measurements are in differential magnitudes. All times were corrected to HJD.

## 11.3 Minimum Timings and O-C Diagram

All observed times of minimum were determined via the method of Kwee-van Woerden, and shown in Table 11-1 with errors, employed filter, and type (primary or secondary eclipse). All previously published times of minima available were compiled and assigned a weight that was inversely proportional to its error. In cases where no error was given, a value of ±.005 days was assumed. A linear least-squared fit to all data was performed and a value of 0.79046274 days obtained for the period. The new ephemerides was thus calculated to be

$T_{pri}$ (HJD) = 2453381.52840069 + 0.79046274··E, where E is the epoch.

An O-C diagram created with the new ephemeris is shown in Figure 11-1.

Table 11-1: Observed Times of Minima for YY Cet

| $T_{min}$ (HJD) | Error (±) | Filter | Type |
|---|---|---|---|
| 2454052.631027 | 0.000564 | B | Pri |
| 2454052.631421 | 0.000728 | V | Pri |
| 2454052.631320 | 0.000584 | R | Pri |
| 2454052.631268 | 0.000273 | I | Pri |

### 11.4 Light Curves and Modeling

All data was compiled into B, V, R, and I light curves and simultaneously solved using the ELC program. Although only about 80% of the curve was obtained, since only parts of the shoulders are missing an accurate model is able to be obtained. In order to set the scale of the system to an appropriate value, the separation was fixed to a value of 5.0$R_\odot$ as found by McFarlane (1986). The values of mass ratio, inclination, each component's fill factor, and each components temperature were allowed to vary. The final solutions allowed for a grid resolution of 1600 points for each component and 360 points per model light curve. The observed light curves with model fits are shown in Figure 11-2.

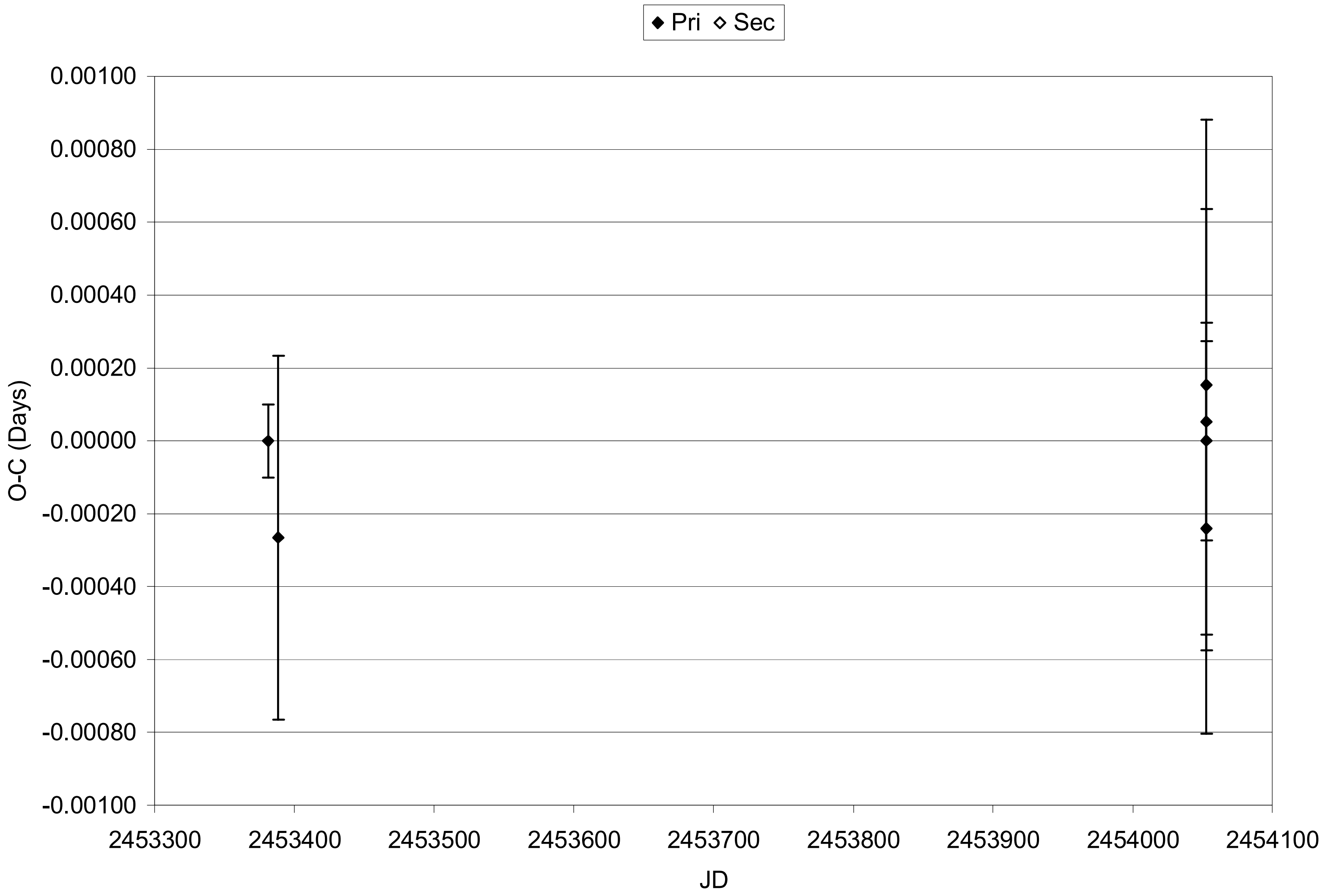

Figure 11-1: O-C Diagram for YY Cet

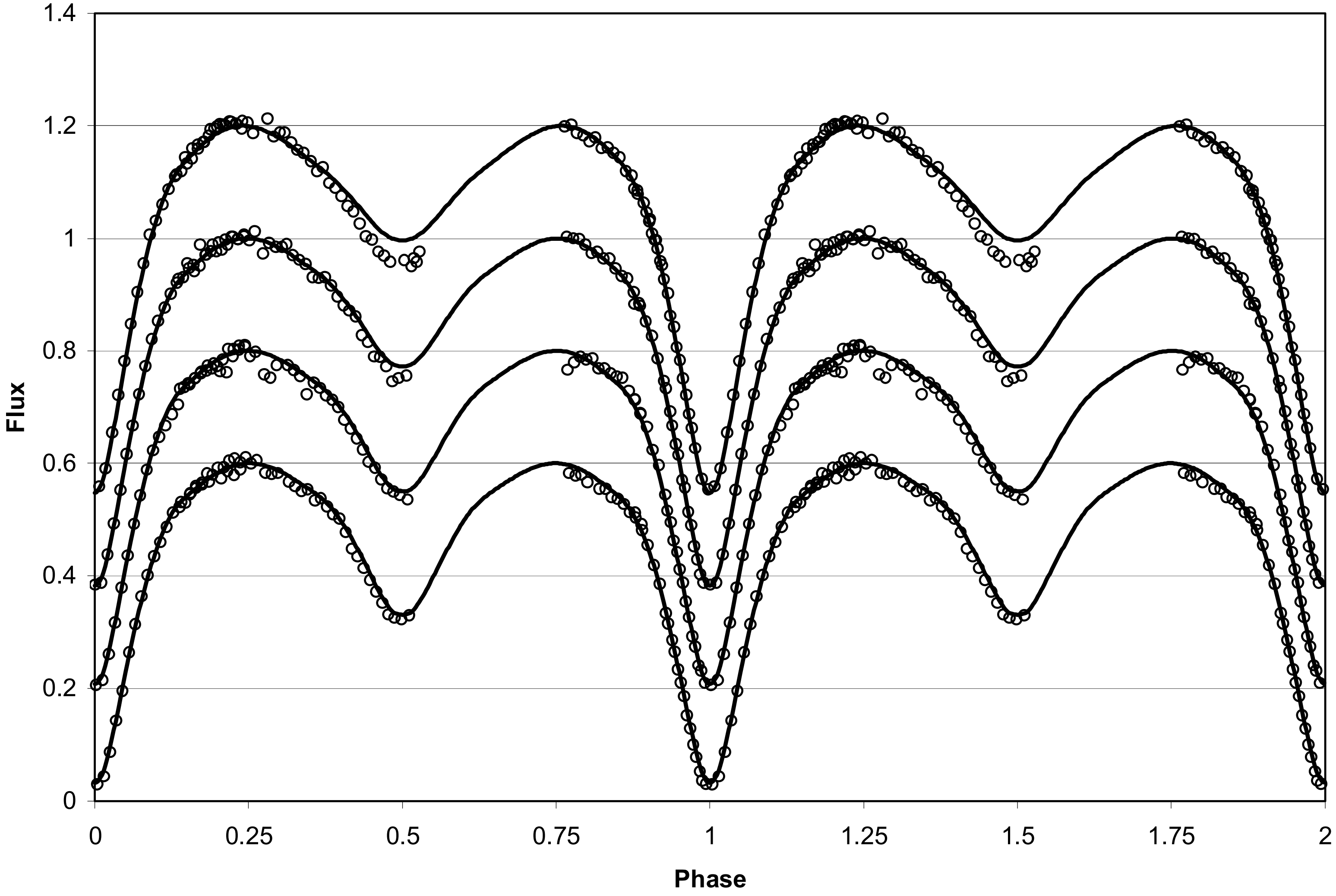

Figure 11-2: Top to bottom: Observed B, V, R, and I light curves (open circles) for YY Cet with model fits (solid lines).

## 11.5 Results

The results for the orbital parameters of YY Cet by modeling the composite light curves is shown in Table 11-2, and a geometrical model of the system shown in Figure 11-3.

Table 11-2: Orbital Parameter Solutions for YY Cet

| Parameter | Value |
|---|---|
| Inclination | 84.50° |
| Mass Ratio ($M_{pri}/M_{sec}$) | 1.89 |
| Primary Temperature | 6089K |
| Secondary Temperature | 4593K |
| Fill Factor of Primary | 0.955 |
| Fill Factor of Secondary | 0.872 |
| Fractional Radius of Primary | 0.428 |
| Fractional Radius of Secondary | 0.260 |

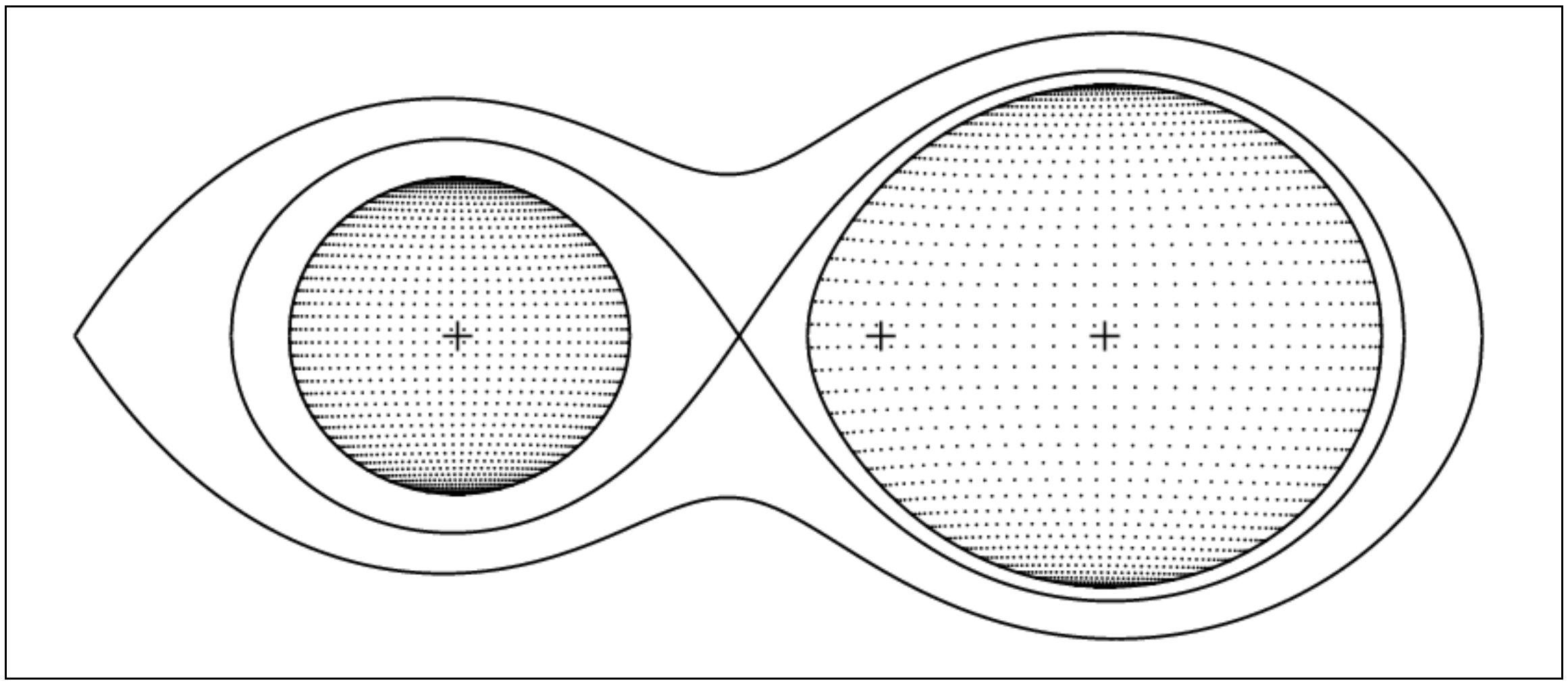

Figure 11-3: A geometrical model of the YY Cet system. The crosses represent the center of each star and the barycenter of the system. The inner solid line represents the Roche Lobe for each star. The primary, hotter star is on the right.

## 11.6 Discussion

McFarlane et al. (1986) derived i = 87°, and q ($m_{pri}/m_{sec}$) = 1.96, which agrees somewhat with the values derived in this analysis. The derived surface temperatures however are significantly different, with McFarlane et. al (1986) deriving $T_{pri}$ = 7500K and $T_{sec}$ = 5314K. As McFarlane (1986) derived spectral types of A8V and late G for the two components, and as the derived mass from radial velocities are in agreement, it is most likely that this study has underestimated the temperatures due to either inadequate observations of the secondary minima or errors in applying the model atmospheres.

The major conflict however is that McFarlane (1986) finds that the secondary component is just filling its Roche Lobe and the primary is not, while this study finds the primary component is almost filling its Roche lobe, and the secondary is well within its own. This may be a result of the difference in surface temperatures employed by the two solutions; however, this paper's solution is seemingly more intuitive. McFarlane (1986) suggests, based on his derivation of the secondary being over-sized, that at some point in the past the secondary was the primary component and evolved, transferring mass to what is now the primary. However, he does not elucidate how the components would have become so very widely separated again, or retained such large differences in surface temperature. Thus, if the solution presented in this paper holds, it would be much simpler as the components would have never before interacted via mass transfer, and would fall in the same category as RU Eri, on its way to a UU Lyn type state and then into an A-type W Uma system.

# XI. REFERENCES

Caton, D. 2005, IBVS 5595

Charbonneau, P. 1995, ApJS, 101, 309

Chen, K-Y., Williamon, R. M., Liu, Q., Yang, Y., Lu, L. 1985, AJ, 90, 1855

Erdem, A., Demircan, O., and Güre, M. 2001, A&A, 379, 878

Heckert, P. A. 1995, IBVS, 4224

Jeffery, C. S. 1984, MNRAS, 207, 323

Kopal, Z. 1959, Close Binary Systems (Wiley, New York)

Kholopov, P. N. 1985, General Catalogue of Variable Stars, 4th ed. (Nauka, Moscow), Vols. 1 and 2

Kholopov, P. N. 1987, General Catalogue of Variable Stars, 4th ed. (Nauka, Moscow), Vol. 3

Kjurkchieva, D.P., Marchev, D.V., and Ogloza, A. 2001, A&A, 378, 102

Kukarkin, B. V. 1968, IBVS, 311

Kwee, K. K., and van Woerden, H. 1956, B.A.N., 12, 464

Lucy, L.B. 1975, ApJ, 205, 208

McFarlane, T. M., King, D. J., Hilditch, R. W. 1986, 218, 159

Nakamura, Y., Yamasaki, A., and Kitamura, M. 1984, PASJ, 36, 277

Niarchos, P.G., Hoffman, M., and Duerbeck, H. W. 1996, A&AS, 117, 105

Orosz, J. A., and Hauschildt, P.H. 2000, A&A, 364, 265

Popper, D.M. 1971, ApJ, 169, 549

Pribulla, T. et al. 2000, A&A, 362, 169

Russo, G. 1982, Ap&SS, 81, 209

Sarma, M. B. K., and Sanwal, N.B. 1981, Ap&SS, 74, 41

Schmidt, E. G. 1991, AJ, 102, 1776

Shaw, J. S. 1994, MmSAI, 65, 95

Sowell, J. R. 2007, Private Communication

Sterne, T. E. 1939, MNRAS, 99, 451

Tucker, R., Sowell, J. R., and Williamon, R. M. 2006, AAS, 20915102T

Van Hamme, W. 1993, AJ, 106, 2096

Williamon, R. M. 1974, PASP, 86, 924

Williamon, R. M. 1975, AJ, 80, 976

Wilson, R. E. and Devinney, E. J. 1971, ApJ., 166, 605.

Yamasaki, A., Okazaki, A., and Kitamura, M. 1983, PASJ, 35, 131